\documentclass[10pt,a4paper]{article}
\usepackage{jheppub}
\usepackage{latexsym}
\usepackage{pstricks}
\usepackage{color}

\usepackage{enumerate}

\usepackage{color,amsmath,amssymb,exscale,epsfig}

\input epsf

\newcommand{\be}{\begin{equation}}
\newcommand{\ee}{\end{equation}}
\newcommand{\bea}{\begin{eqnarray}}
\newcommand{\eea}{\end{eqnarray}}
\newcommand{\lbl}[1]{\label{eq:#1}}
\newcommand{ \rf}[1]{(\ref{eq:#1})}
\newcommand{\Br}{\text{Br}}

\newcommand{\GeV}{\,\text{GeV}}

\newcommand{\lapprox}{%
\mathrel{%
\setbox0=\hbox{$<$}
\raise0.6ex\copy0\kern-\wd0
\lower0.65ex\hbox{$\sim$}
}}
\newcommand{\gapprox}{%
\mathrel{%
\setbox0=\hbox{$>$}
\raise0.6ex\copy0\kern-\wd0
\lower0.65ex\hbox{$\sim$}
}}

\def\theequation{\arabic{section}.\arabic{equation}}

\title{On the amplitudes for the CP-conserving $\boldsymbol{K^\pm(K_S)\to\pi^\pm(\pi^0)\ell^+\ell^-}$ rare decay modes }

\author[a]{Giancarlo D'Ambrosio,}
\emailAdd{gdambros@na.infn.it}
\affiliation[a]{INFN-Sezione di Napoli, Via Cintia, I-80126 Napoli, Italia}

\author[b]{David Greynat}
\emailAdd{david.greynat@gmail.com}
\affiliation[b]{No affiliation}

\author[c]{and Marc Knecht}
\emailAdd{knecht@cpt.univ-mrs.fr}
\affiliation[c]{Centre de Physique Th\'{e}orique, CNRS/Aix-Marseille Univ./Univ. du Sud Toulon-Var (UMR 7332)\\
CNRS-Luminy Case 907, F-13288 Marseille Cedex 9, France}

\date\today

\abstract{
The amplitudes for the rare decay modes  $K^\pm\to\pi^\pm\ell^+\ell^-$
and $K_S\to\pi^0\ell^+\ell^-$ are studied with the aim of obtaining predictions
for them, such as to enable the possibility to search for 
violations of lepton-flavour universality in the kaon sector. The issue
is first addressed from the perspective of the low-energy expansion, and a
two-loop representation of the corresponding form factors is constructed,
leaving as unknown quantities their values and slopes at vanishing momentum 
transfer. In a second step a phenomenological determination of the latter
is proposed. It consists of the contribution of the resonant two-pion state
in the $P$ wave, and of the leading short-distance contribution determined
by the operator-product expansion. The interpolation between the two energy 
regimes is described by an infinite tower of zero-width resonances matching 
the QCD short-distance behaviour. Finally, perspectives for future improvements 
in the theoretical understanding of these amplitudes are discussed.
}

\begin{document}

\vspace*{2cm}

\makeatletter
\def\@fpheader{\relax}
\makeatother
\maketitle

\section{Introduction}
\setcounter{equation}{0}

Rare kaon decays have been playing a crucial role 
in flavour physics, and more generally in particle physics. 
Since they proceed through strangeness-changing neutral currents,
they are quite suppressed in the standard model (SM), and thus
offer a window through which we might possibly catch a glimpse of new physics (NP). 
This research subject has been thoroughly reviewed in ref. \cite{Cirigliano:2011ny}, where
a detailed list of references can be found. It is again becoming quite timely nowadays,
thanks to experiments like NA62 at the CERN SPS, planning to collect, already in 2018, 
the data required in order to measure the decay rate of $K^+\to\pi^+ \nu \overline{\nu}$ 
with a precision of 10\,\% of its SM prediction 
{  \cite{Ceccucci:2018bnw,Ceccucci:2018lpb}}, 
and KOTO at J-PARC, aiming at measuring the decay rate of $K_L \to \pi^0 \nu \overline{\nu}$, at around the SM 
sensitivity in a first stage~\cite{Komatsubara:2012pn,Shiomi:2014sfa,KOTO}.

The study of these rare decay modes of the kaon is of particular interest in view of the
present situation in particle physics. Indeed, while no direct evidence for physics beyond the SM 
was found during the first run of the LHC, there are some interesting indirect hints for 
NP in the flavour sector, mainly from neutral-current decays of the $B$ meson into di-lepton channels.
Here, the very recent result on the $  R(K^*)$ ratio measured by LHCb~\cite{Aaij:2017vbb}  
has confirmed the deviations from $\mu/e$ universality in neutral currents of the $b\to s \ell^+ \ell^-$
type, $\ell=e,\mu$, already observed previously in  $B \to K \ell^+ \ell^-$ decays~\cite{Aaij:2014ora},
\begin{equation}
\label{RK}
R(K)=\frac{\Br[B\to K \mu^+\mu^-]}{\Br[B\to K e^+e^-]}=0.745^{+0.090}_{-0.074}\pm 0.036
.
\end{equation}
This result disagrees with the theoretically clean SM prediction 
$R_{\rm SM}(K)=1.0003 \pm 0.0001$~\cite{Bobeth:2007dw,Bordone:2016gaq} by $2.6\sigma$.
Further confirmation of this anomaly also comes from  
$B\to K^* \mu^+\mu^-$,  where an angular observable called $P_5^\prime$~\cite{Descotes-Genon:2013vna}, 
deviates from its SM value with a significance of $2$--$3\sigma$, depending on the way 
hadronic uncertainties  are evaluated \cite{Descotes-Genon:2013wba,Descotes-Genon:2014uoa,Altmannshofer:2014rta,Jager:2014rwa}. 
Typical NP explanations for the measured $B\to K^* \mu^+\mu^-$ observables require, for instance, $Z'$ 
vector bosons~\cite{Sierra:2014nqa,Heeck:2014qea,Gauld:2013qba,Buras:2013qja,Gauld:2013qja,Buras:2013dea,Altmannshofer:2014cfa,Glashow:2014iga,%
Crivellin:2015mga,Dorsner:2015mja,Omura:2015nja,Varzielas:2015joa,Crivellin:2015lwa,Niehoff:2015bfa,Sierra:2015fma,Crivellin:2015era,%
Celis:2015ara,Carmona:2015ena} 
or leptoquarks
\cite{Sakaki:2013bfa,Gripaios:2014tna,Becirevic:2015asa,Varzielas:2015iva,Alonso:2015sja,Calibbi:2015kma,Bauer:2015knc,%
Barbieri:2015yvd,Fajfer:2015ycq,Greljo:2015mma},
in order to generate non-SM contributions to current-current effective interactions like 
$(\bar{s}\gamma_\alpha P_L b)(\bar{\mu}\gamma^\alpha \mu)$.

The rare processes $K^\pm \to \pi^\pm \ell^+\ell^-$ and $K_S \to \pi^0  \ell^+\ell^-$ are
analogous to those mentioned in eq. \eqref{RK}, and appear thus as particularly suitable 
in order to uncover possible violations of lepton-flavour universality (LFUV) in the kaon sector
\cite{Crivellin:2016vjc}.  
The experimental programs of NA62 \cite{NA62_physics} (for charged kaon decays) and of LHCb 
\cite{Bediaga:2018lhg,Junior:2018odx} (for $K_S$ decays) offer
quite interesting prospects in this regard. Trying, in parallel, to improve our theoretical
understanding of these processes therefore constitutes a quite timely undertaking. It is
thus not surprising that efforts in this directions have also become part of the agenda of 
the lattice-QCD community \cite{Isidori:2005tv,Christ:2015aha,Christ:2016mmq}.

From the experimental point of view, the situation has evolved in a rather spectacular manner
during the last two decades [the present situation is briefly summarized in table \ref{table:exp}], 
especially as far as the two $K^\pm \to \pi^\pm \ell^+\ell^-$ channels, where the branching fractions 
are largest, are concerned. These branching fractions have been measured in refs.
\cite{Alliegro:1992pp,Adler:1997zk,Ma:1999uj,Park:2001cv} and the decays have subsequently been
studied more precisely with high statistics in refs.
\cite{Appel:1999yq,Batley:2009aa,Batley:2011zz}, these more recent experiments providing also 
detailed information on the decay distribution. The latest PDG averages for the branching
fractions are~\cite{Olive:2016xmw} 
\begin{align}
  \Br[K^+\to\pi^+e^+e^-] &= (3.00\pm 0.09)\times 10^{-7} \notag , \\
  \Br[K^+\to\pi^+\mu^+\mu^-] &= (9.4\pm 0.6)\times 10^{-8},
\end{align}
where the error on the muonic mode includes a scale factor $S=2.6$ due to the conflict with
the result reported upon in ref. \cite{Adler:1997zk}. 
These values lead to $R_{K^\pm} [{\rm PDG}]=0.313(71)$, where
\be
R_{K^\pm} \equiv \frac{{\rm Br}[K^\pm\to\pi^\pm\mu^+\mu^-]}{{\rm Br}[K^\pm\to\pi^\pm e^+e^-]}
.
\lbl{R_K}
\ee
Taking only the high-precision data collected by the NA48/2 Collaboration \cite{Batley:2009aa,Batley:2011zz}
into account, one obtains instead the almost two times more accurate result $R_{K^\pm} [{\rm NA48/2}]=0.309(43)$.
In the neutral-kaon sector, the observed decay rates are~\cite{Batley:2003mu,Batley:2004wg} 
\begin{align}
\label{K_S_rates}
  \Br[K_S\to\pi^0e^+e^-]_{m_{ee}>0.165\GeV} &= [3.0^{+1.5}_{-1.2} ({\rm stat})\pm 0.2 ({\rm syst}) ]\times 10^{-9} , \notag \\
  \Br[K_S\to\pi^0\mu^+\mu^-] &= [2.9^{+1.5}_{-1.2} ({\rm stat})\pm 0.2 ({\rm syst}) ]\times 10^{-9}.
\end{align}
These measurements have not yet reached the level of precision already available in the case of the charged kaon.

\begin{table}[ht]
\renewcommand{\arraystretch}{1.6}
\begin{center}
\begin{tabular}{|c|c|c|c|}
\hline
exp. & ~ ref. ~ & mode & number of events
\\
\hline\hline
BNL$^*$   & \cite{Alliegro:1992pp} & $K^+\to\pi^+ e^+ e^-$ & $\sim 500$   
\\
\hline
BNL-E865$^*$  &  \cite{Appel:1999yq} & $K^+\to\pi^+ e^+ e^-$ & $10\,300$   
\\
\hline
NA48/2$^*$    &  \cite{Batley:2009aa} & $K^\pm\to\pi^\pm e^+ e^-$ & $7\,263$
\\
\hline\hline
BNL-E787      &  \cite{Adler:1997zk}  &  $K^+\to\pi^+ \mu^+ \mu^-$ & $\sim 200$
\\
\hline
BNL-E865  &  \cite{Ma:1999uj}    & $K^+\to\pi^+ \mu^+ \mu^-$ & $\sim 400$
\\
\hline
FNAL-E871 &  \cite{Park:2001cv}  & $K^\pm\to\pi^\pm \mu^+ \mu^-$ & $\sim 100$
\\
\hline
NA48/2$^*$    &  \cite{Batley:2011zz} & $K^\pm\to\pi^\pm \mu^+ \mu^-$ & $3120$
\\
\hline\hline
NA48/1     &  \cite{Batley:2003mu} & $K_S\to\pi^0 e^+ e^-$  &  7      
\\
\hline\hline
NA48/1     &  \cite{Batley:2004wg} & $K_S\to\pi^0 \mu^+ \mu^-$  &  6         
\\
\hline
\end{tabular}\end{center}
\caption{Experimental situation concerning the decay modes
$K^\pm \to \pi^\pm \ell^+\ell^-$ and $K_S \to \pi^0 \ell^+\ell^-$.
The experiments marked with an asterisk also provide information
on the decay distribution.}
\label{table:exp}
\end{table}

At the theoretical level, the situation for the CP-conserving decays $K^\pm (K_{S}) \rightarrow \pi ^{\pm} 
(\pi ^{0})\ell ^{+}\ell ^{-}$ is less favourable  than for the $K\to\pi\nu{\bar\nu}$ modes.
Indeed, whereas the latter are dominated by short distances, the former are governed by the long-distance 
process $K\to\pi    \gamma ^*  \to \pi   \ell^+ \ell^-$ 
\cite{EPdR87,EPdR88,DAmbrosio:1998gur}, involving a weak form factor, $W_+(z)$ in the case 
of $K^\pm\to\pi^\pm \gamma ^*$, and $W_S(z)$ in the case of $K_S\to\pi^0 \gamma ^*$. These
form factors are given by the matrix elements
of the electromagnetic current between a kaon state and a pion state, in the presence of the weak
interactions, considered at first order in the Fermi constant. The corresponding momentum
transfer squared, the di-lepton invariant mass squared $s=zM_K^2$, is small enough, it ranges from 
$4 m_\ell^2$ to $(M_K - M_\pi)^2$, where $m_\ell$ is the mass of the charged lepton, so that these processes 
can be treated within the framework of chiral perturbation theory (ChPT) 
\cite{Wein79,GassLeut,GassLeut84,Gasser:1984gg}. Conservation of 
the electromagnetic current implies a vanishing contribution at lowest order, ${\mathcal O}(E^2)$ in the chiral expansion.
A non-vanishing form factor is generated at next-to-leading order, both by pion and kaon loops,
as well as by order ${\mathcal O}(E^4)$ counterterms. The main feature of this one-loop representation
of the form factor is the appearance of a unitarity cut on the positive real-$s$ axis, due to the two-pion
intermediate state [there is also a cut due to the opening of the $K{\bar K}$ channel, but { the} later is
located far enough from the kinematic region of interest, so that { its contribution} can be, for all practical
purposes, approximated by a polynomial]. The corresponding absorptive part (discontinuity) is 
given by the product of the pion electromagnetic form factor $F_V^\pi (s)$ with the $P$-wave projection 
$f_1^{K\pi\to\pi^+\pi^-} (s)$ of the weak $K \pi \to \pi^+\pi^-$ amplitude [$(K,\pi)$ stands for either 
$(K^\pm , \pi^\mp)$ or $(K_S, \pi^0)$], both taken at tree level.
These basic properties of the one-loop form factor already suggest some simple ways to improve upon this result 
and to collect some ${\mathcal O}(E^6)$ effects. Specifically, in  ref. \cite{DAmbrosio:1998gur}, the $O(E^2)$ 
$K \pi\to \pi^+\pi^-$ vertex was replaced by the phenomenologically well-known Dalitz-plot expansion of the
$K\rightarrow  \pi \pi^+\pi^-$ amplitude, and a slope was added to the pion form factor. Furthermore, slopes $b_{+,S}$
in the di-lepton invariant mass squared $s$, required by the data, but also by order ${\mathcal O} (E^6)$ counterterm 
contributions, were included in the weak form factors $W_{+,S}(z)$, in addition to the constant terms $a_{+,S}\propto W_{+,S}(0)$, 
already generated by the ${\mathcal O} (E^4)$ counterterms.  
There are, however, several reasons to go beyond the approximations considered by the authors of ref. \cite{ DAmbrosio:1998gur}:
\begin{itemize}
\item The parameters $a_{+,S}$ and $b_{+,S}$ appear merely as phenomenological constants, which have to be fixed from
the available data.  We will provide an update of the situation using the most
recent data on the decay distributions. At this stage, one may already observe that the value \rf{R_K}
obtained for $R_{K^\pm}$ is rather different from the one
quoted in ref. \cite{DAmbrosio:1998gur}, $R_{K^\pm} =0.167(36)$, and based on the older 
data from refs. \cite{Adler:1997zk} and \cite{Alliegro:1992pp}. This quite substantial increase in the value 
of $R_{K^\pm}$ confirms the prediction $R_{K^\pm} \gapprox 0.23$ made by the authors of ref. \cite{DAmbrosio:1998gur}. {  We} will also test
for potential effects of the data on { the value of the curvature} (quadratic slope) of the $K^\pm\rightarrow  \pi^\pm \pi^+\pi^-$ Dalitz plots. 
\item  Final-state rescattering effects are only partially taken into account by the parameterization
of the form factors $W_{+,S}(z)$ proposed in  ref. \cite{DAmbrosio:1998gur}. In particular, 
the one-loop $P$-wave projections of the weak $K \pi \to \pi^+\pi^-$ amplitudes also involve pion 
rescattering in the crossed channels. Since the two pions are in the $P$ wave,
these are not expected to be as large as, for instance, in the case of the $K \to \pi\pi$ amplitudes
\cite{Cirigliano:2011ny}, where the pions are in the $S$ wave.
Nevertheless, they could have an impact on 
the shape of the form factor, and hence on the determinations of $b_+$ and $b_S$. 

\item
The phenomenological values of  $a_+$ and $b_+$ are comparable in size, whereas, from
naive chiral counting, one would expect $b_+$ to be substantially smaller than $a_+$. A
theoretical explanation of $\vert b_+/a_+\vert\sim 1$ is still lacking.
\item 
Let us close this list with the most important point.
The potential detection of any manifestation of LFUV hinges on the possibility to obtain
independent information on $a_+$ and $b_+$, such as to be able to make a prediction for the ratio $R_{K^\pm}$.
This definitely requires to go beyond the low-energy expansion itself.
\end{itemize}

In order to address these issues, we improve the existing theoretical description of the form factors 
$W_+(z)$ and $W_S(z)$ in several ways:
 \begin{itemize} 
 \item 
We provide complete order ${\mathcal O} (E^6)$ representations of these form factors.
They are obtained as the result of a two-step recursive procedure. The first step reproduces the one-loop
representations of ref. \cite{EPdR87}. The second step includes, besides the effects already accounted for
by the representation of ref. \cite{DAmbrosio:1998gur}, all one-loop pion-pion rescattering effects,
both in the pion form factor and in the $K\to 3\pi $ vertex.
\item
We extend the previous representation of $W_+(z)$ beyond the low-energy domain upon using a simple parameterization
of the pion form factor that describes the data over a large energy range including the region of the $\rho (770)$ resonance. 
Since no data are available as far as the $K^+ \pi^- \to \pi^+\pi^-$ scattering amplitude is concerned [existing data 
only concern the decay region, in the form of Dalitz-plot expansions], we proceed by applying a unitarization procedure 
to the ${\mathcal O} (E^4)$ $P$-wave projection of this amplitude. These two items allow us to construct the absorptive 
part of the form factor $W_+(z)$ arising from two-pion intermediate states. The form factor
$W_+(z)$ itself is then recovered through an unsubtracted dispersion relation.
Upon comparing the behaviour of this ``exact'' form factor at small values of $z$ with its
low-energy expansion, we obtain sum rules for the parameters $a_+$ and $b_+$, which we can then
compare to their direct determinations from data.
\item 
We model contributions due to other intermediate states, which occur at higher thresholds, by
an infinite sum of single-resonance states, { with couplings} tuned such as to correctly reproduce the known high-energy
behaviour of the form factor in quantum chromo-dynamics (QCD).
\end{itemize}

{  For completeness, we should mention that there are other theoretical studies \cite{Friot:2004yr,Leskow16,Dubnickova08} of the form
factors $W(z)$ that go beyond the low-energy expansion. In ref. \cite{Friot:2004yr}, a two-parameter representation
for $W_{+}(z)$ is proposed, combining the lowest-order chiral expression with resonance exchanges.
In ref. \cite{Leskow16} the form factors are described within the large-$N_c$ treatment of weak hadronic matrix 
elements, see ref. \cite{Buras:1988kp} and references quoted therein, matching the quadratic cut-off 
dependence of the low-energy contribution with the logarithmic one from the short-distance part.
Finally, the authors of \cite{Dubnickova08} give a representation of the form factors in terms of meson form factors,
but the issue of the matching with the short-distance part is not addressed. For a recent account on the 
theoretical situation of the $K\to\pi\ell^+\ell^-$ amplitudes, see ref. \cite{Portoles:Kaon16}}.

The content of this paper is consequently organized as follows. In section \ref{section:theory},
we first discuss general features of the weak form factors $W_{+,S} (z)$ in QCD, and then address more
specifically their short-distance behaviour. Long-distance properties of the form factors are 
described in section \ref{section:long-dist}, where
we recall the results of the existing one-loop calculations \cite{EPdR87,AnantIm12}, as well
as the beyond-one-loop representation of ref. \cite{DAmbrosio:1998gur}. 
Phenomenological aspects linked to the determination of $a_+$ and $b_+$ from the recent
high-precision data are the subject of section \ref{section:data}. 
Starting from a discussion of the analyticity properties of the form factors, we next
construct (section \ref{sec:2_loop}) a two-loop representation that accounts for all $\pi\pi$ rescattering
effects at this order. We compare this representation of the form factors with the one used in ref. 
\cite{DAmbrosio:1998gur}, and discuss the impact on the determination of the parameters $a_{+,S}$ 
and $b_{+,S}$. 
Section \ref{sect:model} constitutes the main part of the paper as far as the issues
raised above are concerned. There, we construct a dispersive model for the form factor $W_+(z)$. 
The corresponding absorptive part includes the contribution, now not restricted to low energies, of the $\pi\pi$ 
intermediate states, and an infinite sum over {zero-width} resonances, with couplings chosen such
as to provide matching with the short-distance behaviour established previously
in section \ref{section:theory}. Assuming unsubtracted dispersion relations,
we evaluate the parameters $a_{+}$ and $b_{+}$ through the corresponding
sum rules. A final section summarizes our results and provides our conclusions,
as well as critical comments and some perspectives for the future. Numerical values
for various input parameters are { collected} in appendix \ref{app:num}. Technical details related to the content
of Sec. \ref{sec:2_loop} are { displayed} in appendix \ref{app:psi}.

\indent

\section{Theory overview}\label{section:theory}
\setcounter{equation}{0}

In this section, we provide background material concerning the theoretical aspects of the amplitudes 
describing the weak $K \pi \to \ell^+\ell^-$ transitions 
in the standard model, which will also allow us to set up the notation to be
used in this study. We first start with the description within the effective four-fermion effective theory below the
electroweak scale, and turn next to the short-distance behaviour of the form factors in three-flavour QCD.

\subsection{The structure of the form factors in three-flavour QCD}

In the standard model, weak non-leptonic ${\Delta S = 1} $ transitions of hadrons
are described, at a low-energy scale, i.e. below the charm threshold, and at first order in 
the Fermi constant $G_F$,
by an effective lagrangian ${\cal L}_{\Delta S = 1} (x)$ given by 
\cite{Gaillard:1974nj,Altarelli:1974exa,Witten:1976kx,Shifman:1975tn,Wise:1979at,Gilman:1979bc}   
\be
{\cal L}_{\Delta S = 1} (x) = - \frac{G_F}{\sqrt{2}} V_{us}^* V_{ud} 
\sum_{I=1}^{6}
C_I (\nu) Q_I (x ; \nu)
.
\lbl{eff_lag_non-lept}   
\ee
This expression involves the current-current four-quark operators $Q_1$ and $Q_2$, as well
as the QCD penguin operators $Q_3$,...$Q_6$. At lowest order in both the fine-structure 
constant $\alpha$ and the Fermi constant $G_F$, and from a low-energy (long distance) 
point of view, the two CP-conserving transitions
$K^\pm (k)  \to \pi^\pm (p) \ell^+ (p_{\ell^+}\!) \ell^- (p_{\ell^-}\!)$ and 
$K_S \to \pi^0 (p) \ell^+ (p_{\ell^+}\!) \ell^- (p_{\ell^-}\!)$
proceed through the one-photon exchange process $K \to \pi \gamma^*$, 
so that their amplitudes will involve the weak form factors 
$W_{+,S}(z)$, $z=s/M_K^2$, $s \equiv (k-p)^2 = (p_{\ell^+} + p_{\ell^-})^2$,
defined as [we use the notation 
$W(z)$ to stand for either $W_+(z)$ or $W_S(z)$, whenever
the discussion applies to both channels]
\be
\frac{W(z ; \nu)}{16 \pi^2} \times \left[ z (k + p)_\rho - \left(1 - \frac{M_\pi^2}{M_K^2} \right) (k - p)_\rho \right]
=
i \! \int \! d^4 x \,
\langle\pi(p)  \vert T \{  j_\rho  (0) {\cal L}_{\Delta S = 1} (x)  \} \vert K(k) \rangle
.
\lbl{Kpi_FF_1st-order}
\ee 
In terms of these form factors the amplitudes read
\be
{\cal A} (K\to\pi\ell^+\ell^-) =
- e^2 \times {\bar{\rm u}} (p_{\ell^-}\!) \gamma_\rho {\rm v} (p_{\ell^+}\!)
\times (k + p)^\rho \times
\frac{W(z ; \nu)}{16 \pi^2 M_K^2} 
.
\lbl{amplitudes}
\ee
Here, $j_\rho (x)$ stands for the electromagnetic current corresponding to
the three lightest quark flavours [$e_q$ denotes their charges in unit of the
positron charge], 
\be
j_\rho (x) = \sum_{q=u,d,s} e_q ({\bar q} \gamma_\rho q) (x)
.
\ee
Electroweak quark-penguin operators,
as well as the mixing of $Q_1$,...$Q_6$ with them, give contributions of
order ${\mathcal O} (\alpha^2 G_F)$ to the amplitudes, and will not be considered here. 

The four-quark operators
and their Wilson coefficients $C_I (\nu)$ depend on the QCD renormalization scale $\nu$,
and their evolution with respect to this scale is given by the renormalization-group
equations
\be
\nu \frac{d Q_I (x;\nu)}{d \nu} = - \sum_{J=1}^6 {\gamma}_{I,J} (\alpha_s) Q_J (x;\nu) 
,\qquad
\nu \frac{d C_I (\nu)}{d \nu} = + \sum_{J=1}^6 {\gamma}_{J,I} (\alpha_s) C_J (\nu)
.
\lbl{C_and_Q_RG}
\ee
The pure QCD anomalous-dimension matrix ${\hat \gamma} (\alpha_s)$ for three flavours, 
whose matrix elements ${\gamma}_{I,J} (\alpha_s)$ occur in
these equations, is known at leading-order (LO) and next-to-leading (NLO) accuracy,
\be
{\hat \gamma} (\alpha_s) = {\hat \gamma}^{(0)} \frac{\alpha_s}{4 \pi} + 
{\hat \gamma}^{(1)} \left( \frac{\alpha_s}{4 \pi} \right)^2 + \cdots
,
\ee
and the corresponding coefficients ${\gamma}^{(0)}_{I,J}$ and ${\gamma}^{(1)}_{I,J}$ 
can be found in refs. \cite{Gaillard:1974nj,Altarelli:1974exa,Shifman:1975tn,Gilman:1979bc} 
and \cite{Altarelli:1980fi,Buras:1989xd,Buras:1991jm,Buras:1992tc,Ciuchini:1993vr}, respectively.
At this stage, let us make two remarks.

First, we should stress that the form factors $W_{+;S}(z ; \nu)$
defined by eq. \rf{Kpi_FF_1st-order} depend actually on the QCD renormalization scale $\nu$. Indeed,
although the two composite operators involved in this definition are separately finite,
\be
\nu \frac{d}{d \nu} j_\rho (x) =0, \qquad \nu \frac{d}{d \nu} \sum_{I=1}^{6} C_I (\nu) Q_I (x ; \nu) = 0
,
\lbl{renorm}
\ee
the electromagnetic current because it is conserved, the lagrangian ${\cal L}_{\Delta S = 1}$ for 
strangeness changing non-leptonic transitions by explicit renormalization of the four-quark operators
and renormalization group evolution of the Wilson coefficients, as given in eq. \rf{C_and_Q_RG},
their time-ordered product is singular at short distances, and needs to be renormalized \cite{Dib:1988md,Dib:1988js}.
Indeed, in the short-distance analysis of the $K \to \pi \ell^+ \ell^-$ transitions,
two mixed quark-lepton four-fermion operators, having the factorized form of a quark
current times a leptonic current,
\be
Q_{7V} = ( {\bar s}^i d_i )_{V-A} ( {\bar \ell} \ell )_V
,\qquad
Q_{7A} = ( {\bar s}^i d_i )_{V-A} ( {\bar \ell} \ell )_A
,
\ee
are also encountered, and provide an additional contribution
to the effective lagrangian \cite{Witten:1976kx,Gilman:1979ud,Dib:1988md,Dib:1988js}:
\be
{\cal L}_{\Delta S = 1}^{\rm lept} (x) =
- \frac{G_{\rm F}}{\sqrt{2}} V_{us}^* V_{ud} 
\left[
C_{7V} (\nu)  Q_{7V} (x) + C_{7A} Q_{7A} (x)
\right]
.
\lbl{eff_lag_lept}
\ee
The presence of the axial-current operator $Q_{7A}$ reflects the
contribution of the $Z^0$ and of $W$-box diagrams, which also contribute to $Q_{7V}$.
More importantly, however, the operator $Q_{7V}$ as well receives
the contribution from the electromagnetic-penguin type of diagram,
with heavy quarks in the loop, as discussed in \cite{Dib:1988md,Dib:1988js}.
This contribution will induce, already at order ${\mathcal O}(\alpha_s^0)$,
i.e. even before QCD corrections are applied, a dependence on the renormalization scale $\nu$ 
in the Wilson coefficient $C_{7V}$, which can be expressed as 
\be
\nu \frac{d C_{7V} (\nu)}{d \nu} 
=
\frac{\alpha}{\alpha_s (\nu)} \sum_{J=1}^6 \gamma_{J,7} (\alpha_s) C_J (\nu) 
=
\frac{\alpha}{4 \pi} \sum_{J=1}^6 \left[ \gamma_{J,7}^{(0)} + {\cal O}(\alpha_s) \right] C_J (\nu) 
.
\ee
As revealed by their structures, the two operators $Q_{7V}$ and $Q_{7A}$ are finite,
and do not depend on the renormalization scale $\nu$, as long as only QCD corrections are considered,
which will be the case here. 
At lowest order, the coefficients $\gamma_{J,7}$ are given by \cite{Gilman:1979ud,Dib:1988md,Dib:1988js,Flynn:1988ve}  
\be
\gamma^{(0)}_{1,7} = -\frac{16}{9} N_c \quad 
\gamma^{(0)}_{2,7} = -\frac{16}{9}  \quad 
\gamma^{(0)}_{3,7} = +\frac{16}{9} \quad 
\gamma^{(0)}_{4,7} = +\frac{16}{9} N_c \quad 
\gamma^{(0)}_{5,7} = \gamma^{(0)}_{6,7} = 0
\lbl{gamma_{I,7}}
\ee
for three active flavours $u$, $d$, $s$. Their values at next-to-leading order are also available from 
the literature \cite{Buras:1994qa}. The Wilson coefficient $C_{7A}$  does not depend on $\nu$.
Moreover, it is proportional to the very small quantity $\tau \equiv - V_{td} V_{ts}^*/V_{ud} V_{us}^*$,
$\vert \tau \vert \sim 1.6 \cdot 10^{-3}$,
and the contribution of $Q_{7A}$ can be neglected as long as one does not discuss issues related to the violation of CP
or processes that are short-distance dominated.
Keeping only the  contribution from $Q_{7V}$, the amplitudes in eq. \rf{amplitudes} actually read
\bea
{\cal A} (K^\pm\to\pi^\pm\ell^+\ell^-) &=& - e^2 {\bar{\rm u}} (p_{\ell^-}\!) \gamma_\mu {\rm v} (p_{\ell^+}\!)\nonumber\\
&&\times (k+p)^\mu \left[ \frac{W_+(z ; \nu)}{16 \pi^2 M_K^2} \nonumber
+ 
\frac{G_{\rm F}}{\sqrt{2}} V_{us}^* V_{ud} \times \frac{C_{7V} (\nu)}{4 \pi \alpha} f_+^{K^\pm\pi^\mp} (z M_K^2)  \right]
,
\nonumber\\
\\
\lbl{amplitudes_full}
{\cal A} (K_S\to\pi^0\ell^+\ell^-) &=& - e^2 {\bar{\rm u}} (p_{\ell^-}\!) \gamma_\mu {\rm v} (p_{\ell^+}\!) \nonumber\\
&&\times (k+p)^\mu \left[ \frac{W_S(z ; \nu)}{16 \pi^2 M_K^2}  
-
\frac{G_{\rm F}}{\sqrt{2}} V_{us}^* V_{ud} \times \frac{C_{7V} (\nu)}{4 \pi \alpha} f_+^{K_S\pi^0} (z M_K^2)  \right]
.
\eea
They involve the form factors $f_+^{K\pi} (s)$, defined as [the plus sign applies
for $(K,\pi)=(K^\pm,\pi^\mp)$, and the minus sign for $(K,\pi)=(K_S,\pi^0)$, in
agreement with the phase convention chosen in eq. \rf{Kpi_FF_1st-order}]
\be
\langle \pi (p) \vert \left( {\bar s} \, \gamma_\mu  d \right) \! (0)  \vert K (k) \rangle
=
\pm [(k + p )_\mu f_+^{K\pi} (s) + (k-p)_\mu f_-^{K\pi} (s) ]
,
\ee
and normalized to $f_+^{K\pi} (0)=1$ in the limit where the up, down
and strange quarks have equal masses.
The contribution from $Q_{7A}$, which we have omitted,
would also involve the form factors $f_-^{K\pi} (s)$, but multiplied by the lepton mass
and the leptonic pseudoscalar density.
The amplitudes being observables, they should no longer depend
on the scale $\nu$. This means that the scale dependence coming from the short-distance
singularity of the form factor $W (z ; \nu)$ within three-flavour QCD
has to cancel the $\nu$-dependence of the Wilson coefficient $C_{7V} (\nu)$ generated at the electroweak
scale,
\be
\frac{d W_{+,S}(z)}{d \nu} = 0,
\quad
W_{+,S} (z) \equiv W_{+,S} (z ; \nu) \pm 16 \pi^2 M_K^2 \left( \frac{G_{\rm F}}{\sqrt{2}} V_{us}^* V_{ud} 
\right) \frac{C_{7V} (\nu)}{4\pi\alpha}  \,f_+^{K\pi} (z M_K^2)
.
\lbl{F_renorm}
\ee
We will come back to this issue in greater detail in section \ref{sect:short-dist} below.

Second, notice that throughout we are
working within the framework provided by pure QCD with three flavours of light quarks. 
Knowledge on the manner how three-flavour QCD is embedded into the full standard model
is not required. In particular, the four-quark operators evolve, at all scales, 
according to the three-flavour matrix of anomalous dimensions ${\hat \gamma} (\alpha_s)$
[truncated, in practice, at NLO].
Actually, from this point of view, the only input from the SM which is required,
besides, of course, the structure of the effective lagrangian itself,
as given by Eqs. \rf{eff_lag_non-lept} and \rf{eff_lag_lept},
are the "initial" values of the Wilson coefficient $C_I(\nu_0)$, $I=1,\ldots 6$, and $C_{7V} (\nu_0)$
at some scale $\nu_0$ slightly above 1 GeV, but in any case below the charm threshold.

\subsection{Properties of the form factors at high momentum transfer}\label{sect:short-dist}

\begin{figure}[ht]
\center\epsfig{figure=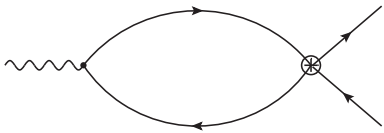,height=2.5cm}
\caption{The diagram corresponding to the leading short-distance contribution to the operator-product
expansion of the time-ordered product in eq. \rf{OPE}.
The external lines correspond to the insertion of the current $j_\mu (x)$
(wiggly line) and of the quark fields $d$ and ${\bar s}$. The circled vertex
materializes the insertion of a four-quark operator $Q_I$, $I=1,\ldots ,6$.\label{fig:OPE}}
\end{figure}

This subsection is devoted to the discussion of the behaviour of the 
form factors $W_{+,S} (z;\nu)$ at high momentum transfer, $z\to -\infty$,
within the framework of three-flavour QCD.
For this purpose, we consider the short-distance properties [in what follows,
$q_\mu$ is an Euclidean four-vector, $q^2 < 0$, whose components become all simultaneously
large] of the time-ordered product of the electromagnetic current with the
various four-quark operators $Q_I$ that appear in ${\cal L}_{\Delta S = 1}$,
\be
\lim_{q \to \infty} i \int d^4 x ~ e^{i q \cdot x} T \{  j^\mu \! (x) Q_I (0 ; \nu) \}
.
\lbl{OPE}
\ee
One can identify several contributions with the appropriate quantum numbers 
to the corresponding operator-product expansion (OPE).
The leading contribution occurs at order ${\cal O} (q^2)$, and consists of the term
\be
(q^\mu q^\nu - q^2 \eta^{\mu\nu}) [ {\bar s} \gamma_\nu (1 - \gamma_5) d]
.
\ee
It is shown in figure \ref{fig:OPE}, 
and corresponds to a perturbative intermediate state, made up
by a light-flavour quark-antiquark pair. In the absence of QCD corrections,
only the tree-level four-quark operators are involved. Gluonic corrections also contribute
to this class of short-distance behaviour. They will both renormalize the
four quark operators and build up the Wilson coefficient for the ${\cal O} (q^2)$
term in the OPE.
At order ${\cal O} (q)$, one encounters several possibilities, e.g. [$D_\tau$ denotes the
QCD covariant derivative]
\be
\{ (q^\tau \eta^{\mu\lambda} - q^\lambda \eta^{\mu\tau}) \, , \, \epsilon^{\mu\nu\lambda\tau} q_\nu \}
\times
\{ [{\bar s} \gamma_\lambda (1 - \gamma_5) (D_\tau d)] \, , \, [(D_\tau {\bar s}) \gamma_\lambda (1 - \gamma_5) d] \}
,
\ee
and so on.
We will only consider the leading contribution, at order ${\cal O} (q^2)$,
and without QCD corrections, although we briefly comment on the latter below.
%

%
In the case of the two current-current operators $Q_1$ and $Q_2$,
it is then enough to study the leading short-distance behaviour of
[$i,j,k,l$ denote colour indices]
\be
\lim_{q \to \infty} i \int d^4 x ~ e^{i q \cdot x}
T \{  ({\bar u} \gamma^\mu u) (x) [({\bar s}^i u_j)_{V - A} ({\bar u}^k d_l)_{V - A}]  (0) \}
\times T^{\,jl}_{{\! I} \ ik}
,
\ee
with $T^{\,jl}_{{\! 1} \ ik} = \delta_i^l \delta_k^j$ for $Q_1$, and $T^{\,jl}_{{\! 2} \ ik} = \delta_i^j \delta_k^l$ for $Q_2$.
The evaluation of the loop diagram is straightforward. With dimensional regularization, one finds
\bea
&&
\lim_{q \to \infty} i 
{\lefteqn
\int ~~~ d^D x ~ e^{i q \cdot x}
T \{ ({\bar u} \gamma^\mu u) (x)
[({\bar s}^i u_j)_{V - A} ({\bar u}^k d_l)_{V - A}] (0) \}  
}
\nonumber\\
&&=
\delta_j^k 
({\bar s}^{i \alpha} d_{l \beta})  
\left[
\gamma^\rho (1 - \gamma_5)\gamma^\sigma \gamma^\mu \gamma^\tau  \gamma_\rho (1 - \gamma_5)
\right]_{\alpha}^{\,\beta} \nonumber\\
&& \hspace*{4cm}\times
\frac{1}{4} \frac{1}{D-1} \left[ (2-D) q_\sigma q_\tau - q^2 \eta_{\sigma\tau} \right] 
\nu_{\mbox{\tiny MS}}^{4-D} J (q^2) + {\mathcal O}(q)
,
\lbl{OPE_Q1_and_Q2}
\eea
with [$\nu$ stands for the renormalization scale in the ${\overline{\rm MS}}$ scheme,
$\nu_{\mbox{\tiny MS}} \equiv \nu e^{\gamma_E/2}/\sqrt{4\pi}$, and $D=4-2\varepsilon$]
\be
\nu_{\mbox{\tiny MS}}^{4-D}
 J (q^2)  =
\frac{1}{(4 \pi)^{D/2}} \frac{[\Gamma (1-\varepsilon)]^2 \Gamma(\varepsilon)}{\Gamma(2-2\varepsilon)} 
\left(  \frac{\nu_{\mbox{\tiny MS}}^2}{-q^2} \right)^\varepsilon
=
\frac{1}{(4\pi)^2} \left[ \frac{1}{\varepsilon} + 2
- \ln\left(\frac{-q^2}{\nu^2}\right) \right]
+ {\mathcal O}(D-4) 
.
\ee
The discussion of the penguin operators $Q_3$ and $Q_4$ actually
turns out to be quite similar to the one for the operators $Q_1$ and $Q_2$. 
Indeed, from the expressions of these operators, 
\be
Q_3 = ( {\bar s}^i d_i )_{V-A} \sum_{q=u,d,s} ( {\bar q}^j q_j )_{V-A}
,
\quad
Q_4 = ( {\bar s}^i d_j )_{V-A} \sum_{q=u,d,s} ( {\bar q}^j q_i )_{V-A}
,
\lbl{Q3_Q4}
\ee
one sees that there are two ways to perform the contraction with the current $j^\mu$
shown in figure \ref{fig:OPE}. The first one consists in contracting with
the quark bilinears occurring in the sum over flavours. From the point of view of the colour
structures, this contraction corresponds to $T^{\,jl}_{{\! 1} \ ik}$ for $Q_3$,
and to $T^{\,jl}_{{\! 2} \ ik}$ for $Q_4$. However, since the quark masses are irrelevant
for the leading short-distance behaviour, this leads 
to the sum of three identical contributions, weighted by the corresponding quark charges.
But since $e_u + e_d + e_s = 0$, this weighted sum vanishes. There only remains to consider
the second possibility, where the contraction is done with the $d$ (or ${\bar s}$)
quark from the term in front of the flavour sum, and with the ${\bar d}$ (or $s$)
quark from the second (or third) term of this sum. But this amounts to the
same computation as before for $Q_1$ and for $Q_2$. The colour structure now corresponds
to $T^{\,jl}_{{\! 2} \ ik}$ for $Q_3$, and to $T^{\,jl}_{{\! 1} \ ik}$ for $Q_4$. Furthermore,
instead of multiplying by $e_u$, one multiplies by $e_d + e_s = - e_u$.
Finally, for the remaining operators $Q_5$ and $Q_6$ one obtains a vanishing result.
This is due to their $(V-A)\otimes (V + A)$ structure, which makes the factor arising 
from the Dirac matrices vanish, whether one uses the naive dimensional regularization (NDR)
\cite{Chanowitz:1979zu} or the 't~Hooft-Veltman (HV) \cite{tHooft:1972tcz,Breitenlohner:1977hr} scheme.
The product of Dirac matrices in eq. \rf{OPE_Q1_and_Q2} can also be simplified, but this time the result 
will depend on which scheme is being used. After minimal subtraction, one obtains
\begin{multline}
\lim_{q \to \infty} i
\int d^4 x ~ e^{i q \cdot x}
T \{ j^\mu (x)
Q_I (0) \}  
=
[ q^\mu q^\rho - q^2 \eta^{\mu\rho} ]
\times {\bar s} \gamma_\rho (1 - \gamma_5) d\\
\times \frac{1}{4 \pi} \left[ \xi_{00}^I - \xi_{01}^I  \ln\frac{- q^2}{\nu^2} \right]
+ \,
{\cal O} (q)
,\qquad~
\lbl{result_OPE_Q}
\end{multline}
with $\xi_{00}^{5,6} = \xi_{01}^{5,6} =0$,
\be
\xi_{01}^1 = \frac{1}{4 \pi} \, \frac{8}{9} \, N_c , \quad \xi_{01}^2 = \frac{1}{4 \pi} \, \frac{8}{9} ,
\quad \xi_{00}^1 =
\frac{1}{4 \pi} \times
\left\{
\begin{array}{l}
\frac{16}{27} \, N_c ~~ {\rm NDR} \\
\\
\frac{40}{27} \, N_c ~~ {\rm HV}
\end{array}
\right. 
,
\quad \xi_{00}^2 =
\frac{1}{4 \pi} \times
\left\{
\begin{array}{l}
\frac{16}{27}  ~~ {\rm NDR} \\
\\
\frac{40}{27} ~~ {\rm HV}
\end{array}
\right.
,
\lbl{xi_1_2}
\ee 
and
\be
\xi_{01}^3 = - \frac{1}{4 \pi} \, \frac{8}{9} , \quad \xi_{01}^4 = - \frac{1}{4 \pi} \, \frac{8}{9} \, N_c ,
\quad \xi_{00}^3 =
\frac{1}{4 \pi} \times
\left\{
\begin{array}{l}
- \frac{16}{27} ~~ {\rm NDR} \\
\\
- \frac{40}{27} ~~ {\rm HV}
\end{array}
\right. 
,
\quad \xi_{00}^4 =
\frac{1}{4 \pi} \times
\left\{
\begin{array}{l}
- \frac{16}{27} \, N_c  ~~ {\rm NDR} \\
\\
- \frac{40}{27} \, N_c ~~ {\rm HV}
\end{array}
\right.
.
\lbl{xi_3_4}
\ee
Note that in the presence of QCD corrections, this expression becomes
\begin{multline}
\lim_{q \to \infty} i
\int ~~~ d^4 x ~ e^{i q \cdot x}
T \{ j^\mu (x)
Q_I (0; \nu) \}  
=
[ q^\mu q^\rho - q^2 \eta^{\mu\rho} ] \times [{\bar s} \gamma_\rho (1 - \gamma_5) d](0)\\
\times \frac{1}{4 \pi} \xi_I (\alpha_s ; \nu^2/q^2)
+ \,
{\cal O} (q)
,\qquad~
\lbl{result_OPE_QCD}
\end{multline}
where the general form of the Wilson coefficient reads
\be
\xi_I (\alpha_s \, ; \nu^2/q^2) =
\sum_{p\ge 0} \sum_{r =0}^{p+1}
\xi^I_{pr}
\alpha_s^p (\nu) \ln^r (-\nu^2/q^2)
.
\ee
The result of the OPE with the complete lagrangian \rf{eff_lag_non-lept} then writes as
\bea
\lim_{q \to \infty} i
{\lefteqn
\int ~~~ d^4 x ~ e^{i q \cdot x}
T \{ j^\mu (x)
{\cal L}_{\Delta S = 1} (0) \}  
}
&&=   
\left(- \frac{G_{\rm F}}{\sqrt{2}} V_{us}^* V_{ud} \right)
[ q^\mu q^\rho - q^2 \eta^{\mu\rho} ] \times {\bar s} \gamma_\rho (1 - \gamma_5) d
\nonumber\\
&&
\ 
\times \frac{1}{4 \pi} \sum_{I=1}^4 C_I (\nu) \, \xi_I (\alpha_s \, ; \nu^2/q^2)
+ \,
{\cal O} (q)
.\qquad~
\lbl{result_OPE}
\eea

The preceding short-distance analysis tells us that the OPE of the time-ordered product 
of the (three-flavour) electromagnetic current with ${\cal L}_{\Delta S = 1}$ is dominated
by the same axial-current operator ${\bar s} \gamma_\rho (1 - \gamma_5) d$ that also
appears in the expression of $Q_{7V}$. Furthermore, it also exhibits a short-distance singularity, 
which is renormalized by minimal subtraction,
leaving over a dependence on the ${\overline{\rm MS}}$ renormalization scale $\nu$. 
At the level of the dimensionally regularized form factor itself, this translates into
\be
\lim_{z\to - \infty} W_{+,S}(z ; \nu) = \pm 16 \pi^2 M_K^2 \left( \frac{G_{\rm F}}{\sqrt{2}} V_{us}^* V_{ud} \right) 
\times \frac{f_+^{K\pi} (z M_K^2)}{4 \pi} \sum_{I=1}^4 C_I (\nu) \, \xi_I (\alpha_s \, ; \nu^2/zM_K^2)
\lbl{high-s}
\ee
and implies that the scale-dependence of the form factors is given by
\be
\nu \frac{dW_{+,S}(z ; \nu)}{d \nu} = \pm 16 \pi^2 M_K^2 \left( \frac{G_{\rm F}}{\sqrt{2}} V_{us}^* V_{ud} \right) 
\times \frac{f_+^{K\pi} (z M_K^2)}{4 \pi} 
\nu \frac{d }{d \nu} \sum_I C_I (\nu) \, \xi_I (\alpha_s \, ; \nu^2/zM_K^2)
.
\ee
Turning now toward the condition \rf{F_renorm}, we find that it is
indeed satisfied at order ${\mathcal O} (\alpha_s^0)$, 
where it reads
\begin{multline}
\nu \frac{d}{d \nu} \left[ 
\frac{W_{+,S}(z ; \nu)}{16 \pi^2 M_K^2}  
\pm
\frac{G_{\rm F}}{\sqrt{2}} V_{us}^* V_{ud} \times \frac{C_{7V} (\nu)}{4 \pi \alpha} {  f_+^{K\pi}} (z M_K^2)\right] \\
= 
\pm \left( \frac{G_{\rm F}}{\sqrt{2}} V_{us}^* V_{ud} \right) \frac{  f_+^{K\pi} (z M_K^2)}{4 \pi}
\sum_{I=1}^6 \left( \frac{\gamma^{(0)}_{I,7}}{4 \pi} + 2 \xi_{01}^I \right) C_I (\nu)
,
\end{multline}
since the
comparison with eq. \rf{gamma_{I,7}} shows that
\be
\xi_{01}^I = - \frac{1}{4\pi} \cdot \frac{1}{2} \gamma^{(0)}_{I,7}
.
\lbl{xi_01}
\ee
One may actually turn the preceding argument around, and, starting with
the requirement that eq. \rf{F_renorm} be satisfied, obtain information
on the  Wilson coefficients $\xi_I (\alpha_s \, ; \nu^2/s)$ on
the right-hand sides of Eqs. \rf{result_OPE}. Indeed, from
\begin{align}
\label{eq:xiho}
\nu \frac{d}{d \nu} \xi_I (\alpha_s \, ; \nu^2/s) &=
2 \sum_{p\ge 0} \sum_{r =0}^{p+1}
p \xi^I_{pr} \beta(\alpha_s) \alpha_s^{p-1} (\nu) \ln^r (-\nu^2/s) \nonumber\\
&\;\;+ 2 \sum_{p\ge 0} \sum_{r =0}^{p+1}
r \xi^I_{pr} \alpha_s^{p} (\nu) \ln^{r-1} (-\nu^2/s)
,
\end{align}
one deduces that { 
\bea
&&
\hspace{-0.35cm}
{\lefteqn{
\nu \frac{d}{d \nu} \left[ 
\frac{W_{+,S}(z ; \nu)}{16 \pi^2 M_K^2}  
\pm
\frac{G_{\rm F}}{\sqrt{2}} V_{us}^* V_{ud} \times \frac{C_{7V} (\nu)}{4 \pi \alpha} f_+^{K\pi} (z M_K^2)\right] 
=   }}
\nonumber\\
&& 
\hspace{-0.35cm}
=
\pm \left( \frac{G_{\rm F}}{\sqrt{2}} V_{us}^* V_{ud} \right) \frac{f_+^{K\pi} (z M_K^2)}{4 \pi}
\sum_{I=1}^6 \left( \frac{\gamma^{(0)}_{I,7}}{4 \pi} + 2 \xi_{01}^I \right) C_I (\nu)
\nonumber\\
&&
\pm \left( \frac{G_{\rm F}}{\sqrt{2}} V_{us}^* V_{ud} \right) \frac{f_+^{K\pi} (z M_K^2)}{4 \pi} 
\alpha_s (\nu)  \nonumber\\
&&\hspace*{1cm}\times \sum_{I=1}^6 C_I (\nu)\left\{
\frac{\gamma^{(1)}_{I,7}}{(4 \pi)^2} + 2 \xi_{11}^I +  \sum_{J=1}^6 \frac{\gamma^{(0)}_{IJ}}{4 \pi} \xi_{00}^J
+ \ln \left(-\frac{\nu^2}{s} \right)
\left[
4 \xi_{12}^I + \sum_{J=1}^6 \frac{\gamma^{(0)}_{IJ}}{4 \pi} \xi_{01}^J 
\right]
\right\}
+ {\mathcal O} (\alpha_s^2)
.
\nonumber\\
&&
\eea
}
The combination on the left-hand side will thus be scale independent
at NLO provided that in addition to \rf{xi_01} the relations
\be
\xi_{11}^I =
- \frac{1}{2} \frac{\gamma^{(1)}_{I,7V}}{(4 \pi)^2} 
- \frac{1}{2}  \sum_{J=1}^6 \frac{\gamma^{(0)}_{IJ}}{4 \pi} \xi_{00}^J
\qquad
\xi_{12}^I = - \frac{1}{4} \sum_{J=1}^6 \frac{\gamma^{(0)}_{IJ}}{4 \pi} \xi_{01}^J 
\ee
hold. Performing the calculation gives, for instance,
\be
\xi_{12}^I = \frac{1}{(4\pi)^2} \, \frac{4}{27} \, \left(N_c - \frac{1}{N_c} \right)
\times
\left(
0 \, , \, - 8 \, , \, + 11 \, , \, N_f \, , \, 0 \, , \, N_f
\right),
\ee
where the number of active flavours is $N_f=3$. The coefficients $\xi_{p0}^I$
are not constrained by this type of argument. They can only be determined by an explicit 
calculation of QCD corrections to the diagram of figure \ref{fig:OPE}, which we 
will however not attempt to perform here.

Before closing this section, 
let us briefly leave our three-flavour world, and consider how the preceding discussion is modified
in the presence of a fourth quark flavour, which corresponds to the situation
considered in lattice calculations \cite{Isidori:2005tv,Christ:2015aha,Christ:2016mmq}.
For $m_c < \nu < m_b$, the effective lagrangian reads
\bea
{\cal L}_{\Delta S = 1} & = &
- \frac{G_{\rm F}}{\sqrt{2}} 
V_{us}^* V_{ud}
\bigg\{
(1 - \tau) \Big[ C_1 (\nu) \left( Q_1  (x ; \nu) - Q_1^{(c)}  (x ; \nu) \right) + 
C_2 (\nu) \left( Q_2  (x ; \nu) - Q_2^{(c)}  (x ; \nu) \right)  \Big]
\nonumber\\
&&\qquad\qquad\qquad
+ \,
\tau
\sum_{I=1}^6 C_I (\nu) Q_I  (x ; \nu) 
\bigg\}
,\qquad
\tau \equiv - \frac{V_{ts}^* V_{td}}{V_{us}^* V_{ud}}
.
\eea
The operators appearing in this expression are
\be
Q_1^{(q)} = ( {\bar s}^i q_j )_{V-A} ( {\bar q}^j d_i )_{V-A}
\ \quad
Q_2^{(q)} =  ( {\bar s}^i q_i )_{V-A} ( {\bar q}^j d_j )_{V-A}
,
\ee
with the understanding that $Q_1^{(u)} \equiv Q_1$, $Q_2^{(u)} \equiv Q_2$
and, moreover, that in the QCD penguin operators the sums over the quarks $q$,
as in eq. \rf{Q3_Q4}, span the whole range of still active flavours, i.e. $q=u,d,s,c$ in the present case.
Likewise, the electromagnetic current is the one corresponding to four active flavours.
Repeating the same exercise as before, one now obtains [recall that all quarks are considered to be massless]
\bea
&&
\lim_{q \to \infty} i
{\lefteqn
\int ~~~ d^4 x ~ e^{i q \cdot x}
T \{ j_\rho (x)
{\cal L}_{\Delta S = 1} (0) \}  
}
=
\nonumber\\
&&
\ =  
\left(- \frac{G_{\rm F}}{\sqrt{2}} 
V_{us}^* V_{ud} \right)
({\bar s}^{i \alpha} d_{l \beta}) (0)
\left[
\gamma^\rho (1 - \gamma_5)\gamma^\sigma \gamma^\mu \gamma^\tau  \gamma_\rho (1 - \gamma_5)
\right]_{\alpha}^{\,\beta} 
\times 
\frac{1}{4} \frac{1}{D-1} \left[ (2-D) q_\sigma q_\tau - q^2 \eta_{\sigma\tau} \right] \nonumber\\
&&\qquad
\times
\nu_{\mbox{\tiny MS}}^{4-D} J (q^2)\bigg\{ T^{\,jl}_{{\! 1} \ ik} \left[ C_1 (\nu) (e_u - (1-\tau)e_c) 
+ \tau C_3 (\nu) (e_u + e_d + e_s + e_c) + \tau C_4 (\nu) (e_d + e_s) \right]
\nonumber\\
&&\qquad
\ + \, T^{\,jl}_{{\! 2} \ ik} \left[ C_2 (\nu) (e_u - (1-\tau)e_c) 
+ \tau C_3 (\nu) (e_d + e_s) + \tau C_4 (\nu) (e_u + e_d + e_s + e_c) \right]
\bigg\}
\delta_j^k 
+ {\mathcal O}(q)
\nonumber\\
&&
\ =  
\left(- \frac{G_{\rm F}}{\sqrt{2}} 
V_{us}^* V_{ud} \right)
({\bar s}^{i \alpha} d_{l \beta}) (0)
\left[
\gamma^\rho (1 - \gamma_5)\gamma^\sigma \gamma^\mu \gamma^\tau  \gamma_\rho (1 - \gamma_5)
\right]_{\alpha}^{\,\beta} 
\times 
\frac{1}{6} \frac{1}{D-1} \left[ (2-D) q_\sigma q_\tau - q^2 \eta_{\sigma\tau} \right] \nonumber\\
&&\qquad
\times \nu_{\mbox{\tiny MS}}^{4-D} J (q^2)\tau
\left\{ \left[ C_1 (\nu) + C_3 (\nu) - C_4 (\nu) \right] T^{\,jl}_{{\! 1} \ ik} 
+ \left[ C_2 (\nu) - C_3 (\mu) + C_4 (\nu) \right] T^{\,jl}_{{\! 2} \ ik}  \right\}
\delta_j^k 
+ {\mathcal O}(q).\nonumber\\
&&
\lbl{OPE_result_charm}
\eea
The scale dependence is now proportional to the small quantity $\tau$, defined after eq. \rf{gamma_{I,7}}. 
To the extent that CP-violating effects are not considered, these contributions can be safely omitted for all
practical purposes. But strictly speaking, the absence of a short-distance singularity in the form factor holds
only, in the case of four active flavours, within this approximation. 
This picture is consistent with the fact that, above the charm threshold, the Wilson coefficient 
$C_{7V}$ of the operator $Q_{7V}$ is also proportional to $\tau$ \cite{Buras:1994qa}.

\section{Low-energy expansion of the weak form factors}\label{section:long-dist}
\setcounter{equation}{0}

In this section, we recall the properties of the form factors $W_{+,S}(z)$
from the point of view of their low-energy expansion. We first give their expressions
at one loop and discuss some of their properties. We consider next the regime where $z\ll 1$ ($s \ll M_K^2$), 
so that the pion loops remain as the only sources of non-analyticity. In this regime, we then
briefly discuss the description of the form factors proposed in ref. \cite{DAmbrosio:1998gur}.

\subsection{The form factors at one loop}\label{subsec:1loop}

Since the $K^\pm\to\pi^\pm\gamma^*$ and $K_S\to\pi^0\gamma^*$ transition form factors vanish at lowest 
order \cite{EPdR87}, the low-energy expansions of the form factors $W_{+,S}(z)$ start at order one loop.
These one-loop expressions have been computed 
in ref. \cite{EPdR87} as far as the octet component is concerned. The contribution of the 
27-plet has only been worked out more recently, in ref. \cite{AnantIm12}. 
In a notation slightly different from the one used in these references, the resulting expressions read
\begin{multline}
W_{+,S;{\rm 1L}} (z) = G_F M_K^2 a_{+,S}^{\mbox{\scriptsize CT-1L}}
+ \frac{8\pi^2}{3} \, \frac{M_K^2}{M_\pi^2}  \bigg\{
\alpha_{+,S}^{\rm tree} \bigg[ \frac{z - 4 \frac{M_\pi^2}{M_K^2}}{z} \, {\bar J}_{\pi\pi} (z M_K^2) - \frac{1}{48 \pi^2} \bigg]
\\
+ {\tilde\alpha}_{+,S}^{\rm tree} \bigg[ \frac{z - 4}{z} \, {\bar J}_{KK} (z M_K^2) - \frac{1}{48 \pi^2} \bigg]
\bigg\}
.
\lbl{W_1loop}
\end{multline}
The function ${\bar J}_{\pi\pi}$ is defined as ($\lambda(x,y,z)$ denotes the K\"{a}llen function) 
\be
{\bar J}_{ab} (s) = \frac{s}{16\pi^2} \int_{s_{ab}}^\infty
\frac{dx}{x} \, \frac{1}{x-s-i0} \, \frac{\lambda^{1/2}_{ab} (x)}{x}
,\quad
s_{ab} \equiv (M_a + M_b)^2
, \quad \lambda_{ab} (s) \equiv \lambda (s, M_a^2 , M_b^2)
,
\lbl{Jbar_def}
\ee
for $M_a=M_b=M_\pi$. Explicit expressions for this function are given in refs. \cite{GassLeut,Gasser:1984gg}.
The first term in eq. \rf{W_1loop} gathers the contributions from the ${\cal O}(E^4)$ counterterms,
\be
a_{+,S}^{\mbox{\scriptsize CT-1L}} 
= \left( - \frac{1}{\sqrt{2}} V_{us}^* V_{ud} \right) \left( g_8 w_{+,S}^{(8)} + g_{27} w_{+,S}^{(27)} \right)  
,
\ee
which can be expressed in terms of low-energy constants defined in ref. \cite{Ecker:1992de} and chiral 
logarithms [$\mu_\chi$ denotes the renormalization scale in the effective low-energy theory] as
\bea
w_+^{(8)} &=& \frac{64 \pi^2}{3} \left[ N_{14}^r (\mu_\chi) - N_{15}^r (\mu_\chi) + 3 L_9^r (\mu_\chi) \right]
- \frac{1}{6} \ln \frac{M_\pi^2}{\mu_\chi^2} - \frac{1}{6} \ln \frac{M_K^2}{\mu_\chi^2}
,
\nonumber\\
\lbl{w_8}
\\
w_S^{(8)} &=& - \frac{32 \pi^2}{3} \left[ 2 N_{14}^r (\mu_\chi) + N_{15}^r (\mu_\chi) \right]
+ \frac{1}{3} \ln \frac{M_K^2}{\mu_\chi^2}
,
\nonumber
\eea 
and
\bea
w_+^{(27)} &=& - \frac{32}{3} \pi^2 \left[ R_{13}^r (\mu_\chi) - 2 R_{15}^r (\mu_\chi) - 4 L_9^r (\mu_\chi) \right]
+ \frac{13}{18} \ln \frac{M_\pi^2}{\mu_\chi^2} + \frac{13}{18} \ln \frac{M_K^2}{\mu_\chi^2}
,
\nonumber\\
\lbl{w_27}
\\
w_S^{(27)} &=& \frac{32}{3} \pi^2 R_{13}^r (\mu_\chi) 
- \frac{5}{18} \frac{3 M_K^2 - 2 M_\pi^2}{M_K^2 - M_\pi^2} \ln \frac{M_\pi^2}{\mu_\chi^2}
- \frac{1}{18} \frac{M_K^2 - 6 M_\pi^2}{M_K^2 - M_\pi^2} \ln \frac{M_K^2}{\mu_\chi^2}
.
\nonumber
\eea
These combinations of counterterms and chiral logarithms do not depend
on $\mu_\chi$: $dw^{(8,27)}_{+,S}/d\mu_\chi = 0$.
The second and third terms in the expression \rf{W_1loop} come from the
pion and kaon loops, respectively. $\alpha_{+}^{\rm tree}$ ($\alpha_{S}^{\rm tree}$)
corresponds to the $P$-wave projection of the amplitude of the reaction 
$K^+\pi^-\to\pi^+\pi^-$ ($K_S\pi^0\to\pi^+\pi^-$) at order ${\cal O}(E^2)$ 
in the low-energy expansion, and coincides with the linear slope, evaluated at 
lowest order. of the Dalitz plot of the decay $K^\pm\to\pi^\pm\pi^+\pi^-$
($K_S\to\pi^+\pi^-\pi^0$).
The interpretation of the quantity ${\tilde\alpha}_{+}^{\rm tree}$ 
(${\tilde\alpha}_{S}^{\rm tree}$) is similar, but in terms of the amplitude
for the reaction $K^+\pi^-\to K^+ K^-$ ($K_S\pi^0\to K^+ K^-$). In this
case there is no decay region, and ${\tilde\alpha}_{+}^{\rm tree}$ and
${\tilde\alpha}_{S}^{\rm tree}$ rather correspond to subtheshold parameters
in the expansions of the amplitudes for $K^+\pi^-\to K^+ K^-$ and $K_S\pi^0\to K^+ K^-$
around the centres of their respective Dalitz plots [$s=M_K^2+M_\pi^2/3$, $t=u$]. 
In terms of the two constants $g_8$ and $g_{27}$ that describe the $K\to\pi\pi$
amplitudes at lowest order \cite{Cirigliano:2011ny}, one has [for the numerical values, 
see table \ref{tab:numerics} in appendix \ref{app:num}]
\be
{\alpha}_{+}^{\rm tree} = {\tilde\alpha}_{+}^{\rm tree}  =
\left( - \frac{G_F}{\sqrt{2}} V_{us}^* V_{ud} \right) M_\pi^2 \left( g_8 - \frac{13}{3} g_{27} \right)
= -0.36 M_\pi^2 G_F = - 8.16 \cdot 10^{-8}
,
\lbl{alpha_plus}
\ee
and
\bea
{\alpha}_{S}^{\rm tree} &=& \left( - \frac{G_F}{\sqrt{2}} V_{us}^* V_{ud} \right) M_\pi^2 
\left( \frac{5}{3} g_{27}\times \frac{3 M_K^2 - 2 M_\pi^2}{M_K^2 - M_\pi^2} \right)
= -0.24 M_\pi^2 G_F = - 5.36 \cdot 10^{-8}
, 
\nonumber\\
\lbl{alpha_S}
\\
{\tilde\alpha}_{S}^{\rm tree} &=& \left( - \frac{G_F}{\sqrt{2}} V_{us}^* V_{ud} \right) M_\pi^2 
\left( - 2 g_8 + \frac{g_{27}}{3} \times \frac{M_K^2 - 6 M_\pi^2}{M_K^2 - M_\pi^2} \right)
= +1.11 M_\pi^2 G_F = + 25.15 \cdot 10^{-8}
.
\nonumber
\eea
As already mentioned in the introduction, the physical region of the decays
$K^\pm(K_S)\to\pi^\pm(\pi^0)\ell^+\ell^-$ corresponds to $4 m_\ell^2/M_K^2 \le z \lapprox 0.5$.
In this range of values the function ${\bar J}_{KK} (z M_K^2)$ is well described
by its Taylor expansion at $z=0$. Performing this expansion in the contribution
from the kaon loops, and keeping terms at most linear in $z$, allows to rewrite the 
one-loop form factors as
\be
W_{+,S;{\rm 1L}}(z) = G_F M_K^2 (a_{+,S}^{\rm 1L} + b_{+,S}^{\rm 1L} z)
\\+ \frac{8\pi^2}{3} \, \frac{M_K^2}{M_\pi^2} \alpha_{+,S}^{\rm tree}
\left[ \frac{z - 4 \frac{M_\pi^2}{M_K^2}}{z} \, {\bar J}_{\pi\pi} (z M_K^2) + \frac{1}{24 \pi^2} \right]
.
\lbl{1_loop_expanded}
\ee
Notice that $a_{+,S}^{\mbox{\scriptsize 1L}}$ and $b_{+,S}^{\mbox{\scriptsize 1L}}$ correspond
to the values of the form factors and of their slope, respectively, at $z=0$,
\be
G_F M_K^2 a_{+,S}^{\mbox{\scriptsize 1L}}  = W_{+,S;{\rm 1L}} (0) , \quad
G_F M_K^2 b_{+,S}^{\mbox{\scriptsize 1L}}  = W'_{+,S;{\rm 1L}} (0)
- \frac{1}{60} \, 
\left( \frac{M_K^2}{M_\pi^2} \right)^2 \alpha_{+,S}
.
\lbl{intrinsic_a_b_1loop}
\ee
The one-loop expressions  of the constants 
$a_{+,S}^{\mbox{\scriptsize 1L}}$ and $b_{+,S}^{\mbox{\scriptsize 1L}}$ that result from this simple exercise read
\begin{align}
&a_{+,S}^{\mbox{\scriptsize 1L}} = a_{+,S}^{\mbox{\scriptsize CT-1L}} + a_{+,S}^{\mbox{\scriptsize 1L};\pi\pi}
+ a_{+,S}^{\mbox{\scriptsize 1L};{\bar K}K} \nonumber\\
& a_{+,S}^{\mbox{\scriptsize 1L};\pi\pi} = - \frac{{\alpha}_{+,S}^{\rm tree}}{6 M_\pi^2 G_F},
\ a_{+,S}^{\mbox{\scriptsize 1L};{\bar K}K} = - \frac{{\tilde\alpha}_{+,S}^{\rm tree}}{6 M_\pi^2 G_F}
\, \lbl{a_1L} \\
&b_{+,S}^{\mbox{\scriptsize 1L}} = b_{+,S}^{\mbox{\scriptsize 1L};{\bar K}K} 
= \frac{{\tilde\alpha}_{+,S}^{\rm tree}}{60 M_\pi^2 G_F} .\nonumber
\end{align}
Predicting the values of $a_{+,S}^{\rm 1L}$ from their one-loop expressions
require some knowledge of the ${\cal O}(E^4)$ counterterms, in particular 
$N_{14}$ and $N_{15}$, which occur in the dominant octet part. Several proposal 
have been made \cite{Ecker:1992de,Ecker:1990in,Pich:1990mw,DAmbrosio:1997ctq,Cappiello:2011re}
in order to extend the determination of the low-energy 
constants through resonance saturation in the strong sector \cite{Ecker:1989yg,Ecker:1988te}
to the weak sector. Unfortunately, these extensions involve unknown parameters
[see also the discussion at the beginning of Sec. \ref{sec:resonances} below],
thus requiring additional assumptions. This unfortunate situation introduces an uncontrolled 
model dependence in the predictions that can be made within this framework.
The prospects for a phenomenological determination of some of these constants from (real or virtual) 
radiative kaon decays have been discussed in the recent account \cite{Cappiello:2017ilv}. 
In this context, we also recall the ``octet dominance hypothesis" discussed in 
refs. \cite{EPdR87,Friot:2004yr},
which corresponds to the assumption that the (scale independent) combination
$N_{15}^r (\mu_\chi) - 2 L_9^r (\mu_\chi)$ vanishes, thus predicting 
$w_+^{(8)} + w_S^{(8)} \sim\frac{1}{6} \ln \frac{M_K^2}{M_\pi^2}$.

\begin{figure}[b]
\begin{center}
\includegraphics[width=8.5cm]{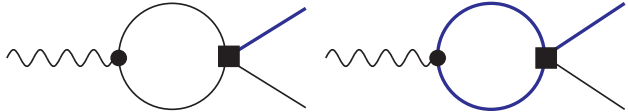} 
\end{center}
\caption{The Feynman diagrams contributing to the non-trivial analyticity
properties of the form factors $W_{+,S;{\rm 1L}}(z)$
at one loop.
The external wiggly line corresponds to the insertion of the 
electromagnetic current $j_\mu (x)$.
Thin lines represent charged pions, and { thicker blue} lines are kaons [only
charged kaons contribute in the loop]. The circular blob stands for a
strong vertex, and the filled square for the weak vertex.
\label{fig:1loop}}
\end{figure}

\indent

This brief description of the structure of the one-loop expressions of the form factors
[we refer the reader to the original articles for further details] brings us to a few remarks:
\begin{itemize}
\item
As already noticed in ref. \cite{EPdR87}, there are substantial differences in the
structures of the two form factors $W_{+;{\rm 1L}}(z)$ and $W_{S;{\rm 1L}}(z)$,
and therefore also between $a_+^{\mbox{\scriptsize 1L}}$ and $a_S^{\mbox{\scriptsize 1L}}$.
In particular, pion loops are suppressed by the $\Delta I = 1/2$ rule in $W_{S;{\rm 1L}}(z)$,
but kaon loops are about three times as important as in $W_{+;{\rm 1L}}(z)$ { (in absolute value)}.
\item
Besides providing tiny slopes $b_{+,S}^{\mbox{\scriptsize 1L}}$, the kaon loops 
also contribute to $a_{+,S}^{\mbox{\scriptsize 1L}}$, see Eqs. \rf{alpha_plus}-\rf{a_1L}:
$a_{+}^{\mbox{\scriptsize 1L};{\bar K}K} = +0.06$, $a_{S}^{\mbox{\scriptsize 1L};{\bar K}K} = -0.19$.
These contributions are proportional to ${\tilde\alpha}_{+}^{\rm tree}$ and
${\tilde\alpha}_{S}^{\rm tree}$, respectively. Higher-order corrections will
move ${\tilde\alpha}_{+,S}^{\rm tree}$ to their phenomenological values ${\tilde\alpha}_{+,S}$,
which are however not known, see the discussion after eq. \rf{w_27}. Assuming, for the sake
of illustration, that this change is of about the same size as the change from ${\alpha}_{+,S}^{\rm tree}$
to ${\alpha}_{+,S}$, i.e. ${\tilde\alpha}_{+}\sim 2.5 {\tilde\alpha}_{+}^{\rm tree}$ and 
${\tilde\alpha}_{S}\sim 1.3 {\tilde\alpha}_{S}^{\rm tree}$ [see Eqs. (\ref{eq:alpha_beta_values}) 
and \rf{alphaS_betaS_values} below],
we would conclude that the kaon loops could contribute to $a_+$ ($a_S$) at the level of $\sim +0.15$ ($\sim -0.25$).
Of course, this argument is at best indicative of the fact that contributions from kaon loops
to $a_{+,S}$, in contrast to $b_{+,S}$, could be substantial. Besides, higher-order corrections will also
produce other effects on the form factors than the simple replacement of  ${\tilde\alpha}_{+,S}^{\rm tree}$
by ${\tilde\alpha}_{+,S}$, but their impact on $a_{+,S}$ are however more difficult to estimate without an 
explicit computation.
\item
It is also possible, and of interest for the sequel, to look at the one-loop form factors in terms 
of their analyticity properties, which follow from the structure of the Feynman diagrams shown
in figure \ref{fig:1loop}. These properties can be expressed in the form of a once-subtracted dispersion 
relation 
\be
W_{+,S;{\rm 1L}} (z) = G_F M_K^2 a_{+,S}^{\rm 1L} + \frac{z M_K^2}{\pi}
\int_0^\infty \frac{dx}{x} \frac{{\rm Im}\,W_{+,S;{\rm 1L}} (x/M_K^2)}{x-zM_K^2-i0}
,
\ee
with the discontinuity along the positive real-$z$ axis provided by one-loop unitarity.
Since at this order the only intermediate states are $\pi^+\pi^-$ and $K^+ K^-$,
see figure \ref{fig:1loop}, it reads 
\begin{align}
&\frac{1}{16\pi^2 M_K^2} {\rm Im}\,W_{+,S;{\rm 1L}} (s/M_K^2) = \nonumber\\
& \theta (s - 4 M_\pi^2) \times  \frac{s-4M_\pi^2}{s} \lambda^{-1/2}_{K\pi} (s) \times
F_V^\pi (s)\vert_{{\cal O}(E^2)} \times f_1^{K\pi \to \pi^+\pi^-} (s) \vert_{{\cal O}(E^2)} \nonumber\\
&+ \, \theta (s - 4 M_K^2) \times 
\frac{s-4M_K^2}{s} \lambda^{-1/2}_{K\pi} (s) \times F_V^{K^\pm} (s)\vert_{{\cal O}(E^2)} \times f_1^{K\pi \to K^+ K^-} (s) \vert_{{\cal O}(E^2)},
\lbl{ImW_1loop}
\end{align}
where $F_V^\pi (s)\vert_{{\cal O}(E^2)} = F_V^{K^\pm} (s)\vert_{{\cal O}(E^2)} = 1$ 
[neutral kaons do not contribute at this order since $F_V^{K^0} (s)\vert_{{\cal O}(E^2)} = 0$],
$\lambda_{K\pi} (s) \equiv \lambda (s , M_K^2 , M_\pi^2)$ [cf. eq. \rf{Jbar_def}], and
\bea
f_1^{K\pi \to \pi^+\pi^-} (s) \vert_{{\cal O}(E^2)} &=& \frac{\alpha_{+,S}^{\rm tree}}{96\pi M_\pi^2}
\times \lambda_{K\pi}^{1/2}(s) \sqrt{1-\frac{4M_\pi^2}{s}}
,
\nonumber\\
f_1^{K\pi \to K^+ K^-} (s) \vert_{{\cal O}(E^2)} &=& \frac{{\tilde\alpha}_{+,S}^{\rm tree}}{96\pi M_\pi^2}
\times \lambda_{K\pi}^{1/2}(s) \sqrt{1-\frac{4 M_K^2}{s}}
,
\eea
are the $P$-wave projections mentioned after eq. \rf{w_27}.
The expressions of ${\rm Im}\,W_{+,S;{\rm 1L}} (s/M_K^2)$ also
show that one subtraction is necessary, but sufficient, in order to
obtain a convergent dispersive integral. Performing this integration leads to
\begin{multline}
W_{+,S;{\rm 1L}} (z) = G_F M_K^2 a_{+,S}^{\mbox{\scriptsize 1L}}
+ \frac{8\pi^2}{3} \, \frac{M_K^2}{M_\pi^2}  \bigg\{
\alpha_{+,S}^{\rm tree} \bigg[ \frac{z - 4 \frac{M_\pi^2}{M_K^2}}{z} \, {\bar J}_{\pi\pi} (z M_K^2) + \frac{1}{24 \pi^2} \bigg]
\\
+ {\tilde\alpha}_{+,S}^{\rm tree} \bigg[ \frac{z - 4}{z} \, {\bar J}_{KK} (z M_K^2) + \frac{1}{24 \pi^2} \bigg]
\bigg\}
.
\lbl{W_1loop_disp}
\end{multline}
Expanding the contribution from the kaon loop as before allows to cast this expression into the form 
given in eq. \rf{1_loop_expanded}. Some information is lost
within this dispersive approach, namely the way how $a_{+,S}^{\mbox{\scriptsize 1L}}$,
which appears here as a mere subtraction constant, decomposes into 
its various components, as displayed in eq. \rf{a_1L}. The issue raised in the 
preceding item of this list, having to do with local contributions from the kaon 
loops, can therefore not be addressed within this dispersive approach.
\item
On the other hand, extending the absorptive parts beyond their lowest-order
expressions \rf{ImW_1loop} provides a starting point for establishing the
structure of the form factors at two loops [beyond two loops, also intermediate
states with more than two mesons need to be considered]. Furthermore, one
may restrict the contributions to the absorptive parts to two-pion states
from the outset, and include the other two-mesons states [${\bar K}K$, but
also, for instance $K\pi$] into the subtraction polynomial, which becomes
a first-order polynomial in $z$ at two loops. The explicit construction of
the two-loop form factors along these lines will be the subject of Section
\ref{sec:2_loop}.
\end{itemize}

\subsection{The form factors beyond one loop}

We now come to the expressions of the form factors considered in ref. \cite{DAmbrosio:1998gur}.
They go beyond the one-loop expressions discussed in the preceding subsection, and read 
\begin{multline}
W_{+,S;{\rm b1L}}(z) = G_F M_K^2 (a_{+,S} + b_{+,S} z)
\\+ \frac{8\pi^2}{3} \, \frac{M_K^2}{M_\pi^2} \left[ \alpha_{+,S} + \beta_{+,S} \frac{M_K^2}{M_\pi^2} (z-z_0)  \right]
\left( 1 + \frac{M_K^2}{M_V^2}  z \right) 
\left[ \frac{z - 4 \frac{M_\pi^2}{M_K^2}}{z} \, {\bar J}_{\pi\pi} (z M_K^2) + \frac{1}{24 \pi^2} \right]
,
\lbl{W_b1L_+S}
\end{multline}
with $z_0 = 1/3 + M_\pi^2/M_K^2 \equiv s_0/M_K^2$ and $M_V=775.5$ MeV.
Their structure is easy to understand from the representation \rf{1_loop_expanded}
of the one-loop form factors, as already described in the introduction. The pion
form factor, which reduces to unity at lowest order in the low-energy expansion,
is extended by the addition of a term linear in $z$, with a slope given by
$M_K^2/M_V^2$. The same kind of modification is also implemented in the 
$K^+\pi^-\to\pi^+\pi^-$ ($K_S\pi^0\to\pi^+\pi^-$) vertex: the parameters $\alpha_+$
and $\beta_+$ ($\alpha_S$ and $\beta_S$) correspond to the linear slope and to one of the
quadratic slopes (curvatures), respectively, describing the Dalitz plot 
of the decay $K^\pm\to\pi^\pm\pi^+\pi^-$ ($K_S\to\to\pi^+\pi^-\pi^0$). 
Expressed in terms of the parameters
introduced in refs. \cite{Holstein:1969uc,Li:1979wa,Kambor:1991ah}, 
$\alpha_+$ and $\beta_+$ read
\be
\alpha_+  = \beta_1 - \frac{\beta_3}{2} + \sqrt{3} \gamma_3
,\quad 
\beta_+ = 2 \left( \xi_1 + \xi_3 - \xi^\prime_3  \right)
.
\ee
Using the values determined from data in ref. \cite{Bijnens:2002vr},
one obtains [errors have been added quadratically, and the numerical
values we use are collected in appendix \ref{app:num} for the reader's convenience]
\be
\alpha_+ = -20.84(74) \cdot 10^{-8}
,\quad
\beta_+ = -2.88(1.08) \cdot 10^{-8}
.
\label{eq:alpha_beta_values}
\ee
These values are quite similar to 
the ones given in ref. \cite{DAmbrosio:1998gur} 
[the authors of this last reference use the notation of ref. \cite{DAmbrosio:1994vba};
ref. \cite{Maiani:1997ui} gives the correspondence between the two sets of parameters] 
and used in order to analyze the data in refs. \cite{Appel:1999yq,Batley:2009aa,Batley:2011zz}.
Notice that the linear slope $\alpha_+$ is determined quite accurately, at less than 4\%,
whereas the curvature $\beta_+$ is only known with a relative precision of about 35\%.
Likewise, the parameters $\alpha_S$ and $\beta_S$ read
\be  
\alpha_S = - \frac{4}{\sqrt{3}} \gamma_3
= - 6.81(74) \cdot 10^{-8}
,\quad   
\beta_S = \frac{8}{3} \xi^\prime_3
= - 1.5(1.1) \cdot 10^{-8}
.
\lbl{alphaS_betaS_values}
\ee
The numerical values again follow from ref. \cite{Bijnens:2002vr}, see App. \ref{app:num}. They
differ somewhat from those that were available to the authors of ref. \cite{DAmbrosio:1998gur},
$\alpha_S = -5.5(5) \cdot 10^{-8}$ and $\beta_S =+0.5(1.3) \cdot 10^{-8}$.

While the form factors $W_{+,S;{\rm b1L}}(z)$ capture some contributions that
arise at order two loops, they do account for all two-loop corrections. This issue
will be discussed in detail in Section \ref{sec:2_loop} below.
Notice also that the relations given in eq. \rf{intrinsic_a_b_1loop} generalize in a straightforward
manner to the representation \rf{W_b1L_+S},
\be
G_F M_K^2 a_{+,S}  = W_{+,S;{\rm b1L}} (0) , \quad
G_F M_K^2 b_{+,S}  = W'_{+,S;{\rm b1L}} (0)
- \frac{1}{60} \, 
\left( \frac{M_K^2}{M_\pi^2} \right)^2 \left( \alpha_{+,S} - \beta_{+,S} \frac{s_0}{M_\pi^2}  \right)
.
\lbl{intrinsic_a_b}
\ee
In the sequel, we will also consider other representations of $W_{+,S}(z)$
than the one-loop or beyond-one-loop ones. For
each of these representations, we will define corresponding parameters $a_{+,S}$ and $b_{+,S}$
upon extending the relations \rf{intrinsic_a_b} to the form factor under consideration.

\indent

\section{Extraction of $\boldsymbol{a_{+,S}}$ and $\boldsymbol{b_{+,S}}$ from recent data}\label{section:data}
\setcounter{equation}{0}

\begin{figure}[ht]
\begin{center}
\includegraphics[width=6.5cm]{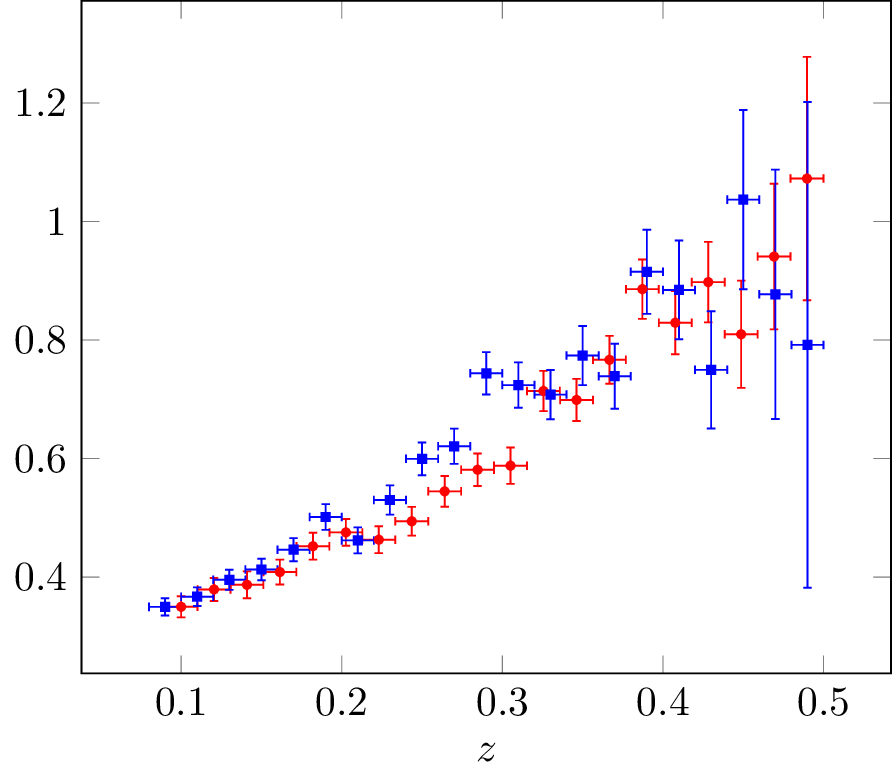} ~\hspace*{1cm}~ \includegraphics[width=6.5cm]{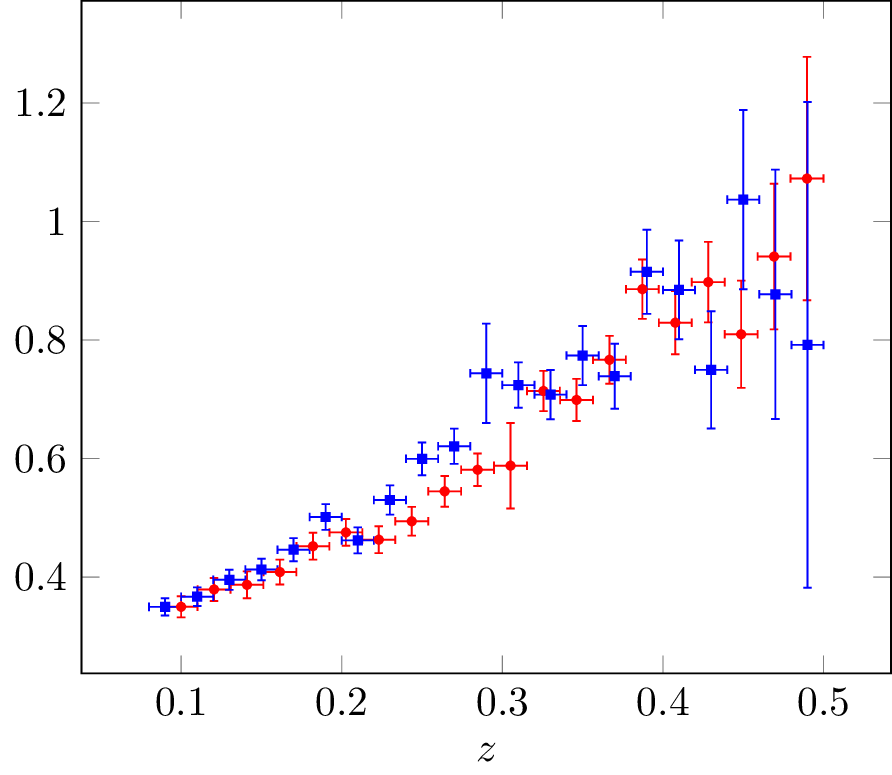} 
\end{center}
\vspace{-0.75cm}
\caption{The left-hand plot shows the data for $\vert W_+(z) \vert^2$, in units
of $10^{-11}$, in the
electron mode from the NA48/2 experiment \cite{Batley:2009aa}
(filled squares, in blue) and from the BNL-E865 experiment \cite{Appel:1999yq} (filled circles, in red). 
The right-hand plot shows the same data, but with
the error bars of two points, one for each experiment, around $z=0.3$, rescaled by a factor of 2.35.
Throughout, only the statistical uncertainties are shown.
\label{fig:data}}
\end{figure}

As already mentioned in the introduction, the main recent feature as far
as the data are concerned is the availability of quite precise information
on the decay distributions for the charged-kaon channels
$K^\pm \rightarrow \pi ^{\pm} \ell ^{+}\ell ^{-}$, see table \ref{table:exp}.
The relation between the decay distribution with respect to the di-lepton
invariant mass squared $s$ and the corresponding form factor is given by
[recall that $\lambda(x,y,z)$ denotes the K\"allen or triangle function, 
whereas $M_K$ stands either for $M_{K^\pm}$ or for $M_{K_S}$]
\be
\frac{d\Gamma_{+,S}}{dz} = \frac{\alpha^2 M_K}{12\pi(4\pi)^4} \, \lambda^{3/2}(1,z,M_\pi^2/M_K^2)\,
\sqrt{1-\frac{4m_\ell^2}{z M_K^2}} \left( 1 + \frac{2m_\ell^2}{z M_K^2}  \right)
\vert W_{+,S}(z) \vert^2 , \quad z\equiv \frac{s}{M_K^2}
.
\ee
Here, we will limit our attention to the results coming from the high-statistics experiments 
\cite{Appel:1999yq,Batley:2009aa} for the electron mode,
and \cite{Batley:2011zz} for the muon mode. For the electron channel, 
the experimental data for $\vert W_+(z) \vert$ are shown on figure \ref{fig:data}.
The NA48/2 data \cite{Batley:2009aa,Batley:2011zz} are available on the HEPData repository \cite{HEPData,HEPData2}\footnote{Only statistical uncertainties are provided, systematic uncertainties are not included.}.
Data points for the E865 experiment \cite{Appel:1999yq} do not seem to be available on a 
public repository.
For both experiments, the uncertainties given for the individual data points
are statistical only. For the systematic uncertainties in the determinations
of $a_+$ and $b_+$, we refer the reader to the experimental articles quoted in table \ref{table:exp}.
These data have been confronted with various ans\"atze for the form factor $W_+(z)$, 
from a simple constant plus linear (in $z$) type of parameterization, 
$W_{\rm lin}(z) = G_F M_K^2 f_0 (1 + \delta z)$, with however little theoretical content,
to theoretically more elaborate models \cite{DAmbrosio:1998gur,Friot:2004yr,Leskow16,Dubnickova08}.
In the present study, we will only consider the ``beyond one loop" form factor 
of ref. \cite{DAmbrosio:1998gur}. 

Fitting the expression \rf{W_b1L_+S} of the form factor, with the values \rf{alpha_beta_values}
as input, to the data of ref. \cite{Appel:1999yq}, we obtain
\be
\begin{cases}
a_+ = - 0.589(13)_{\rm stat}(1)_{\alpha_+}(5)_{\beta_+} \\
b_+ = - 0.646(54)_{\rm stat}(16)_{\alpha_+}(2)_{\beta_+}
\end{cases},\;
\chi^2/{\rm d.o.f} =11.6/18~~~[e^+e^-,~{\rm E865~data}],
\lbl{fit_electron_E865}
\ee
in reasonable agreement, although with somewhat larger uncertainties, with the values given in table 1 of ref. \cite{Appel:1999yq}.
Repeating the exercise with the data of ref. \cite{Batley:2009aa} leads to the rather surprising outcome
\be
\begin{cases}
a_+ = + 0.491(12)_{\rm stat}(1)_{\alpha_+}(5)_{\beta_+}\\
b_+ = + 1.691(57)_{\rm stat}(16)_{\alpha_+}(2)_{\beta_+}
\end{cases},\;
\chi^2/{\rm d.o.f} = 28.3/19~~~[e^+e^-,~{\rm NA48/2~data}].
\lbl{fit_electron_NA48}
\ee
Thus, whereas the BNL-E865 data favour the ``negative" solution \cite{DAmbrosio:1998gur},
$b_+ \lapprox a_+ < 0$, the NA48/2 data rather point toward the ``positive" one,
$b_+ > a_+ >0$, which, with $b_+$ about three times as large as $a_+$, is more difficult 
to understand from a theoretical point of view, as already explained in the introduction. 
Looking however for a second minimum in the NA48/2 data, we indeed find one, corresponding 
to the negative solution, with only a slightly higher value of $\chi^2$,
\be
\begin{cases}
a_+ = -0.585(12)_{\rm stat}(1)_{\alpha_+}(5)_{\beta_+}\\
b_+ = - 0.779(54)_{\rm stat}(16)_{\alpha_+}(2)_{\beta_+}
\end{cases},\;\chi^2/{\rm d.o.f} = 33.0/19~~~[e^+e^-,~{\rm NA48/2~data}].
\lbl{fit_electron_NA48_neg}
\ee
These values then also agree with those quoted in table 2 of ref. \cite{Batley:2009aa}.
Incidentally, a second minimum of the $\chi^2$ function corresponding to the positive solution 
is also present in the BNL-E865 data, but with a value $\chi^2/{\rm d.o.f} = 42.0/18$
it is clearly much less favoured in this case. The same feature is also present in the
data collected by the NA48/2 collaboration in the muon mode \cite{Batley:2011zz}:
the two solutions read

\be
\begin{cases}
a_+ = + 0.384(40) , \quad b_+ = + 2.081(147) ,\quad \chi^2/{\rm d.o.f} = 12.1/15\\
a_+ = -0.598(39) , \quad b_+ = - 0.768(144) , \quad\chi^2/{\rm d.o.f} = 15.2/15
\end{cases} ,\;[\mu^+\mu^-,~{\rm NA48/2~data}].
\lbl{fit_muon_NA48}
\ee
The two results \rf{fit_electron_E865} and
\rf{fit_electron_NA48_neg} are quite compatible as far as $a_+$ is concerned,
while the agreement is somewhat less good for $b_+$. One might contemplate the possibility of
a combined fit of the two data sets in the electron mode. While the negative solution is 
clearly favoured by this combined fit, the quality of the latter is not very good: $\chi^2/{\rm d.o.f} \sim 62/39$ 
[for the positive solution, we find $\chi^2/{\rm d.o.f} \sim 95/39$].
Looking more closely at the data, we observe that each individual data point of one experiment
is compatible, at the 1$\sigma$ level, with at least one data point of the other one, except for 
two points, one for each experiment, which are not compatible, again at the 1$\sigma$ level, with 
any point of the other experiment. These two points are located slightly at the left [NA48/2] and 
on the right [BNL-E865] of the value $z=0.3$, see figure \ref{fig:data} [one might also observe that 
$z=0.32$ corresponds to the threshold of the two-pion intermediate state; on the other hand, the
acceptances of the two experiments do not show any peculiar behaviour in the region
of this threshold; it is thus not clear whether one can establish a link with the discrepancy 
between the two data sets at $z\sim 0.3$]. If, for instance, we increase the error bars of these two data points by 
a scaling factor of 2.35, such as to make them compatible, the quality of the fit improves substantially,
giving as outcome [for comparison, the (relative) minimum corresponding
to the positive solution occurs now at $\chi^2/{\rm d.o.f} = 74.3/39$]
\be
\begin{cases}
a_+ = -0.593(9)_{\rm stat}(1)_{\alpha_+}(6)_{\beta_+}\\
b_+ = - 0.675(40)_{\rm stat}(16)_{\alpha_+}(2)_{\beta_+}
\end{cases}, \;\chi^2/{\rm d.o.f} = 45.7/39.
\lbl{fit_electron_NA48+E865}
\ee

\begin{figure}[ht]
\begin{center}
\includegraphics[width=7.5cm]{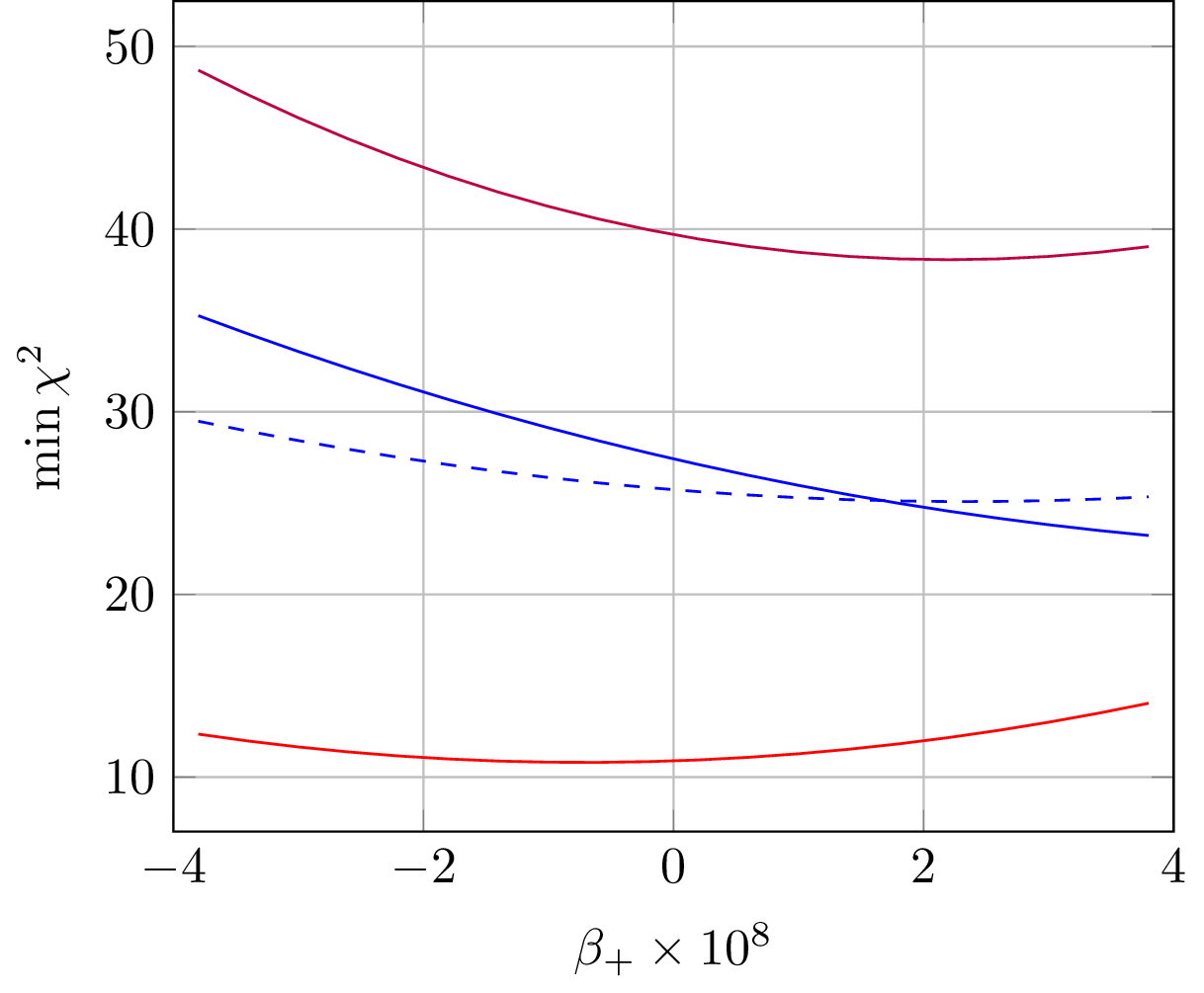} 
\end{center}
\caption{The evolution of the minimum value of the $\chi^2$ function
with $\beta_+$, for the data in the electron mode collected (from bottom to top)
by BNL-E865 \cite{Appel:1999yq}, by NA48/2 \cite{Batley:2009aa}, and for the combination
of the two sets. 
In the case of ref. \cite{Batley:2009aa}, the minimum of the $\chi^2$ function 
corresponding to the positive solution for $a_+$ and $b_+$ is also shown (dashed curve).
\label{fig:chi_2}}
\end{figure}

Up to now, we have analyzed the data with the form factor in eq. \rf{W_b1L_+S},
but for fixed values, as given in eq. \rf{alpha_beta_values}, of the ``external 
parameters" $\alpha_+$ and $\beta_+$. As already observed, the value of $\alpha_+$
is determined quite accurately, the precision on $\beta_+$ being much lower.
Redoing the fits for various values of $\beta_+$ with decreasing values of $\vert\beta_+\vert$, 
see figure \ref{fig:chi_2}, we notice that moving $\beta_+$ a few standard
deviations away from the central value in eq. \rf{alpha_beta_values} somewhat
improves the quality of the fits. The effect is quite mild in the case of
the BNL-E865 data, but somewhat more pronounced in the case of the NA48/2 data. 
This leads us to consider the possibility of fitting simultaneously
$a_+$, $b_+$, and $\beta_+$. Performing this fit on each experiment separately,
we obtain {  the results in table \ref{tab:Fit_3_paramters}} [only the statistical errors are given].

\begin{table}[ht]
{ 
\begin{center}
\begin{tabular}{|c|c|c|c|c|}
\hline
Experiment & $a_+$ & $b_+$ & $ \beta_+ \cdot 10^{8}$ &$\chi^2/{\rm d.o.f}$ \\
\hline
$[e^+e^-,~{\rm E865}]$ & $- 0.573(23)$ & $ - 0.662(57)$ &  $-0.72(2.55)$ & $10.8/17$ \\
\hline
$[e^+e^-,~{\rm NA48/2}]$ & $- 0.535(20)$ & $ -0.771(65)$ &  $+6.52(3.00)$ & $22.4/18$ \\
\hline
$[\mu^+\mu^-,~{\rm NA48/2}]$ & $- 0.40(10)$ & $- 1.10(21)$ &  $+14.3(7.9)$ & $15.2/14$ \\
\hline
\end{tabular}
\end{center}
\label{tab:Fit_3_paramters}
\caption{ Results of the $\chi^2$ fits with three parameters in each experiments. Only the statistical errors are given.}
}
\end{table}

In agreement with the trend shown by figure \ref{fig:chi_2}, the NA48/2 data yield values of $\beta_+$ that tend
to be larger than the value quoted in eq. \rf{alpha_beta_values}, obtained from a global fit
to the data on the Dalitz-plot structure of the $K\to\pi\pi\pi$ decays. One also observes
that the $\chi^2$ function is quite flat, in the vicinity of its minimum, in the $\beta_+$ 
direction, which leads to rather large ranges of values. Finally, the values of $a_+$ and $b_+$
produced by this three-parameter fit to the data come out quite similar to the ones obtained 
from the two-parameter fits, with however, as could be expected, somewhat larger statistical 
uncertainties [the effect is, however, more pronounced in the muon mode].
If we fit the combined NA48/2 and E865 data [after the rescaling of the statistical uncertainties discussed above]
in the electron mode, the 1$\sigma$ interval of values allowed for $\beta_+$ becomes narrower,
\be
\begin{cases}
a_+ = - 0.561(13)\\
b_+ = -0.694(40)\\
\beta_+ \cdot 10^{8}= +2.20(1.92)
\end{cases},\; \chi^2/{\rm d.o.f} = 38.3/38~~~[{\rm NA48/2+E865}],
\ee
but confirms the trend toward larger values of $\beta_+$ than those extracted from the $K\to\pi\pi\pi$ data.

We conclude this study with the following observations:
\begin{itemize}
\item
The NA48/2 data show a marginal preference for the positive solution,
in contrast to the BNL-E865 data, which clearly favour the negative one.
\item
The data in the electron mode from the two experiments are compatible
[up to two data points, one from each experiment, which require a
rescaling of their error bars]
and can be combined. The combined fit clearly favours the negative solution.
\item
All fits show a moderate improvement when somewhat larger values of $\beta_+$
[smaller values of $\vert\beta_+\vert$] are considered. Taking the
rather conservative attitude of letting $\beta_+$ increase by up to two standard 
deviations from its value in eq. \rf{alpha_beta_values},
i.e. $-4\lapprox \beta_+\cdot 10^{8} \lapprox -0.7$, leads to the range of values
[as $\vert\beta_+\vert$ decreases, $\vert a_+\vert$ also decreases, while 
$\vert b_+\vert$ increases]
\be
a_+ = -0.561(9)_{\rm stat}(1)_{\alpha_+} (^{+0.04} _{-0.02})_{\beta_+} ,
\quad b_+ = -0.695(39)_{\rm stat}(17)_{\alpha_+} (2)_{\beta_+} 
.
\ee
\end{itemize}

We finally close this section with a few words about the decay $K_S\to\pi^0\ell^+\ell^-$.
As shown in table \ref{table:exp}, our experimental knowledge of these processes is much more scarce,
and comes from the data collected by the NA48/1 collaboration in refs. \cite{Batley:2003mu} 
(electron mode) and \cite{Batley:2004wg} (muon mode),  and analysed using the expression of the form factor 
from ref. \cite{DAmbrosio:1998gur}, as given in in Eq, \rf{W_b1L_+S}.
A combined analysis of the branching ratios of the two lepton modes
using the values of $\alpha_S$ and of $\beta_S$ quoted in ref. \cite{DAmbrosio:1998gur} 
gives \cite{Batley:2004wg} either $a_S=-1.6^{+2.1}_{-1.8}$, $b_S=10.8^{+5.4}_{-7.7}$ or
$a_S=-1.91^{+1.6}_{-2.4}$, $b_S=-11.3^{+8.8}_{-4.5}$.
Using the more recent values of $\alpha_S$ and of $\beta_S$ given in 
eq. \rf{alphaS_betaS_values}, we obtain instead the two possibilities 
$a_S=-1.29(3.15)$, $b_S=17.8(10.6)$ and
$a_S=1.28(3.16)$, $b_S=-17.6(10.6)$. In both cases, due to the large uncertainties,
the situation is not very conclusive, even as far as the signs of $a_S$ and $b_S$ are concerned.

\indent

\section{The two-loop representation of the form factors}\label{sec:2_loop}

\begin{figure}[b]
\begin{center}
\includegraphics[width=9.5cm]{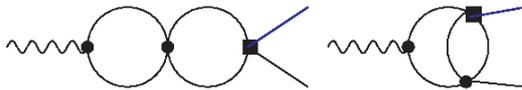} 
\end{center}
\caption{The Feynman diagrams contributing to the non-trivial analyticity
properties of the form factors $W_{+,S}(z)$ at
two loops, restricted to two-pion intermediate states. The meaning of 
the lines and of the vertices is as in figure \ref{fig:1loop}. The second
diagram, with the vertex topology, has a complex absorptive part.
\label{fig:2loop}}
\end{figure}

The discussion at the end of Sec. \ref{subsec:1loop} suggests a dispersive procedure for the 
construction of the form factors, based on the extension at next-to-next-to-leading
order of the relation given in eq. \rf{ImW_1loop}. The successive steps allowing for 
such a construction, based on chiral counting, analyticity and unitarity, have been described in detail 
in refs. \cite{Gasser:1990bv,DescotesGenon:2012gv} 
in the case of the pion form factor, and our task will be to adapt them to the present situation. 

As far as analyticity is concerned, 
the form factors $W_{+,S}(z)$ are analytic in the complex-$z$ plane with cuts 
along the positive real axis, starting at $z=4 M_\pi^2/M_K^2$. The discontinuity
along these cuts is provided by unitarity. For our present purposes, we need only
consider the contributions from two pion intermediate states, which is sufficient 
in order to account for the analyticity properties of the form factors in the range 
of values of $z$ relevant for the processes $K^\pm (K_S) \to \pi^\pm (\pi^0) \ell^+ \ell^-$.
As compared to the case of the treatment of the pion form factor in refs. \cite{Gasser:1990bv,DescotesGenon:2012gv},
we also need to deal with the fact that the kaon becomes an unstable state in the presence 
of weak interactions. This has some general consequences as far as the analyticity properties
are concerned. In particular, the absorptive and dispersive parts 
of $W_{+,S}(z)$ for $z$ real are not real, and $W_{+,S}(z)$ do not satisfy the property of 
real analyticity \cite{Eden:1966dnq},
unlike, for instance, the electromagnetic form factor of the pion or of the kaon. As far as
their analyticity properties are concerned, the form factors $W_{+;S}(z)$ are actually quite akin 
to the form factor $f_+^{\eta\pi}(s)$ describing the isospin-violating $\tau\to\eta\pi\nu_\tau$
second-class transition, which is discussed in detail in ref. \cite{Descotes-Genon:2014tla}. 
Many aspects related to the analyticity properties of $f_+^{\eta\pi}(s)$ can be 
directly transcribed to $W_{+;S}(z)$. For instance, the discussion in ref. \cite{Descotes-Genon:2014tla}
on the absence of anomalous thresholds in $f_+^{\eta\pi}(s)$ applies, {\it mutatis mutandis}, also to $W_{+;S}(z)$.

While the use of dispersive methods might appear as questionable in the presence
of an unstable kaon, one should however recall that
the computation of the form factors to two loops within chiral
perturbation theory rests on the evaluation of Feynman diagrams, which have
well-defined analyticity properties. A careful analysis 
\cite{Bronzan:1963mby,Kacser:1963zz}
shows that these analyticity properties can be reproduced within a dispersive 
framework upon letting the kaon mass squared become a free variable ${\overline M}_K^2$. 
One starts with a value of  ${\overline M}_K^2$ which
lies below the three-pion threshold [the fact that the kaon is also unstable due to its
decay into two pions plays no role in the present discussion], so that the kaon becomes stable
[with respect to its decay into three pions], and
dispersion relations can be implemented. At the end, one moves ${\overline M}_K^2$ above the 
three-pion threshold through an analytic continuation, providing the
kaon mass squared with a small positive imaginary part, ${\overline M}_K^2 \to M_K^2 + i \delta$,
$\delta \to 0_+$. It has in particular been shown \cite{ZdrahalPhD11,Kampf:2018} that this analytic continuation
can be performed without encountering singularities in the case where the masses
of the charged and neutral pions are equal. We will assume this to be the case
in the sequel, isospin-breaking effects being far from our present concern. 
The situation where the difference between the masses of the neutral and charged pions 
is taken into account would anyhow require a separate analysis.

\subsection{Construction of the form factors to two loops}

Having described the overall framework, let us now turn toward 
the explicit implementation of the procedure.
The starting point of the construction is provided by unitarity.
Since the partial-wave projections ${f}^{K\pi \to \pi^+\!\pi^-}_1 \!(s)$
at next-to-leading order have a more complicated analytic structure than at the
lowest order [\citealp{Kennedy:1962apa}, \citealp{Descotes-Genon:2014tla}], it is necessary
to consider first the situation where $M_K < 3 M_\pi$, as discussed previously.
This will be understood to be the case from now on. Then the absorptive part
of the form factor is real [it coincides with its imaginary part], and the
unitarity condition at two loops, restricted to two-pion intermediate states, reads
\bea\lbl{disc_FF_NNLO}
&&\frac{{\rm Abs}\,{W}_{+,S;{\rm 2L}} (s/M_K^2) \vert_{\pi\pi}}{16 \pi^2 M_K^2}  =
\frac{s - 4 M_\pi^2}{s} \, \theta ( s - 4 M_\pi^2 )
{\rm Re} \left[ F_V^{\pi} (s)^* \times \frac{{f}^{K\pi \to \pi^+\!\pi^-}_1 \!(s)}{\lambda_{K\pi}^{1/2} (s)}
\right]_{{\mathcal O} (E^6)}
\nonumber\\
&&=
\frac{s - 4 M_\pi^2}{s} \, \theta ( s - 4 M_\pi^2 )
\times
\Bigg[
{\rm Re} \, F_V^{\pi} (s) \big\vert_{{\mathcal O} (E^4)}
\times
\frac{{f}^{K\pi \to \pi^+\!\pi^-}_1 \!(s) \big\vert_{{\mathcal O} (E^2)}}{\lambda_{K\pi}^{1/2} (s)}
\nonumber\\
&&
+\,
F_V^{\pi} (s) \big\vert_{{\mathcal O} (E^2)}
\times
{\rm Re} \, \frac{{f}^{K\pi \to \pi^+\!\pi^-}_1 \!(s) \big\vert_{{\mathcal O} (E^4)} %
- {f}^{K\pi \to \pi^+\!\pi^-}_1 \!(s) \big\vert_{{\mathcal O} (E^2)}}{\lambda_{K\pi}^{1/2} (s)}
\Bigg]
\\
&&= 
\frac{s - 4 M_\pi^2}{s} \, \theta ( s - 4 M_\pi^2 )
\times
\left[
\frac{\alpha_{+,S}}{96 \pi M_\pi^2} \times {\rm Re} \, F_V^{\pi} (s) \big\vert_{{\mathcal O} (E^4)}  + \psi_{+,S} (s)
\right] \times  \sigma_\pi (s) 
\nonumber
,
\eea
where we have written, for $s>4 M_\pi^2$,
\begin{align}
& \frac{{f}^{K\pi \to \pi^+\!\pi^-}_1 \!(s) \big\vert_{{\mathcal O} (E^2)}}{\lambda_{K\pi}^{1/2} (s)} = 
\frac{\alpha_{+,S}}{96 \pi M_\pi^2} \times \sigma_\pi (s) 
,\nonumber\\
& {\rm Re} \, \frac{{f}^{K\pi \to \pi^+\!\pi^-}_1 \!(s) \big\vert_{{\mathcal O} (E^4)}}{\lambda_{K\pi}^{1/2} (s) } = 
\left[ \frac{\alpha_{+,S}}{96 \pi M_\pi^2}  + \psi_{+,S} (s) \right] \times  \sigma_\pi (s) 
,
\lbl{psi1_defined}
\end{align}
where $\sigma_\pi(s)=\sqrt{1-\frac{4M_\pi^2}{s}}$. 

\begin{figure}[ht]
\begin{center}
\includegraphics[width=5.5cm]{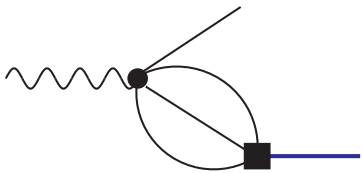} 
\end{center}
\caption{A Feynman diagram that does not contribute to the 
absorptive parts of the form factors $W_{+,S}(z)$ at
two loops, but which gives imaginary parts to $a_{+,S}$
and to $b_{+,S}$. The meaning of 
the lines and of the vertices is as in figure \ref{fig:1loop}.
\label{fig:2loop_bis}}
\end{figure}

The structure of the absorptive parts of the form factors ${W}_{+,S;{\rm 2L}}$ displayed in 
eq. \rf{disc_FF_NNLO} corresponds to the analyticity properties of the two-loop Feynman
diagrams shown in figure \ref{fig:2loop}.
Notice that whereas the partial-wave projection ${f}^{K \pi \to \pi^+\!\pi^-}_1 \!(s)$
is proportional to $\lambda_{K\pi}^{1/2} (s)$ at lowest-order, this is no longer the case at
next-to-leading order. Furthermore, the prescription of endowing $M_K^2$ with an infinitesimal positive 
imaginary part then also specifies a suitable determination of $\lambda_{K\pi}^{1/2} (s)$.
The pion form factor at next-to-leading order can be written as \cite{GassLeut,Gasser:1990bv,DescotesGenon:2012gv}
[as far as notation is concerned, we follow the last of these references]
\be
{\rm Re} \, F_V^{\pi} (s) \big\vert_{{\mathcal O} (E^4)} = 1 
+ a_V^\pi s
+ 16 \pi \varphi_{1;\pi\pi}^{+-;+-} (s) {\rm Re} \, {\bar J}_{\pi\pi} (s)
,
\ee
with
\be
\varphi_{1;\pi\pi}^{+-;+-} (s) = \frac{\beta}{96 \pi} \, \frac{s - 4 M_\pi^2}{F_\pi^2}       
,
\quad
a_V^\pi = \frac{1}{6} \left( \langle r^2 \rangle_V^\pi + \frac{\beta}{24 \pi^2 M_\pi^2} \right)
,
\lbl{LO_pi-pi_P-wave}
\ee
where ${\varphi}_{1;\pi\pi}^{+-;+-} (s)$ is the lowest-order
$P$-wave projection of the amplitude for $\pi^+\pi^- \to \pi^+\pi^-$ scattering, 
$\beta$ being identified with the slope of this amplitude in its expansion around
the center of its Dalitz plot, whereas $\langle r^2 \rangle_V^\pi$ denotes the mean 
square of the charge radius of the pion, and
\be
\theta ( s - 4 M_\pi^2 ) \, {\rm Re} \, {\bar J}_{\pi\pi} (s) = \frac{8 \pi}{\sigma_\pi (s)} {\rm Im} \, {\bar J}_{\pi\pi}^2 (s)
,~~
{\rm Im} \, {\bar J}_{\pi\pi} (s) = \frac{\sigma_\pi (s)}{16 \pi} \, \theta ( s - 4 M_\pi^2 )
.
\ee
Then one has
\bea
&&
{\lefteqn{
\frac{s - 4 M_\pi^2}{s \lambda_{K\pi}^{1/2} (s)} \, \theta ( s - 4 M_\pi^2 )
\times
{\rm Re} \, F_V^{\pi} (s) \big\vert_{{\mathcal O} (E^4)}
\times
{f}^{K\pi \to \pi^+\!\pi^-}_1 \!(s) \big\vert_{{\mathcal O} (E^2)}}}
\nonumber\\
&&\qquad\quad
= \,
\frac{1}{6} \, \frac{\alpha_{+,S}}{M_\pi^2} \, \frac{s - 4 M_\pi^2}{s} 
\left( 1 + a_V^\pi s \right) {\rm Im} \, {\bar J}_{\pi\pi} (s) 
+
\frac{1}{72} \, \frac{\beta \cdot \alpha_{+,S}}{M_\pi^2}
\frac{(s - 4 M_\pi^2)^2}{s F_\pi^2} 
\, {\rm Im}\,{\bar J}_{\pi\pi}^{\,2} (s) 
.~~\quad~
\lbl{disc_FF_NNLO_part1}
\eea
It is interesting to notice that the quantity $\lambda_{K\pi}^{1/2} (s)$ does no 
longer occur in the right-hand side of this last formula. There remains then to compute
\be
\sigma_\pi (s)  \, \frac{s - 4 M_\pi^2}{s } \,  \theta ( s - 4 M_\pi^2 )
\times \psi_{+,S} (s)
,
\lbl{disc_FF_NNLO_part2}
\ee
where $\psi_{+,S} (s)$ describes the one-loop correction to ${\rm Re}\,{f}^{K\pi \to \pi^+\!\pi^-}_1 \!(s)$, see eq. \rf{psi1_defined}.
This computation, while straightforward, constitutes the most involved part of the calculation of the
absorptive parts of the form factors as given by the right-hand side of eq. \rf{disc_FF_NNLO}.
The details of this calculation will not be given here, but are collected, for the interested reader, 
in appendix  \ref{app:psi}, together with the notation. 
Here we simply display the final expressions of the form factors at two-loop accuracy in the low-energy expansion,
\bea\lbl{FF_2loop}
&&\!\!\!
W_{+,S;{\rm 2L}} (s/M_K^2)
=
G_F M_K^2 \!  \left( a_{+,S} + b_{+,S} \frac{s}{M_K^2} \right)\nonumber\\
&&+ \frac{8 \pi^2}{3} \frac{M_K^2}{M_\pi^2} \!
\left[ \alpha_{+,S} \! \left( 1 + a_V^\pi s \right) 
 + \beta_{+,S}  \frac{s-s_0}{M_\pi^2}
\right]   \! 
\left[ \frac{s - 4 M_\pi^2}{s} \, {\bar J}_{\pi\pi} (s) + \frac{1}{24 \pi^2} \right]
\nonumber\\
&&
+ \, \frac{8 \pi^2}{3} \frac{M_K^2}{M_\pi^2} 
\left[ \Delta \alpha_{+,S} 
 - \Delta \beta_{+,S} \frac{s_0}{M_\pi^2} \right]
\left[ \frac{s - 4 M_\pi^2}{s} \left( {\bar J}_{\pi\pi} (s) - \frac{1}{96 \pi^2} \frac{s}{M_\pi^2} \right)
+ \frac{1}{240 \pi^2} \frac{s}{M_\pi^2} \right]
\nonumber\\
&&
+ \,
\Delta \beta_{+,S} \frac{s}{M_\pi^2}   
\left[ \frac{s - 4 M_\pi^2}{s} \, {\bar J}_{\pi\pi} (s) + \frac{1}{24 \pi^2} \right]
+ \,
\frac{4 \pi^2}{9} \, \beta \cdot \alpha_{+,S} \frac{M_K^2}{ F_\pi^2}
\left[
\frac{(s - 4 M_\pi^2)^2}{s M_\pi^2} \,{\bar J}_{\pi\pi}^{\,2} (s) - \frac{1}{576 \pi^2} \frac{s}{M_\pi^2} 
\right]
\nonumber\\
&&
-\,
\frac{1}{16 \pi^2 F_\pi^2}
\times \sum_{i=0}^3  \left[
{\bar{\mathfrak K}}_i^{(\lambda ; 0)} (s) {\mathfrak p}_i (s) 
- {\bar{\mathfrak K}}_i^{(\lambda ; 0)\prime} (0) [ \Delta_{K\pi} {\mathfrak p}^{(-1)}_i  + s {\bar{\mathfrak p}}_i (0) ]
- \frac{s}{2} {\bar{\mathfrak K}}_i^{(\lambda ; 0)\prime\prime} (0) \Delta_{K\pi} {\mathfrak p}^{(-1)}_i
\right]
\nonumber\\
&&
-\,
\frac{1}{16 \pi^2 F_\pi^2}
\times \sum_{i=0}^3  
\frac{\Delta_{K\pi}^2}{M_\pi^2} \, {\mathfrak q}_i
\left[
\frac{{\bar{\mathfrak K}}_i^{(\lambda ; 1)} (s) }{s} - {\bar{\mathfrak K}}_i^{(\lambda ; 1)\prime} (0)
- \frac{s}{2} \, {\bar{\mathfrak K}}_i^{(\lambda ; 1)\prime\prime} (0)
\right]
\!,
\eea
and comment on this result with a few remarks:
\begin{itemize}
\item

{  The origin of the coefficients $\Delta \alpha_{+,S}$ and $\Delta \beta_{+,S}$ is discussed after eq. \rf{P_antisym} and their expression 
given explicitly in Eqs. \rf{Delta_alpha-beta_+} and \rf{Delta_alpha-beta_S}.}
\item
The absorptive parts of the functions ${\bar{\mathfrak K}}_i^{(\lambda ; 0,1)} (s)$,
$i=2,3$, which are given by the functions ${\mathfrak k}_i (s)$ defined in eq. \rf{frak_k},
develop an imaginary part for $s>4 M_\pi^2$ when $M_K$ takes it physical value. This feature 
is at the origin of some of the general properties of the form factors discussed at the 
beginning of this subsection, like the loss of real analyticity. More specifically, the latter 
follows from the analyticity properties of the second type of Feynman diagrams, with the 
vertex topology, depicted in figure \ref{fig:2loop}, and which produce the
contributions involving the functions ${\bar{\mathfrak K}}_i^{(\lambda ; 0,1)} (s)$
for $i=2,3$.
\item
For the same reasons, the constants $a_{+,S}$ and $b_{+,S}$ are in general complex. 
Their imaginary parts are generated for the first time at the two-loop level, and
are thus chirally suppressed. They arise through 
Feynman graphs with the vertex topology, but also through Feynman
graphs without absorptive parts, of the type shown in figure \ref{fig:2loop_bis}.
Moreover they will be proportional to the phase space for the $K\to\pi\pi\pi$ transition, 
which is also small. Due to this double suppression, these imaginary parts can be neglected 
in practice, and will, in particular, not impinge on the analysis in Section \ref{section:data}, 
given the present [and, probably, future] statistical uncertainties of the data.
\item
Uncommon features, like circular cuts in the complex plane, intersecting
the usual unitarity and left-hand cuts on the real axis, also show up in 
the analyticity properties of the partial-wave projections ${f}^{K\pi \to \pi^+\!\pi^-}_1 \!(s)$
computed from the one-loop amplitudes. Their existence follows 
from the general analysis made in ref. \cite{Kennedy:1962apa}, and they are discussed 
more specifically in refs. \cite{ZdrahalPhD11, Buras:1991jm} and \cite{Descotes-Genon:2014tla}.
\end{itemize}

\subsection{Comparing $\boldsymbol{W_{+,S;{\rm 2L}} (z)}$ and $\boldsymbol{W_{+,S;{\rm b1L}} (z)}$ }

\begin{figure}[ht]
\begin{center}
\includegraphics[width=10.5cm]{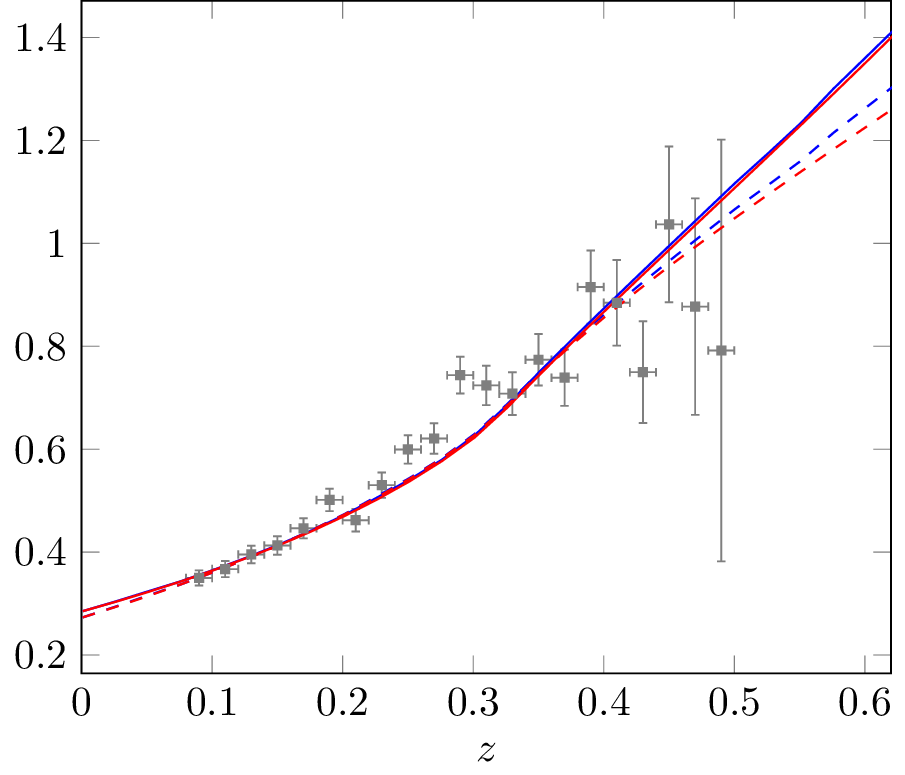} 
\end{center}
\caption{Comparison between $\vert W_{+;{\rm 2L}}(z)\vert^2$ (upper solid and dashed lines, in blue)
and $\vert W_{+;{\rm b1L}}(z)\vert^2$ (lower solid and dashed lines, in red) for
$a_+ = -0.593$,  $b_+= -0.675$,  $\beta_+ = -2.88 \cdot 10^{-8}$ (solid lines),
and for $a_+ = -0.580$,  $b_+= -0.683$,  $\beta_+ = -0.85 \cdot 10^{-8}$ (dashed lines).
The data points shown on the figure are those of ref. \cite{Batley:2009aa}.
\label{fig:comp}}
\end{figure}

The one-loop expression for $W_+ (z)$ does not give a very good description
of the data. If the latter are fitted, as in Sec. \ref{section:data}, with
$W_{+;{\rm 1L}} (z)$, keeping $a_+$ as a free variable, the quality of the fit deteriorates tremendously. Typically,
the value at the minimum of the $\chi^2$  function is in the range between 300 
and 500. The representation $W_{+;{\rm b1L}} (z)$ thus constitutes a real
improvement in the description of the data. Likewise, one may legitimately
ask oneself to which extent the full two-loop expression $W_{+;{\rm 2L}}(z)$
constructed above would further modify this picture.

In comparing the full two-loop representations \rf{FF_2loop} of the form factors 
with the expressions \rf{W_b1L_+S} of ref. \cite{DAmbrosio:1998gur},
we notice that the latter are essentially reproduced by the first line of eq. \rf{FF_2loop},
which follows if in eq. \rf{disc_FF_NNLO} we restrict the pion form factor and the 
$P$-wave projections to their polynomial parts:
\be
{\rm Re} \, F_V^\pi (s)\vert_{{\cal O}(E^4)} \to 1 + a_V^\pi s 
,\quad
{\rm Re} \, f_1^{K\pi \to \pi^+\pi^-} (s) \vert_{{\cal O}(E^4)} \to 
\frac{M_K^2}{M_\pi^2} \left[ \alpha_{+,S} + \beta_{+,S} \frac{s-s_0}{M_\pi^2}   \right]
,
\ee
taking, for $a_V^\pi$, the estimate $a_V^\pi = 1/M_V^2$ given by vector-meson dominance.
What is missing in order to completely reproduce the expressions of the form factors
$W_{+,S;{\rm b1L}} (z)$ is the term proportional to the product $\beta_{+,S} \times a_V^\pi$,
which does not occur in eq. \rf{FF_2loop}. This term actually represents a contribution
of order three loops, whence its absence from eq. \rf{FF_2loop}. Furthermore, if we go once more 
through the fits in Section \ref{section:data} with this terms omitted from eq. \rf{W_b1L_+S}, 
the outcomes for $a_{+,S}$ and $b_{+,S}$ are barely changed, so that, from a phenomenological point of view,
this term is not important in the region of $z$ corresponding to the phase space of
the $K\to\pi\ell^+\ell^-$ transitions.

We may now proceed with the quantitative comparison between the full two-loop expressions given by eq. \rf{FF_2loop}
and the form factors of ref. \cite{DAmbrosio:1998gur}. This comparison is shown
in figure \ref{fig:comp} for the modulus squared of $W_{+;{\rm 2L}}(z)$ and
$W_{+;{\rm b1L}}(z)$, for two sets of values for the parameters $a_+$, $b_+$
and $\beta_+$. This difference remains quite small, and
in any case well below the present statistical uncertainties of the data.
This feature is shared for other values of the parameters taken in the ranges discussed
in Sec. \ref{section:data}. Differences between $W_{+;{\rm 2L}}(z)$ and
$W_{+;{\rm b1L}}(z)$ are visible only for values of $z$ above $0.4$, i.e. toward
the upper end of phase space, where the statistical uncertainties are also largest.
The comparison of $W_{S;{\rm 2L}}(z)$ and of $W_{S;{\rm b1L}}(z)$ leads to similar
conclusions.
The determinations of the parameters $a_{+,S}$ and $b_{+,S}$ from data will thus not be
modified in any substantial way if, instead of $W_{+,S;{\rm b1L}}(z)$, one uses, 
in Sec. \ref{section:data}, the full two-loop amplitudes $W_{+,S;{\rm 2L}}(z)$.

\indent

\section{$\boldsymbol{W_+(z)}$ beyond the low-energy expansion}\label{sect:model}
\setcounter{equation}{0}

While the two-loop representations of the form factors, or the
truncated versions thereof proposed in ref. \cite{DAmbrosio:1998gur},
provide an appropriate description of the experimental data in terms
of two sets of parameters $a_{+,S}$ and $b_{+,S}$, the latter cannot
be predicted within the low-energy framework considered in Sections \ref{section:theory}
and \ref{sec:2_loop}.  In order to obtain predictions for them, it is
necessary to set up a phenomenological description of the form factors 
that goes beyond the low-energy framework. If such a description of
the form factors becomes available, the values of the constants $a_{+,S}$ 
and $b_{+,S}$ can then be obtained through the definitions given in eq. \rf{intrinsic_a_b}.
The purpose of the present section is to proceed
with such a phenomenological construction of the form factor $W_+(z)$.

Building on the results obtained so far,
we will propose a simple model for the form factor $W_+(z)$ that accounts for rescattering effects in the two-pion intermediate
state beyond the framework set by the low-energy expansion, and, at the same time, provides a matching to the short-distance 
behaviour investigated in Sec. \ref{sect:short-dist}. 
Consequently, our model will consist of three parts,
\be
W_+ (z) = W_+^{\pi\pi} (z) + W_+^{\rm res} (z ; \nu) + W_+^{\rm SD} (z ;\nu)
.
\lbl{model}
\ee
Before going into the details of their respective evaluations, let us briefly describe the physical content of each of these parts.

As already mentioned, the first part describes the contribution from
the two-pion intermediate state to $W_+(z)$. It is constructed upon assuming, 
in analogy with the electromagnetic form factor of the pion $F_V^{\pi} (s)$
\cite{Ecker:1989yg,Guerrero1997}, that it is given by an unsubtracted dispersion integral,
\be
W_+^{\pi\pi} (z) = \int_{4 M_\pi^2}^\infty dx \frac{\rho_+^{\pi\pi} (x)}{x-zM_K^2-i0}
.
\lbl{W_pi-pi_disp}
\ee
The absorptive part consists of the two-pion spectral density $\rho_+^{\pi\pi} (s)$,
and is obtained upon inserting a two-pion intermediate state in the 
representation of the form factor given in eq. \rf{Kpi_FF_1st-order},
\be
\rho_+^{\pi\pi} (s) = 
16 \pi^2 M_K^2 \times \frac{s - 4 M_\pi^2}{s} \, \theta ( s - 4 M_\pi^2 )
\times
F_V^{\pi *} (s) 
\times
\frac{  {f}^{K^\pm \pi^\mp \to \pi^+ \pi^-}_1 \!(s)}{\lambda^{1/2}_{K\pi} (s)}
.
\lbl{disc_FF_pi-pi_P-wave}
\ee
In order to evaluate this absorptive part,
we require two ingredients: a representation of the pion form factor $F_V^{\pi} (s)$, and
a representation of the $P$-wave projection $f_1^{K\pi\to\pi\pi}(s)$, which both extend to the whole
energy range set by the cut singularity of $W_+(z)$. We will provide  an explicit realization of these two quantities
in Subsection \ref{sec:two-pions} below. For the time being, let
us just notice that for the convergence of the unsubtracted dispersive integral it is 
sufficient that their product is bounded by, say, a constant for large values of $s$.

Above this lowest threshold, several intermediate states will 
contribute to the discontinuity of $W_+(z)$ in the 1-GeV region and beyond. 
Considering only two-meson intermediate states, the next thresholds 
will come from $K^+\pi^-$ or $K^0 \pi^0$ intermediate states, 
followed by $K^0{\bar K}^0$, $K^+ K^-$, and so on. 
As the energy increases, the number of possible exclusive intermediate states grows, 
and they eventually merge into the inclusive contribution provided by the QCD continuum, 
as discussed in Section \ref{sect:short-dist}. We will describe this process
in terms of an infinite tower of equally-spaced zero-width resonances
\be
W_+^{\rm res} (z ; \nu) =
\frac{f^{K^\pm\pi^\mp}_+ (z M_K^2)}{4 \pi} \int_{M^2}^\infty dx \frac{\rho^{\rm res}_+ (x ; \nu)}{x-z M_K^2 -i0}
,
\ee
where $M\sim 1$ GeV is the scale of the lowest resonance occurring in this tower.
The main task facing us will be to find an appropriate Regge-type resonance model,
\be
\rho^{\rm res}_+ (s ; \nu) \propto \sum_{n=1}^\infty M^2 \mu_n (\nu) \delta (s-nM^2)
,
\ee
which reproduces the correct QCD short-distance properties of 
the form factor. This issue will be adressed in Section \ref{sec:resonances}.
In particular, the dependence of $W_+^{\rm res} (z ; \nu)$ on the short-distance scale 
$\nu$ has to match the same dependence that appears in the third part on 
the right-hand side of eq. \rf{model}, simply given by the factorized 
contribution coming from the $Q_{7V}$ operator,
\be
W_+^{\rm SD} (z ;\nu) = 16 \pi^2 M_K^2 \left( \frac{G_{\rm F}}{\sqrt{2}} V_{us}^* V_{ud} 
\right) \frac{C_{7V} (\nu)}{4\pi\alpha}  \,{  f_+^{K^\pm \pi^\mp} } (z M_K^2)
.
\lbl{W_+_SD}
\ee
This last contribution is evaluated in Section \ref{sec:Q7V}.

\subsection{The contribution from the two-pion state}\label{sec:two-pions}

\begin{figure}[ht]
\begin{center}
\includegraphics[width=6.5cm]{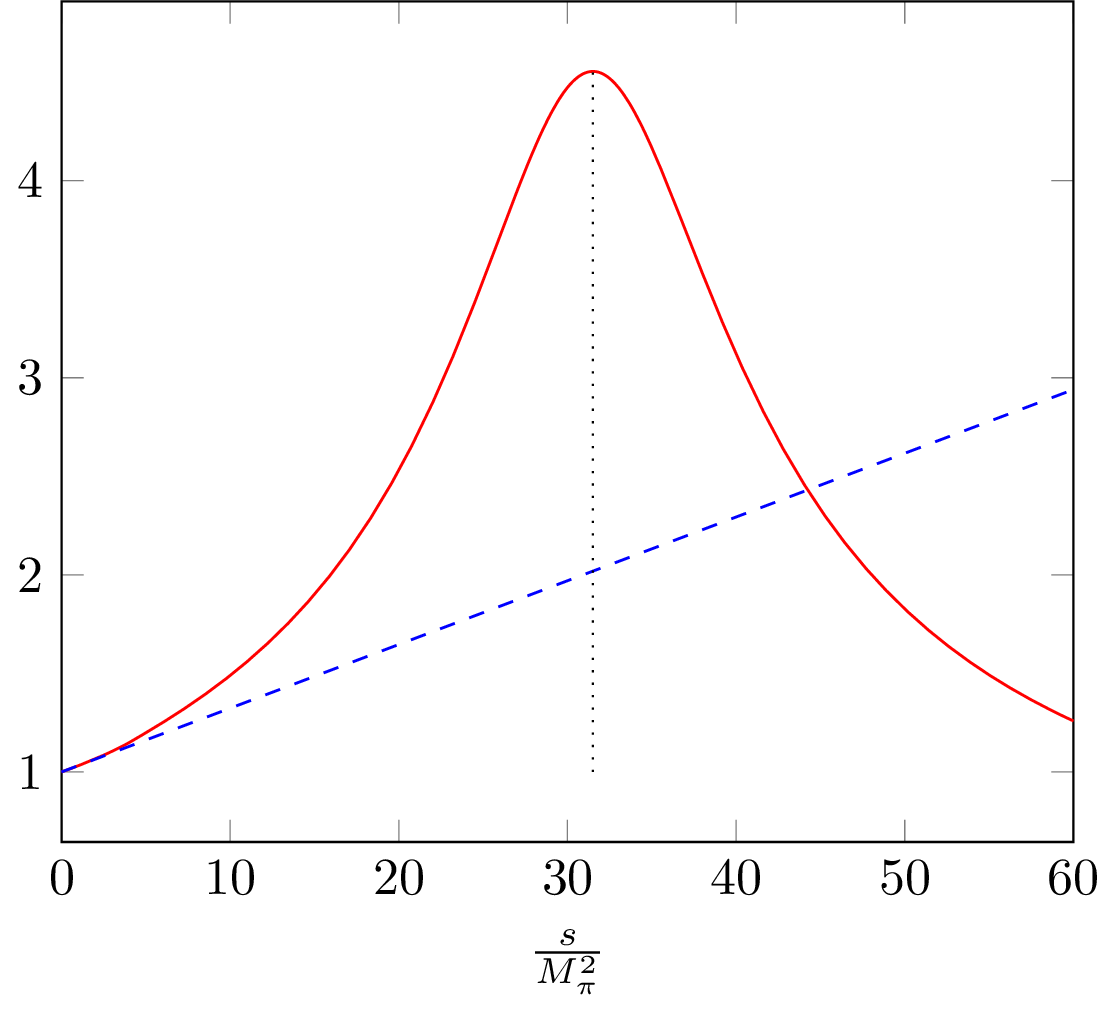} ~~~~~ \includegraphics[width=6.5cm]{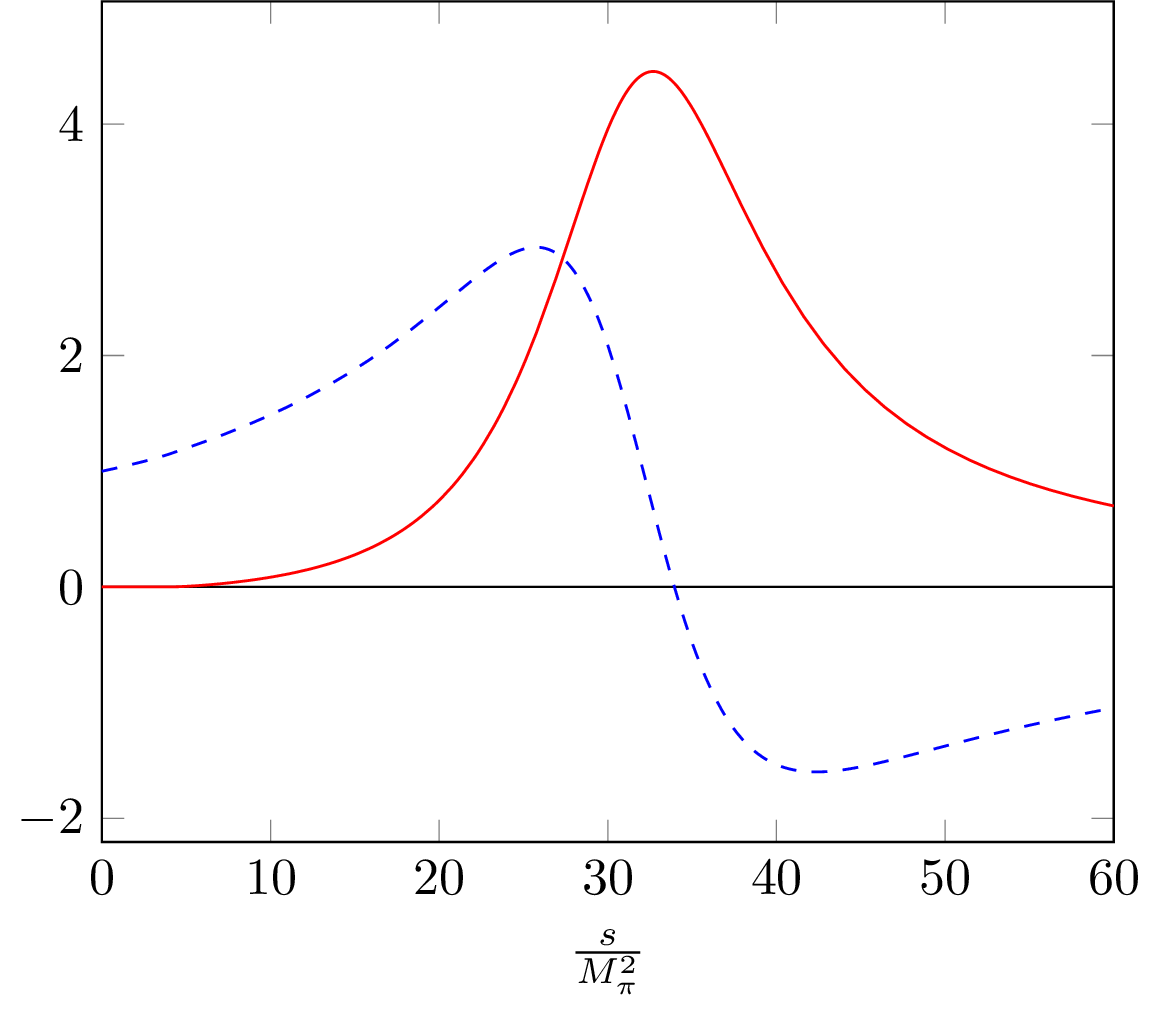} 
\end{center}
\caption{The left-hand plot shows the absolute value $\vert F_V^\pi (s) \vert$ at one loop
[dashed line] and its IAM unitarized version [solid line], as a function of $s/M_\pi^2$, for $s\ge 4 M_\pi^2$ and
for $M_V=M_\rho=775$ MeV [notice that $M_\rho^2 = 31 M_\pi^2$] and $\beta=1.11$. The right-hand plot shows 
the real [dashed line] and imaginary [solid line] parts of the IAM unitarized pion form factor.
\label{fig:F_V_IAM}}
\end{figure}

Assuming we have a representation of the form \rf{W_pi-pi_disp} at our disposal,
the contributions $a_+^{\pi\pi}$ and $b_+^{\pi\pi}$ from the two-pion intermediate 
state  to the coefficients $a_+$ and $b_+$, respectively, are obtained through two 
sum rules that follow from the definitions given in eq. \rf{intrinsic_a_b},
\be
G_F M_K^2 a_+^{\pi\pi} = W_+^{\pi\pi} (0) = 
\int_0^\infty \frac{dx}{x} \,\rho_+^{\pi\pi} (x)
,\quad
G_F M_K^2 b_+^{\pi\pi} = W_+^{\pi\pi\prime} (0) =
M_K^2 \int_0^\infty \frac{dx}{x^2} \,\rho_+^{\pi\pi} (x)
.
\lbl{aplus_bplus_SR} 
\ee
As far as the convergence of the integral in \rf{W_pi-pi_disp} is concerned,
it depends on the behaviour of both the pion form factor $F_V^{\pi} (s)$ and 
the partial wave projection ${f}^{K \pi \to \pi^+\!\pi^-}_1 \!(s)$ for large
values of $s$.  As already mentioned, in 
order to evaluate the sum rules in eq. \rf{aplus_bplus_SR}, we need representations 
of the pion form factor $F_V^{\pi} (s)$ and of the partial-wave projection  
$f_1^{\pi^+\pi^- \to K^+ \pi^-} (s)$ that extend beyond their low-energy expansions. 
There exist several "unitarization" procedures that precisely allow to do this. 
One of them, the inverse-amplitude method (IAM) \cite{Truong:1988zp,Dobado:1989qm}, 
gives quite reasonable results when applied to $\pi\pi$ scattering in the $P$-wave
and to the pion form factor \cite{Beldjoudi:1994yi,Hannah:1996tp} in the region of the $\rho$ meson. 
The result, 
\be
F_V^\pi (s) = \frac{1}{  \displaystyle 1 - \frac{s}{M_V^2} 
- \frac{\beta}{6 F_\pi^2} (s - 4 M_\pi^2) \, {\bar J}_{\pi\pi} (s) }
,
\lbl{FV_IAM}
\ee
obtained starting from the one-loop low-energy expression of $F_V^\pi (s)$, 
provides a rather simple representation of $F_V^\pi (s)$. Its phase is the 
phase of the $P$-wave projection of the $\pi\pi$ scattering amplitude, as required 
by Watson's final-state theorem, and it reproduces the one-loop expression of $F_V^\pi (s)$ 
for small values of $s$. Furthermore, it exhibits a resonant behaviour
at the expected value of the energy squared, i.e. $s \sim M_\rho^2$. 
It is quite easy to understand in simple terms how this last property emerges
from the representation \rf{FV_IAM}. Neglecting at first the contribution
from the pion loops, materialized by the term proportional to $\beta$ in the
denominator, this expression reduces to the well-known VMD form
$F_V^\pi (s) \vert_{\rm VMD} = M_V^2/(M_V^2 - s)$. It has the expected pole at 
$s=M_V^2\sim M_\rho^2$, and the real part of the pion-loop contribution will slightly
change the real part of the pole position, while the imaginary part of the pion loop
will move it from the real axis to the second Riemann sheet, leading to the
resonant behaviour shown in figure \ref{fig:F_V_IAM}.

\begin{figure}[ht]
\begin{center}
\includegraphics[width=4.5cm]{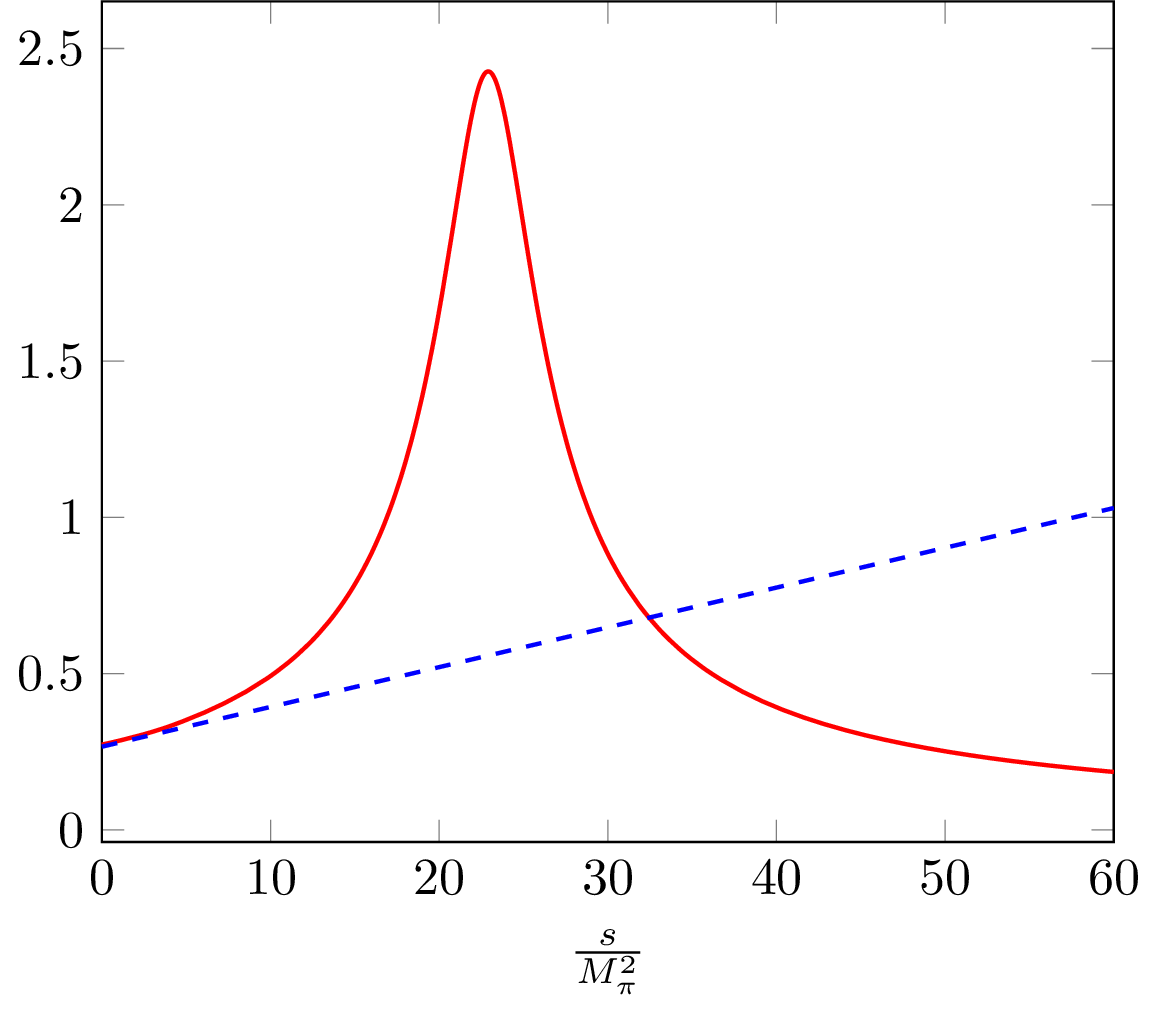} ~~~~~ \includegraphics[width=4.5cm]{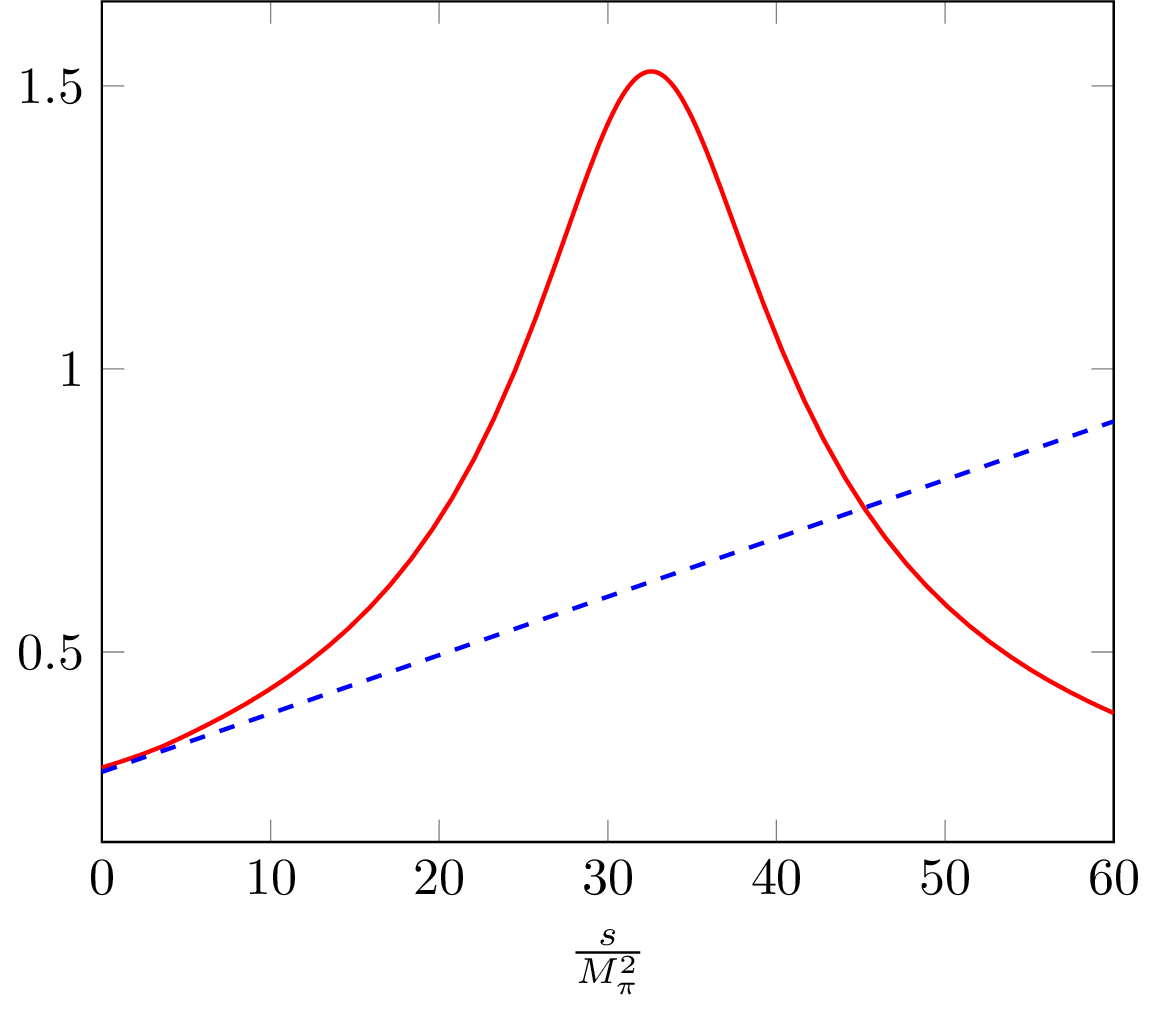} %
 ~~~~~ \includegraphics[width=4.5cm]{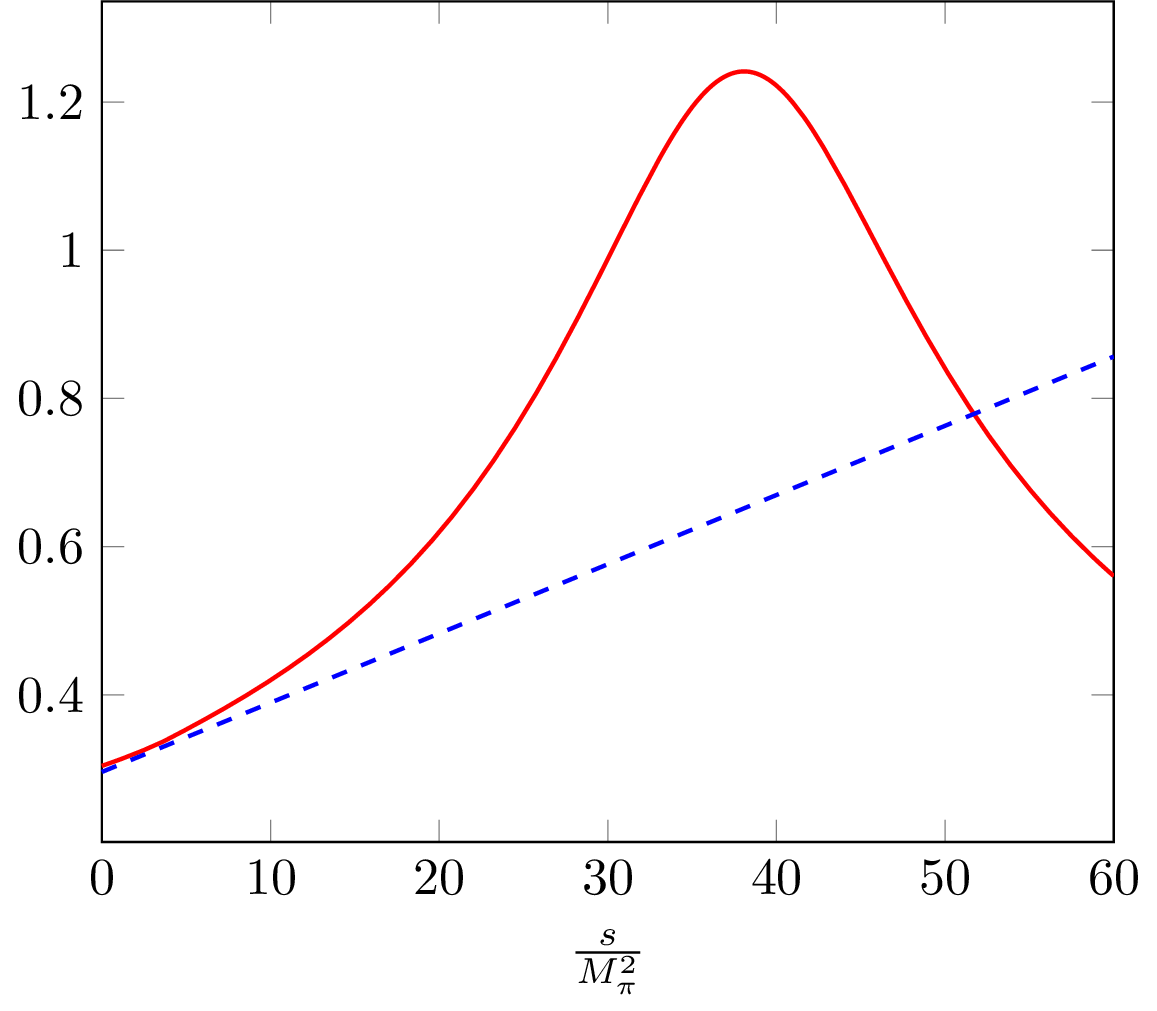} 
\end{center}
\caption{The absolute value of $f_1^{\pi^+\pi^- \to K^+ \pi^-} (s) / \lambda_{K\pi}^{1/2}(s) \sqrt{1-\frac{4 M_\pi^2}{s}}$
as given in eq. \rf{f1_Kplus_IAM} [Breit-Wigner shaped curves, in red] vs. the one-loop counterpart [straight lines, in blue], 
as a function of $s/M_\pi^2$, for $s\ge 4 M_\pi^2$ and
for $\alpha_+ = - 20.84 \cdot 10^{-8}$, $\beta=1.11$, whereas $\beta_+$ is increased from its central
value in eq. \ref{eq:alpha_beta_values} by 1.5 standard deviation [$\beta_+ = -1.26 \cdot 10^{-8} $, left-hand plot],
1.88  standard deviation [$\beta_+ = -0.85 \cdot 10^{-8} $, the value retained in the text, central plot], 
and 2 standard deviations 
[$\beta_+ = -0.72\cdot 10^{-8} $, right-hand plot].
\label{fig:f1_Kplus_IAM_imp}}
\end{figure}

We next turn toward the partial-wave projection $f_1^{K^+ \pi^- \to \pi^+\pi^-} (s)$,
to which we wish to apply the same procedure. Starting from its one-loop expression obtained
in App. \ref{app:psi}, and neglecting the contributions from the circular cuts,
the IAM-unitarized version of $f_1^{K^+ \pi^- \to \pi^+\pi^-} (s)$ 
reads [$\psi_+^{\rm loop} (s)$ is defined in  eq. \rf{Psi1_loop}] 
\begin{multline}
f_1^{K^+ \pi^- \to \pi^+\pi^-} (s) \\
= \frac{\left[{\displaystyle{\frac{\alpha_+}{96\pi M_\pi^2} }} \right]^2%
\times \lambda_{K\pi}^{1/2}(s) \sqrt{1-\frac{4 M_\pi^2}{s}}}{
{\displaystyle{\frac{1}{96\pi M_\pi^2} }} \bigg[ \alpha_+ - \beta_+ \frac{s-s_0}{M_\pi^2} \bigg]
- \psi_+^{\rm loop}(s) + {\rm Re}\,\psi_+^{\rm loop}(s_0) + (s-s_0) {\rm Re}\, \psi_+^{\rm loop~\prime}(s_0)
}
. 
\end{multline}
We will further simplify this expression upon keeping only the contribution of the
unitarity or right-hand cut of $\psi_+^{\rm loop}(s)$ into account, i.e.
the first term on the right-hand side of eq. \rf{Psi1_loop}. The remaining terms 
give only a small correction in comparison.
We thus end up with
\begin{multline}
f_1^{K^+ \pi^- \to \pi^+\pi^-} (s) \\
= \frac{{\displaystyle{\frac{\alpha_+}{96\pi M_\pi^2} }} %
\times \lambda_{K\pi}^{1/2}(s) \sqrt{1-\frac{4 M_\pi^2}{s}} }{
1 - {\displaystyle{\frac{\beta_+}{\alpha_+} \frac{s - s_0}{M_\pi^2} }} 
- {\displaystyle{\frac{\beta}{6} \frac{s-4M_\pi^2}{F_\pi^2} }}
\left[{\bar J}_{\pi\pi} (s) - {\rm Re} \,{\bar J}_{\pi\pi} (s_0)\right] 
+ {\displaystyle{\frac{\beta}{6} \frac{s_0-4M_\pi^2}{F_\pi^2} }}
(s-s_0) {\rm Re} \,{\bar J}_{\pi\pi}' (s_0)
}
.~~~
\lbl{f1_Kplus_IAM}
\end{multline}
If we transpose the discussion that follows eq. \rf{FV_IAM} to the representation
\rf{f1_Kplus_IAM}, we observe that, when the contributions from the pion loops 
are discarded, the ``bare" pole is located at a value of $s$ given by 
$s_{\rm pole}=s_0 + (\alpha_+/\beta_+) M_\pi^2$. For the values given in eq. \rf{alpha_beta_values},
i.e.  $\alpha_+/\beta_+ \sim 7$, this gives $s_{\rm pole} \sim 12 M_\pi^2$. In order to
obtain a pole located at $s_{\rm pole} \sim M_\rho^2$, one needs a higher value
of the ratio $\alpha_+/\beta_+$, say $\alpha_+/\beta_+ \sim 25$. In view
of the error bars of the numerical values of $\alpha_+$ and $\beta_+$, this
can be achieved most economically upon increasing $\beta_+$ by about two standard deviations 
from its central value in eq. \rf{alpha_beta_values}. Assigning even part of the effect to an 
increase of the absolute value of $\alpha_+$ would represent a much more significant
deviation from its value in \rf{alpha_beta_values}. At this stage one might recall
the discussion in Sec. \ref{section:data}, where it was already noticed that upon
keeping $\beta_+$ as a free variable to be fitted to the data, the outcome was favouring values
deviating from the one in \rf{alpha_beta_values} by a similar amount. It would certainly appear as 
somewhat far-fetched to ground the reason for preferring larger values of $\beta_+$ on the
simple characteristics of the IAM-unitarized partial wave $f_1^{K^+ \pi^- \to \pi^+\pi^-} (s)$.
Nevertheless, the concordance, on this issue, with the analysis presented in Sec. \ref{section:data}
is definitely interesting and noteworthy\footnote{ For the sake of completeness, let us mention that the Dalitz plot 
of the $K^+ \rightarrow \pi^+ \pi^+\pi^-$ transition has been measured with high precision by the NA-48/2 
Collaboration \cite{Batley:2007md}. These results were published after ref. \cite{Bijnens:2002vr}. 
Using only ref. \cite{Batley:2007md} does not allow for a separate determination of $\alpha_+$ and $\beta_+$ 
but only of the ratio $\alpha_+/\beta_+$. In terms of the Dalitz plot parameters $g,h$ and $k$ of 
ref. \cite{Batley:2007md}, one obtains $\alpha_+/\beta_+ \sim -g/(h-3k) = 6.53(1)$ in agreement 
with the value obtained from eq. \eqref{eq:alpha_beta_values}.}. In figure \ref{fig:f1_Kplus_IAM_imp}, we illustrate the evolution of 
$\vert f_1^{K^+ \pi^- \to \pi^+\pi^-} (s) \vert$ as given in eq.  \rf{f1_Kplus_IAM} 
for different values of $\beta_+$ and for $s\ge 4 M_\pi^2$. Notice also that the
approximation considered in eq. \rf{f1_Kplus_IAM} preserves Watson's final-state theorem,
and the phases of $F_V^\pi (s)$ and of $ f_1^{K^+ \pi^- \to \pi^+\pi^-} (s)$
should be identical. For the choice $\beta=1.11$ [this value of $\beta$ is itself
chosen such as to match the phase of the pion form factor to the phase of the
$P$-wave projection of the $\pi\pi$ amplitude, both obtained upon unitarization
of their one-loop expressions by the IAM method], this is the case for $\beta_+= -0.85 \cdot 10^{-8} $.

The numerical evaluation of the two sum rules in eq. \rf{aplus_bplus_SR} 
for these values of $\beta$ and of $\beta_+$ then gives
\be
a_+^{\pi\pi} = -1.58,\quad  b_+^{\pi\pi} = -0.76
.
\lbl{a_+_and_b_+_values}
\ee
In this approach, the overall negative sign of these numbers is driven by
the negative sign of $\alpha_+$. The result for $b_+$ comes out relatively 
close to the values extracted from the data in Sec. \ref{section:data}, whereas the 
absolute value of $a_+$ is about twice as large. Of course, in both cases 
there are other contributions which we need to discuss before being in a position
of making a more definite statement.

Let us finally notice that the same approach can also be implemented in order to evaluate the 
contribution from the two-pion state to $a_S$ and $b_S$. It suffices to make the appropriate replacements
in eq. \rf{f1_Kplus_IAM}. We first notice that a ratio $\alpha_S/\beta_S \sim 25$ lies (almost) 
within reach for the values indicated in eq. \rf{alphaS_betaS_values}, e.g. $\alpha_S \sim -7.5 \cdot 10^{-8}$, 
$\beta_S \sim -0.3\cdot 10^{-8}$. The resulting values of $a_S^{\pi\pi}$ 
and $b_S^{\pi\pi}$ would then also be negative, and about three times smaller in absolute value 
than those obtained for $a_+^{\pi\pi}$ and $b_+^{\pi\pi}$ in eq. \rf{a_+_and_b_+_values}.

\subsection{Intermediate states with higher thresholds}\label{sec:resonances}

\begin{figure}[ht]
\center\epsfig{figure=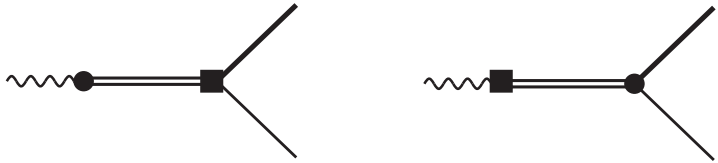,height=3.0cm}
\caption{The diagrams for the exchange of {  zero-width} resonances (double line).
The external lines correspond to the insertion of the current $j_\mu (x)$
(wiggly line) and to the kaon (thick line) and pion (thin line). The circular
blob denotes a strong coupling vertex, and the square box a weak coupling vertex.
The diagram on the left corresponds to the exchange of resonances 
with $J^{PC}=1^{--}$, $I=1$, $S=0$, like the $\rho$ or the $\phi (1020)$ meson. 
The resonances exchanged in the diagram on the right
have quantum numbers $J^{P}=1^{-}$, $I=1/2$, $S=\pm 1$, like the $K^*(892)$. \label{fig_Res}}
\end{figure}

In Sec. \ref{sect:short-dist}, we have established the high-energy behaviour of
the dimensionally renormalized form factor, see eq. \rf{high-s}.
Reproducing this behaviour is necessary in order to obtain an amplitude that
does no longer depend on the short-distance renormalization scale $\nu$
once the contribution from the local factorized operator $Q_{7V}$ is added.
This behaviour is, however, not reproduced by the contribution of the $\pi\pi$ intermediate state
that we have just studied. It has therefore to come from the remaining infinite number
of intermediate states, with higher and higher thresholds, and which eventually
end up into the perturbative contribution of quarks and gluons at short distances.
One might attempt to describe some of these additional intermediate states in a
similar treatment as the one adopted here for the two-pion states. This holds,
in particular, for the next thresholds, due to $K\pi$ and to ${\bar K}K$ intermediate states. 
We leave such improvements for future work. Here, we will adopt a simpler point of view,
where the additional intermediate states are described by zero-width resonance
states. Such a picture would naturally emerge, for instance, from the perspective of the 
limit where the number of colours $N_c$ becomes large \cite{tHooft74,Witten79}. 
Contributions of this type are depicted on 
figure \ref{fig_Res}, and would produce a form factor with the following expression
\be
W_+^{\rm res} (z ; \nu) =
\sum_{V=\phi\cdots} \frac{f_V {\widetilde g}_V}{s-M_V^2+i0}
\, +
\sum_{V=K^*\cdots} \frac{g_V {\widetilde f}_V}{s-M_V^2+i0}
.
\ee
The first sum runs over resonances with quantum numbers $J^{PC}=1^{--}$, $I=1$, $S=0$.
It starts here with the $\phi (1020)$, since the $\rho$ meson is 
already contained in the contribution of the $\pi\pi$ intermediate states that
we have discussed previously. The second sum runs over the resonances with quantum
numbers $J^{P}=1^{-}$, $I=1/2$, $S=\pm 1$, and starts with the $K^* (892)$. The strong
couplings $f_V$ and $g_V$ are fixed, for instance, by the widths of the decays like
$\phi\to e^+ e^-$ and $K^* \to K\pi$, respectively. Information on the weak couplings
${\widetilde f}_V$ and ${\widetilde g}_V$ is, however, not available. This represents
the usual difficulty in obtaining reliable estimates of the low-energy constants in 
the weak sector through resonance saturation 
\cite{Ecker:1992de,Ecker:1990in,Pich:1990mw,DAmbrosio:1997ctq,Cappiello:2011re}. 
Another difficulty lies in the fact
that the correct short-distance behaviour \rf{high-s} can only be recovered upon considering
an infinite number of resonances \cite{Witten79}. This second difficulty can be dealt with
upon using available techniques \cite{Greynat:2013cta,Shifman:2000jv,Rafael:2012sy}
to construct resonance models with spectra of the Regge-type, and with residues
that can be tuned such as to build up harmonic sums that can often be resummed exactly 
and, moreover, reproduce the prescribed asymptotic behaviour.
We will present such a construction in the case of interest here.
%
%

In order to set the stage, let us go back to the calculation in Section \ref{sect:short-dist}.
It teaches us that a dispersive representation of the type
\be
W_+ (z) = \frac{f^{K^\pm\pi^\mp}_+ (z M_K^2)}{4 \pi} \int dx \frac{\rho_M (x)}{x-z M_K^2 -i0}
,
\quad
\rho_M (s) = A \theta(s - M^2) \, \sqrt{1-\frac{M^2}{s}} 
,
\lbl{disp_int_dim}
\ee
would fulfill the required conditions.
Here $A$ is a normalization constant, and $M$ is a mass scale which corresponds 
to the onset of the perturbative continuum due to the quark loop, and which we would 
like to identify with a typical resonance scale $M\sim 1$ GeV. Of course, the above 
dispersive integral does not converge, and we first need to consider the dimensionally 
regularized version of the spectral density $\rho_{M} (s)$, namely \cite{Gasser:1998qt}
\be
\rho_{M} (s ; D) = A (D) (4\pi)^{2-\frac{D}{2}} \left(\frac{M^2}{\nu_{\mbox{\tiny MS}}^2}\right)^{\frac{D}{2}-2}
\frac{\Gamma\left( \frac{3}{2} \right)}{\Gamma\left( \frac{D}{2}-\frac{1}{2} \right)}
\left( \frac{s}{M^2} - 1 \right)^{\frac{D}{2} - 2} \, \sqrt{1-\frac{M^2}{s}} \, \theta(s-M^2)
,
\ee
where the function $A(D)$ is only constrained by the condition $A(4)=A$, $A(D) = A + (D-4)A' + \cdots$,
and $A'$ parameterizes the scheme-dependence arising from the Dirac matrices in the calculation of Section \ref{sect:short-dist}. 
The divergence of the dispersive integral is then contained in the value of the integral at $z=0$,
\bea
\int \! dx \, \frac{\rho_{M} (x ; D)}{x} &=& 
A (D) (4\pi)^{2-\frac{D}{2}} \! \left(\frac{M^2}{\nu_{\mbox{\tiny MS}}^2}\right)^{\frac{D}{2}-2} \!\!
\frac{\Gamma\!\left( \frac{3}{2} \right)}{\Gamma\!\left( \frac{D}{2}-\frac{1}{2} \right)}
 \frac{\Gamma\!\left( \frac{D}{2}-\frac{1}{2} \right) \Gamma\! \left(2 - \frac{D}{2}  \right)}{\Gamma\!\left( \frac{3}{2} \right)}\nonumber \\
&=& A (D) (4\pi)^{2-\frac{D}{2}}\left(\frac{M^2}{\nu_{\mbox{\tiny MS}}^2}\right)^{\frac{D}{2}-2} \! \Gamma \! \left(2 - \frac{D}{2}  \right)
\nonumber\\
&=& A \left[ \frac{-2}{D-4} - \gamma_E + \ln(4\pi) - \ln \frac{M^2}{\nu_{\mbox{\tiny MS}}^2} - 2 A'/A + \cdots \right]
.
\lbl{prop_z=0}
\eea
After renormalization in the ${\overline{\rm MS}}$ scheme, 
one therefore finds
\begin{multline}
W_+ (z ; \nu) =  f^{K^\pm\pi^\mp}_+ (z M_K^2)   \\
\times \frac{1}{4 \pi} 
\bigg[
z M_K^2  \int \frac{dx}{x} \, \theta(x - M^2) \, \sqrt{1-\frac{M^2}{x}} \frac{A}{x-z M_K^2-i0}
- A \ln \frac{M^2}{\nu^2} - 2 A'
\bigg]
.
\end{multline}
The limit of large space-like values of $z$ gives
\be
\lim_{z\to - \infty} W_+ (z ; \nu) / f^{K^\pm\pi^\mp}_+ (z M_K^2)
=
 \frac{1}{4 \pi}\bigg[ 2 (A - A' - A \ln 2) - A \ln \frac{-zM_K^2}{\nu^2} \bigg]
.
\lbl{asymptotic}
\ee
The short-distance behaviour given in eq. \rf{high-s} is then recovered
in this case with the choices
\begin{align}
A &= 16 \pi^2 M_K^2 \left( \frac{G_{\rm F}}{\sqrt{2}} V_{us}^* V_{ud} \right) \sum_I C_I(\nu) \xi_{01}^I 
, \nonumber\\
A' &=  - 16 \pi^2 M_K^2 \left( \frac{G_{\rm F}}{\sqrt{2}} V_{us}^* V_{ud} \right) \sum_I C_I(\nu) 
\left[ \frac{1}{2} \xi_{00}^I - (1 - \ln2)  \xi_{01}^I \right]
.
\lbl{A_and_Aprime}
\end{align}
Actually, the dispersive integral \rf{disp_int_dim} with $\rho_M (x;D)$ can be done explicitly $( w\equiv -z M_K^2/M^2)$:
\be
\int dx \, \frac{\rho_{M} (x ; D)}{x - zM_K^2} =
A (D) (4\pi)^{2-\frac{D}{2}}\left(\frac{M^2}{\nu_{\mbox{\tiny MS}}^2}\right)^{\frac{D}{2}-2} 
\, \Gamma\left( 2-\frac{D}{2}  \right) \times { }_2 \mathrm{F}_1 \left(   
\begin{tabular}{c}
$1$ , $2-\frac{D}{2}$
\\
$\frac{3}{2}$
\end{tabular} 
\Bigg\vert  -w \right).
\ee
For the definition and properties of the hypergeometric function ${_2}\mathrm{F}_1$,
we refer the reader to ref. \cite{DLMF}.
The properties \rf{prop_z=0} and \rf{asymptotic} can then be directly recovered from those of 
this function, see e.g. \cite[eq. 15.8.2]{DLMF}, 
\be
\hspace*{-0.5cm}{_2}\mathrm{F}_1 \!\left( \left. \begin{matrix}
1 \,,\, 2-\frac{D}{2} \\ \frac{3}{2} \end{matrix} \right\vert -w \right)
= \left\{
\begin{tabular}{l}
$1 + \mathcal{O}(w)$~~~[$w \rightarrow 0$] \\
\\
$\frac{\pi^{\frac{3}{2}}}{2 \sin \left(\pi \frac{4-D}{2}\right)}
\frac{1}{\Gamma\left(\frac{4-D}{2}\right)\Gamma\left(\frac{D-1}{2}\right)} \; w^{-\frac{4-D}{2}}
\left[1 + \mathcal{O}(w^{-1}) \right] $
~~~[$w \rightarrow +\infty$]
\end{tabular}
\right.
.
\ee

It is possible to reproduce the salient properties of the simple model discussed above through 
a Regge-type resonance model of the form 
\bea
\rho^{K\pi}_{\rm res}(s ; D) &=& 
A (D) (4\pi)^{2-\frac{D}{2}}\left(\frac{M^2}{\nu_{\mbox{\tiny MS}}^2}\right)^{\frac{D}{2}-2} 
\, \Gamma\left( 2-\frac{D}{2}  \right)
\sum_{n\ge 1} M^2 \mu_n (D) \delta (s - n M^2)
,
\nonumber\\
\\
\int dx \, \frac{\rho^{K\pi}_{\rm res} (x ; D)}{x + w M^2} &=&
A (D) (4\pi)^{2-\frac{d}{2}}\left(\frac{M^2}{\nu_{\mbox{\tiny MS}}^2}\right)^{\frac{D}{2}-2} 
\, \Gamma\left( 2-\frac{D}{2}  \right) 
\sum_{n\ge 1} \frac{\mu_n (D)}{(n  + w)} 
,
\nonumber
\eea
provided one can find a set of functions $\mu_n (D)$ that satisfies the following requirements:
\begin{itemize}
\item
$\sum_{n\ge 1} \frac{\mu_n (D)}{n} = 1$
\item
$ \mu_n (D) \underset{D \rightarrow 4}{=} (D-4) {\bar\mu}_n + {\cal O} \left((D-4)^2 \right)$
\item
$\xi(w) \equiv \sum_{n\ge 1} \frac{{\bar\mu}_n}{n(n+w)}$ converges
\item
$\xi(w) \underset{w \rightarrow +\infty}{\sim} \ln w$
\end{itemize}
As we now show, a solution, by far not unique, to this list 
of requirements can then be constructed in the form
\be
\mu_n (D) =  \mathsf{a} (D) \, n^{\frac{D-4}{2}} 
+ \mathsf{b}(D) \, n^2 \left(\frac{D}{2}-1\right)^n  
,
\lbl{mu_ansatz}
\ee
for suitably chosen functions $\mathsf{a} (D)$ and $\mathsf{b} (D)$.
Indeed, starting from the inverse Mellin representation 
\begin{equation}
\frac{1}{1+\frac{w}{n}} = \int \limits_{c_1 - i\infty}^{c_1 + i\infty} \! \frac{d u}{2i\pi}  
\left(\frac{w}{n}\right)^{-u} \;\frac{\pi}{\sin \pi u}\;,
\end{equation}
valid for $0<c_1<1$, and making use of the sums
\be
\sum_{n=1}^\infty n^{\frac{D-4}{2}+u-1} = \zeta\!\left(\frac{6-D}{2}-u\right)
,\qquad
\sum_{n=1}^\infty \left(\frac{D}{2}-1\right)^n n^{u+1} = {\rm Li}_{-1-u} \left(\frac{D}{2}-1\right)
,
\ee
one obtains
\begin{multline}
\lbl{eq:MBrho}
\sum_{n=1}^\infty \left[ \mathsf{a}\, n^{\frac{D-4}{2}} + \mathsf{b}\, n^2 \left(\frac{D}{2}-1\right)^n  \right]  \frac{1}{n+w} \\
= \int \limits_{c_2 - i\infty}^{c_2 + i\infty} 
\! \frac{d u}{2i\pi} \; w^{-u} \;\frac{\pi}{\sin \pi u} \left[ \mathsf{a}\, \zeta\!\left(\frac{6-D}{2}-u\right) 
+  \mathsf{b} \, {\rm Li}_{-1-u} \left(\frac{D}{2}-1\right)\right]\;,
\end{multline}
with $0<c_2<\frac{4-D}{2}$. Here $\zeta (x)$ is the Riemann $\zeta$-function and ${\rm Li}_y(x)$ denotes the 
polylogarithm function, defined as ${\rm Li}_y (x)=\sum_{n>0} x^n n^{-y}$ for $\vert x \vert <1$
and arbitrary complex $y$, and by analytic continuation for other values of $x$.
The integrand in the relation \rf{eq:MBrho} has a pole at $u=0$, coming from the pre-factor only,
since the terms inside the square brackets are well behaved at $u=0$, with  ${\rm Li}_{-1} (x)= x/(1-x)^2$.
According to the \textit{Converse Mapping Theorem} \cite{FGD95,Friot:2005cu}, from 
this pole at $u=0$, located on the left of the fundamental strip $0<c_2<\frac{4-D}{2}$, 
one deduces that
\begin{equation}
\sum_{n=1}^\infty \left[ \mathsf{a}\, n^{\frac{D-4}{2}} + \mathsf{b} \, n^2 \left(\frac{D}{2}-1\right)^n  \right]  
\frac{1}{n+w} \underset{w \rightarrow 0}{\sim} \mathsf{a}\, \zeta\!\left(\frac{6-D}{2}\right) 
+ 2 \mathsf{b} \,\frac{D-2}{(D-4)^2}
.
\end{equation}
Since $\zeta (3 - D/2 -u) \to -1/(u-2+D/2)+\cdots$ for $u\to 2-D/2$, while
${\rm Li}_{-1-u} (D/2-1)$ remains finite as this limit is taken, the integrand
has also a pole at $u=\frac{4-D}{2}$, due to the first term in the square brackets.
In this case, the pole being located on the right of the fundamental strip $0<c_2<\frac{4-D}{2}$,
the \textit{Converse Mapping Theorem} allows us to state that
\begin{equation}
\sum_{n=1}^\infty \left[ \mathsf{a}\, n^{\frac{D-4}{2}} + \mathsf{b} \, n^2 \left(\frac{D}{2}-1\right)^n  \right]  
\frac{1}{n+w} \underset{w \rightarrow \infty }{\sim} \mathsf{a}\, \frac{\pi}{\sin \left(\pi \frac{4-D}{2}\right)} \; w^{-\frac{4-D}{2}} 
.
\end{equation}
We thus conclude that we are able to build a Regge-type resonance model, 
defined by eq. \rf{mu_ansatz}, with
\begin{equation}
\mathsf{a} (D) = \frac{\frac{\sqrt{\pi}}{2}}{\Gamma\left(\frac{4-D}{2}\right)\Gamma\left(\frac{D-1}{2}\right)}\;\;\text{and}\;\; 
\mathsf{b} (D) = \frac{1}{2} \, \frac{(D-4)^2}{D-2} 
\left[1 - \frac{\frac{\sqrt{\pi}}{2}\zeta\!\left(\frac{6-D}{2}\right)}{\Gamma\left(\frac{4-D}{2}\right)\Gamma\left(\frac{D-1}{2}\right)}\right]
,
\end{equation} 
and which satisfies all the required properties. The part of the integral that remains 
finite in the limit $D\to 4$ can be resummed explicitly, and we find
\begin{equation}
\int d x \, \frac{\rho_{\rm res} (x ; D)}{x-zM_K^2} \\=
A(D)(4\pi)^{\frac{4-D}{2}} \left(\frac{M^2}{\nu^2_\text{MS}}\right)^{\frac{D-4}{2}}\, 
\Gamma\left(\frac{4-D}{2}\right)
-  
A \left[ \gamma_E + \psi(1+w) \right]
+ {\cal O} (D-4)
,
\lbl{resummed}
\end{equation}
where the di-gamma function $\psi$ arises through the sum
\be
\sum_{n=1}^\infty \frac{w}{n(n+w)} =  \gamma_E + \psi(1+w) 
.
\ee
Actually, considering more generally the poles when $u$ equals a negative integer, 
the \textit{Converse Mapping Theorem} gives
\begin{equation}
\sum_{n=1}^\infty \left[ \mathsf{a}(D)\, n^{\frac{D-4}{2}} + \mathsf{b}(D) \, n^2 \left(\frac{D}{2}-1\right)^n  \right]  
\frac{1}{n+w} \ \underset{w \rightarrow 0}{\sim} \ \sum_{p=0}^N (-1)^p \mathsf{c}_p(D) w^p + {\mathcal O} (w^{N+1})
,
\end{equation}
with the coefficients $\mathsf{c}_p(D)$ given by $\mathsf{c}_0(D) = 1$ and, for $p>0$, by
\be
\mathsf{c}_p(D) = \mathsf{a} (D)\, \zeta\!\left(\frac{6 + 2p -D}{2}\right) 
+ \mathsf{b} (D) \, {\rm Li}_{p-1} \left( \frac{D}{2} - 1 \right)
\qquad [p\ge 1]
.
\ee
In the limit where $D\to 4$, one obtains
\be
\Gamma\left( \frac{4-D}{2}\right) \mathsf{c}_p(D)
\to \zeta (p+1)
,
\lbl{Taylor}
\ee
which indeed corresponds to eq. \rf{resummed}.

Summarizing, we end up, after minimal subtraction, with the expression
\begin{multline}
\label{eq:Wres}
W_+^{\rm res} (z ; \nu)  =
  \frac{f^{K^\pm\pi^\mp}_+ (z M_K^2) }{4 \pi} \\
\times
16 \pi^2 M_K^2 \left( \frac{G_{\rm F}}{\sqrt{2}} V_{us}^* V_{ud} \right) \sum_I C_I(\nu) 
\bigg\{  \xi_{00}^I  -  \xi_{01}^I
\left[  \ln \frac{M^2}{\nu^2} + \psi \left(1-z \,\frac{M_K^2}{M^2} \right) \right]
\bigg\}
.
\end{multline}
Accordingly, the corresponding contributions to $a_+$ and $b_+$ read [cf.  eq. \rf{intrinsic_a_b}],
\be
G_F M_K^2 a_+^{\rm res} (\nu) = W_+^{\rm res} (0;\nu),\quad
G_F M_K^2 b_+^{\rm res} (\nu) = W_+^{{\rm res}\,\prime} (0;\nu)
,
\ee
i.e.
\bea
a_+^{\rm res} (\nu)  &=&  \frac{f^{K^\pm\pi^\mp}_+ (0) }{4 \pi} \times
16 \pi^2  \left( \frac{1}{\sqrt{2}} V_{us}^* V_{ud} \right) \sum_I C_I(\nu) 
\bigg\{  \xi_{00}^I  -  \xi_{01}^I
\left[  \ln \frac{M^2}{\nu^2} - \gamma_E \right]
\bigg\}
,
\nonumber\\
b_+^{\rm res} (\nu)  &=&  
\frac{f^{K^\pm\pi^\mp}_+ (0) }{4 \pi} \times
16 \pi^2 \left( \frac{1}{\sqrt{2}} V_{us}^* V_{ud} \right) 
\, \frac{\pi^2}{6} \, \frac{M_K^2}{M^2} \sum_I C_I(\nu) \,\xi_{01}^I 
\nonumber\\
& & \!\!\!\!\! + \,
\frac{f^{K^\pm\pi^\mp}_+ (0) }{4 \pi} \times \lambda_+\,\frac{M_K^2}{M_\pi^2} \nonumber\\
&& \hspace*{2cm}\times
16 \pi^2 \left( \frac{1}{\sqrt{2}} V_{us}^* V_{ud} \right) \sum_I C_I(\nu) 
\bigg\{  \xi_{00}^I  -  \xi_{01}^I
\left[  \ln \frac{M^2}{\nu^2} - \gamma_E \right]
\bigg\} 
.
\lbl{aplus_bplus_res} 
\eea

The expression of $b_+^{\rm res} (\nu)$ involves the slope at the origin $\lambda_+$ of the 
form factor $f^{K^\pm\pi^\mp}_+ (s)$. It is defined in eq. \rf{K_pi_FF}, and numerical 
values for both $f^{K^\pm\pi^\mp}_+ (0)$ and $\lambda_+$ are given in eq. \rf{K_pi_FF_values}.

We draw attention here to the fact that only the order $\mathcal{O}(\alpha_s^0)$ contribution to the short 
distance behaviour of $W(s;\nu)$ is reproduced by the resonance model. This means that eq. \eqref{eq:Wres} 
satisfies both \eqref{eq:F_renorm} and \eqref{eq:high-s} at order $\mathcal{O}(\alpha_s^0)$ only, which implies 
that some scale dependence will remain. We could improve on this aspect upon including $\mathcal{O}(\alpha_s)$ 
corrections, which are almost completely determined from the renormalization-group analysis in Sec. \ref{sect:short-dist}. 
We would then need to build a resonance model that reproduces both the correct $\ln(-s/\nu^2)$ and $\ln^2(-s/\nu^2)$ 
behaviours at high energy. We leave such an improvement for a future study.

\subsection{The contribution from the factorized $\boldsymbol{Q_{7V}}$ operator}\label{sec:Q7V}

It remains to evaluate the last contribution, due to $W_+^{\rm SD} (z ;\nu)$. From its definition
in eq. \rf{W_+_SD}, combined with eq. \rf{intrinsic_a_b}, we obtain right away
\begin{align}
a_+^{\rm SD} (\nu) &= \frac{f^{K^\pm\pi^\mp}_+ (0) }{4 \pi} \times
16 \pi^2 \left( \frac{1}{\sqrt{2}} V_{us}^* V_{ud} \right) 
\frac{C_{7V} (\nu)}{\alpha} \nonumber\\
b_+^{\rm SD} (\nu) & = \frac{f^{K^\pm\pi^\mp}_+ (0) }{4 \pi} \times \lambda_+\,\frac{M_K^2}{M_\pi^2}\times
16 \pi^2 \left( \frac{1}{\sqrt{2}} V_{us}^* V_{ud} \right)
\frac{C_{7V} (\nu)}{\alpha}  
.
\end{align}
Values for $C_{7V}$ can be found in ref. \cite{Buras:1994qa}.
In particular, we use $C_{7V}^{({\rm NDR,HV})}(1~{\rm GeV})/\alpha =(-0.037,0.000)$.
Here we have set $\tau=0$, i.e. we have identified $C_{7V}$ with $z_{7V}$, and we
have chosen $\Lambda^{(4)}_{\overline{\rm MS}} = 300~{\rm MeV}$. In order to
investigate the dependence of the sums $a_+^{\rm res} (\nu) + a_+^{\rm SD} (\nu)$ and 
$  b_+^{\rm res} + b_+^{\rm SD} (\nu)$ on the short-distance scale $\nu$, we use the
lowest-order evolution { equations}, neglecting the mixing with the penguin operators, i.e.
\be
\frac{C_{7V}(\nu)}{\alpha} = \frac{C_{7V}(\nu_0)}{\alpha}
+ \frac{16}{99}
\left[ 1 - \left( \frac{\alpha_s^{(3)} (\nu)}{\alpha_s^{(3)} (\nu_0)} \right)^{-11/9}  \right] \frac{C_+ (\nu_0)}{\alpha_s (\nu_0)}
- \frac{8}{45}
\left[ 1 - \left( \frac{\alpha_s^{(3)} (\nu)}{\alpha_s^{(3)} (\nu_0)} \right)^{-5/9}  \right]  \frac{C_- (\nu_0)}{\alpha_s (\nu_0)}
,
\ee
where $C_\pm(\nu) = C_2(\nu) \pm C_1(\nu)$ and 
\begin{equation}
C_+(\nu) = \left( \frac{\alpha_s^{(3)} (\nu)}{\alpha_s^{(3)} (\nu_0)} \right)^{-2/9}  C_+ (\nu_0) \;,\;\; C_-(\nu) = 
\left( \frac{\alpha_s^{(3)} (\nu)}{\alpha_s^{(3)} (\nu_0)} \right)^{4/9}  C_- (\nu_0) .
\end{equation}
Here $\alpha_s^{(3)}$ stands for the running QCD coupling for $N_f=3$ active flavours,
\be
\alpha_s^{(N_f)} (\nu) = \frac{12\pi}{(33 - 2 N_f) t^{(N_f)}(\nu)} , \quad t^{(N_f)} (\nu) \equiv 2 \ln(\nu/\Lambda^{(N_f)}_{\overline{\rm MS}})
,
\ee
and for the numerical evaluation, we use $\Lambda^{(3)}_{\overline{\rm MS}} = 340~{\rm MeV}$
and the following input values \cite{Buras:1994qa}
\be
C_+ (1~{\rm GeV}) = 
\left\{
\begin{tabular}{ll}
0.771 & NDR 
\\
0.735 & HV
\end{tabular}
\right.
,\quad
C_- (1~{\rm GeV}) = 
\left\{
\begin{tabular}{ll}
1.737 & NDR 
\\
1.937 & HV
\end{tabular}
\right.
\ee

\subsection{Evaluations of $\boldsymbol{a_+}$ and $\boldsymbol{b_+}$}

\begin{figure}[ht]
\begin{center}
\includegraphics[width=7cm]{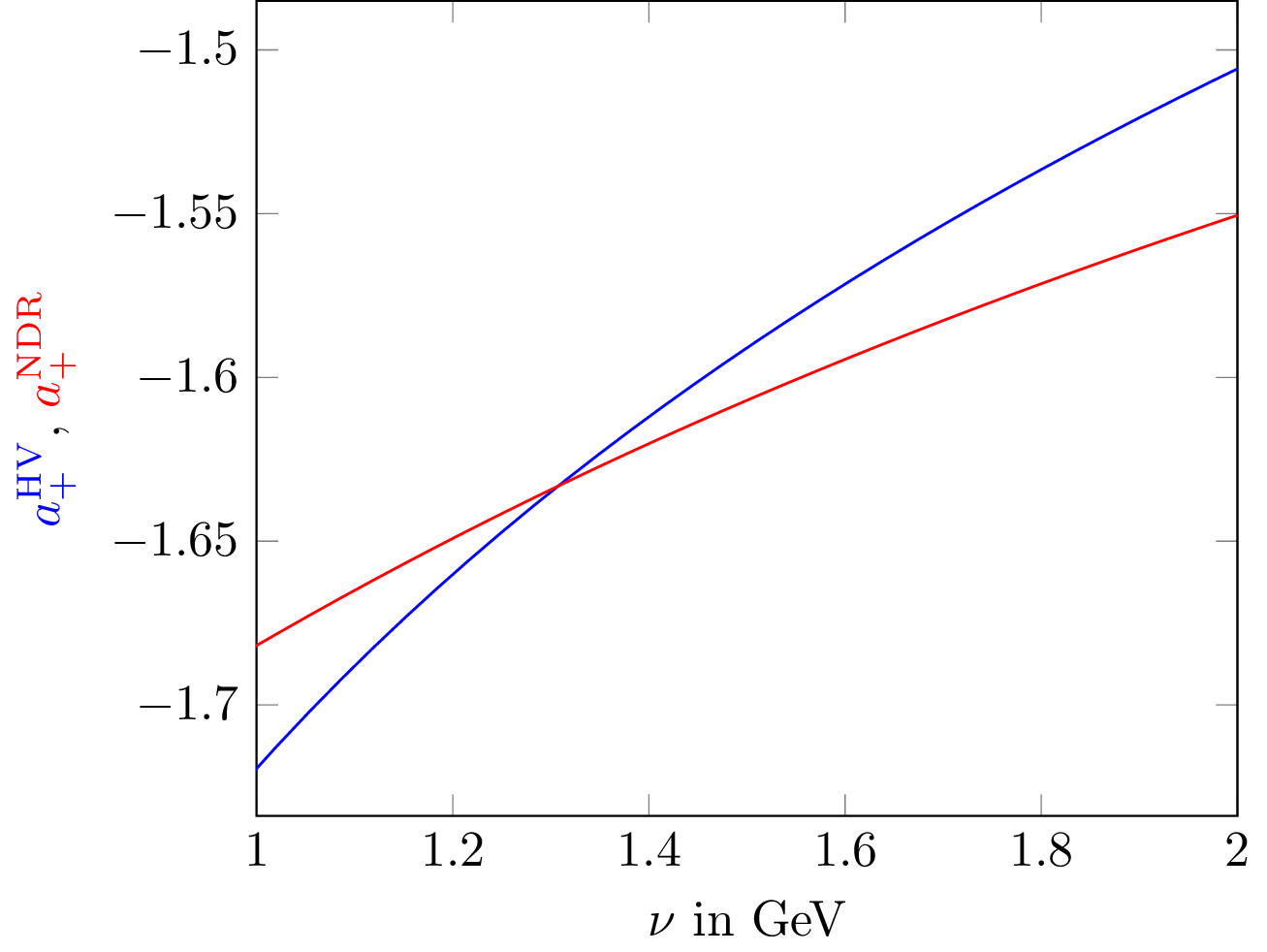} ~\hspace*{0.6cm}~ \includegraphics[width=7cm]{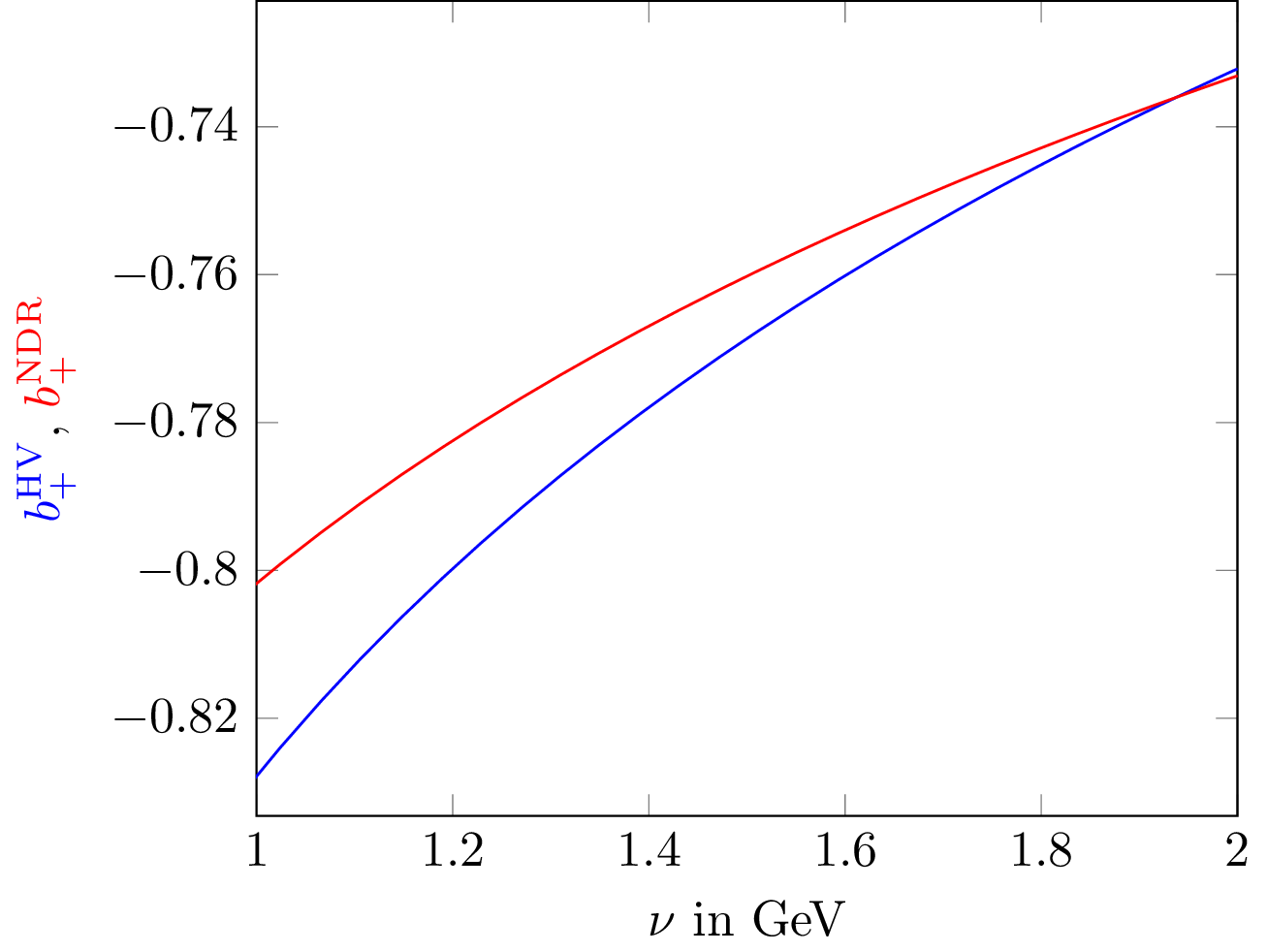}
\end{center}

\caption{  The evolution of  $a_+$ and $b_+$ with respect to $\nu$ in both NDR and HV schemes, for $M=1$ GeV. \label{Fig:11}}
\end{figure}

Collecting the various contributions to $a_+$ and to $b_+$ from the model proposed in this section, 
we end up, according to eq. \rf{intrinsic_a_b}, with the following expressions :
\bea
a_+ &=& 
\int_0^\infty \frac{dx}{x} \,\frac{\rho_+^{\pi\pi} (x)}{G_F M_K^2} \nonumber\\
&& \hspace*{1cm}+
\frac{f^{K^\pm\pi^\mp}_+ (0) }{4 \pi} \times
16 \pi^2  \left( \frac{1}{\sqrt{2}} V_{us}^* V_{ud} \right) 
\bigg\{ \frac{C_{7V} (\nu)}{\alpha}  + \sum_I C_I(\nu) 
\bigg[  \xi_{00}^I  -  \xi_{01}^I
\left(  \ln \frac{M^2}{\nu^2} - \gamma_E \right)
\bigg]
\bigg\}
,
\nonumber\\
b_+ &=&
\int_0^\infty \frac{dx}{x^2} \,\frac{\rho_+^{\pi\pi} (x)}{G_F}
+\frac{f^{K^\pm\pi^\mp}_+ (0) }{4 \pi} \times
16 \pi^2 \left( \frac{1}{\sqrt{2}} V_{us}^* V_{ud} \right) 
\, \frac{\pi^2}{6} \, \frac{M_K^2}{M^2} \sum_I C_I(\nu) \,\xi_{01}^I 
\nonumber\\
& & + \,
\frac{f^{K^\pm\pi^\mp}_+ (0) }{4 \pi} \times \lambda_+\,\frac{M_K^2}{M_\pi^2}\times
16 \pi^2  \left( \frac{1}{\sqrt{2}} V_{us}^* V_{ud} \right) 
\bigg\{ \frac{C_{7V} (\nu)}{\alpha}  + \sum_I C_I(\nu) 
\bigg[  \xi_{00}^I  -  \xi_{01}^I
\left(  \ln \frac{M^2}{\nu^2} - \gamma_E \right)
\bigg]
\bigg\}
\nonumber\\
& & - \,
\frac{1}{60} \, 
\left( \frac{M_K^2}{M_\pi^2} \right)^2 {  \alpha_{+} \left( 1 - \frac{\beta_{+}}{\alpha_{+}} \frac{s_0}{M_\pi^2}  \right)}
.
\label{aplus_bplus_tot} 
\eea
\noindent  
Numerically, for $ 1\, {\rm GeV} \leqslant \nu \leqslant 2\, {\rm GeV}$, $M=1\,{\rm GeV}$,  
and considering only contributions from $C_1$ and $C_2$, one obtains 
\begin{align}
&a_+ = -1.58 + \begin{cases} [-0.10,0.03] & {\rm NDR} \\ [-0.14,0.07] & {\rm HV} \end{cases}  & & b_+ = -0.76+ 
\begin{cases} [-0.04,0.03] & {\rm NDR} \\ [-0.07,0.03] & {\rm HV} \end{cases} .
\end{align}
The residual $\nu$-dependence of $a_+$ and $b_+$ is depicted in figure \ref{Fig:11}.

This result should mainly be viewed as a first serious attemp to evaluate $a_+$ and $b_+$.
Improvements, for instance on the description of $W_+^{\pi\pi} (z)$, or the inclusion of QCD
corrections in $W_+^{\rm res} (z)$, are clearly required before realistic error bars can be
assigned to the values displayed in eq. \eqref{aplus_bplus_tot}. We come back to these issues in the next section.
Nevertheless, we find it encouraging that the outcome of the rather simple approach followed
in the present section comes rather close to the values obtained from the experimental data
in Sec. \ref{section:data}. This is particularly true for $b_+$.

\section{Summary, conclusions, outlook}\label{summary}
\setcounter{equation}{0}

In this final section, we wish to summarize the content of this article,
going successively through the main aspects of the issues that have been 
addressed, roughly following the list of items given at the end of the introduction.
For each item, we provide conclusions and/or critical remarks, as well as an outline
of perspectives for future improvements. 

\subsection{Extracting $\boldsymbol{a_{+,S}}$ and $\boldsymbol{b_{+,S}}$ from recent data}

Our first concern was to establish the values of the constants $a_{+,S}$ and 
$b_{+,S}$ that are provided by recent experimental data. In the case of $K^\pm\to\pi^\pm e^+e^-$,
we have shown that a combined fit to the data on the decay distribution 
from the two high-statistics experiments BNL-E865 and NA48/2 clearly favours 
the solution where $a_+$ and $b_+$ are both negative, with $\vert a_+ \vert$ and 
$\vert b_+ \vert$ comparable in size. 

We have also performed fits to the data keeping, in addition to
$a_+$ and $b_+$, the curvature $\beta_+$ of
the $K^\pm\to\pi^\pm\pi^+\pi^-$ Dalitz plot as a free parameter, and
have found that somewhat smaller (in absolute value) values than
those obtained by direct determinations from $K\to\pi\pi\pi$ data
are preferred.

In order to make comparisons between different approaches more convenient, we have
introduced intrinsic definitions of the coefficients $a_{+,S}$ and $b_{+,S}$ in terms
of the values of the form factors and of their slopes at the origin $z=0$.

\subsection{The low-energy expansion of the form factors to two loops}

The extraction of $a_{+,S}$ and $b_{+,S}$ from data was done using the expressions
$W_{+,S;{\rm b1L}}(z)$ of the form factors given in ref. \cite{DAmbrosio:1998gur}.
Our second concern was then to establish whether two-loop corrections not accounted
for by these expressions might have an effect on these determinations.

We have provided the full two-loop expression of the form factors, taking only the
singularities due to two-pion intermediate states into account, in analogy with 
ref. \cite{DAmbrosio:1998gur}. These two-loop
expressions are based on the complete one-loop expression of the partial-wave 
projections $f_1^{K\pi\to\pi^+\pi^-}(s)$, which include also $\pi\pi$ rescattering 
in the crossed channels. These features are not accounted for by the expressions of
$W_{+,S;{\rm b1L}}(z)$ given in ref. \cite{DAmbrosio:1998gur}.

From a numerical point of view, these effects turn out to be quite small in the region
of $z$ corresponding to the phase space for the the $K\to\pi\ell^+\ell^-$
decays. In practice, one may thus use the simpler form $W_{\rm b1L}(z)$ in order 
to analyze the data, instead of the full two-loop expression $W_{\rm 2L}(z)$, without 
significant impact on the determinations of $a_{+,S}$ and $b_{+,S}$. This also
strongly suggests that still higher-order corrections are quite small and
are likely not { to} modify the picture in any substantial way.

We have also pointed out that the existing one-loop calculations suggest that
kaon loops could possibly have a sizeable effect on $a_+$ and $a_S$. However,
making a quantitative statement on this issue would require a complete two-loop 
calculation in the usual framework of three-flavour chiral perturbation theory,
including the computation of those Feynman graphs that do not {exhibit} non-trivial
analyticity properties.

\subsection{Contribution from the two-pion state}

We have addressed the phenomenological evaluation of the contribution from 
the two-pion state to $a_{+}$ and $b_{+}$ upon writing an unsubtracted dispersion 
relation for the form factor $W_+(z)$, from which sum rules for $a_{+}$ and $b_{+}$
can be obtained. The absorptive part of the dispersion relation is provided by the 
electromagnetic form factor of the pion $F_V^\pi(s)$ and by the $P$-wave projection 
$f_1^{K\pi\to\pi^+\pi^-}(s)$ of the $K^\pm\pi^\mp\to\pi^+\pi^-$ amplitude.
For { these}, we have used a very {  simple} approach, where these two quantities are constructed
through unitarization with the inverse amplitude method of their one-loop expressions 
in the chiral expansion. Nevertheless, this simple description leads to numerical
results that lie in the ballpark of the values extracted from data. Constructing or using more realistic
representations of $F_V^\pi(s)$ and of $f_1^{K\pi\to\pi^+\pi^-}(s)$ is certainly an aspect 
where improvements are possible.

Indeed, there exist in the literature more elaborate representations of the electromagnetic
form factor of the pion that describe data in a wide range of momentum transfer, see for instance
refs. \cite{Gounaris:1968mw,Gasser:1990bv,Guerrero1997,Oller:2000ug,Pich:2001pj,Leutwyler:2002hm,Bruch:2004py,Czyz:2010hj,Lomon:2016eyp}
for a representative sample.
In the case of the partial waves $f_1^{K\pi\to\pi^+\pi^-}(s)$, recourse to a similar data driven description
is unfortunately not possible. Moreover, the simple IAM unitarization procedure we have considered 
does not appropriately account for their full analyticity structure. More involved 
methods, like numerical implementation of the the Khuri-Treiman representation \cite{Khuri:1960zz}, 
which has been used in other instances, see for instance refs. \cite{Descotes-Genon:2014tla,Kambor:1995yc,Anisovich:1996tx}, 
are available and would represent a significant improvement in the description of these partial waves beyond the low-energy region.
It would also be interesting to see whether the problem caused, within the IAM, by the too small value
of $\alpha_+/\beta_+$ persists when a different unitarization method is considered.
Another interesting possibility would be to also include the $K\overline{K}$ intermediate states into 
the dispersive representation, which however requires a two-channel analysis \cite{Albaladejo:2017hhj}.

\subsection{Matching with the short-distance regime}

Whereas the form factors $W_+(z)$ and $W_S(z)$ are clearly dominated by low- or intermediate-energy physics,
the time-ordered product of the electromagnetic current with the lagrangian density for $\vert\Delta S\vert = 1$ 
transitions is singular at short distances and needs to be renormalized. This renormalization is implemented
through the operator $Q_{7V}$ and its Wilson coefficient $C_{7V}(\nu)$. As a result, the form factors behave,
in the asymptotic Euclidean region, as $\sim\ln(-s/\nu^2)$, where $\nu$ is the renormalization scale. At the 
phenomenological level, this short-distance behaviour results from the pile-up of more and more {  complicated} 
intermediate states, with higher and higher thresholds, until the {  region} where the QCD continuum sets in 
is reached. We have described this process through a, necessary infinite, set of zero-width resonances.
In the absence of QCD corrections, we have shown that it is possible to adjust the couplings of these resonances
such as to reproduce the correct high-energy behaviour. Working only at lowest order leaves a rather strong
sensitivity to the subtraction scale. Extending the resonance model in order to account also for the 
order ${\cal O}(\alpha_s)$ QCD effects in the high-energy part,  which are to a large extent known from {renormalization group} argument given in Sec. \ref{sect:short-dist}, would probably reduce this dependence on the short-distance scale.

\subsection{Conclusion}

This study was undertaken with the aim of exploring the possibility to achieve a determination of the
constants $a_{+,S}$ and $b_{+,S}$, describing the decay distribution of the $K^\pm (K_S) \to\pi^\pm (\pi^0)\ell^+\ell^-$
decay modes, such as to assess, through confrontation with present and forthcoming experimental data, 
the amount (if any!) of violation of lepton flavour universality in the kaon sector.
We hope that our study demonstrates that such a determination based on a phenomenological 
approach is possible, with a reasonable amount of theoretical work, 
and that there is room for improvement in several of the aspects that
contribute to it. Such an endeavour would then be complementary to existing and future
efforts to address this issue through numerical simulations of QCD on the lattice
\cite{Isidori:2005tv,Christ:2015aha,Christ:2016mmq}. 
We plan to come back to some of the aspects involved in such a phenomenological determination 
and discussed above in future work.

\indent

\indent

\begin{acknowledgments}

\noindent
This work started while the three authors were attending the ``NA62 Kaon Physics Handbook"
meeting at the Mainz Institute for Theoretical Physics. The authors would therefore like to express 
their gratitude to MITP, its Director and its staff for their warm hospitality and for partial financial support.
{  M. K. and D.G. thank the INFN-Sezione di Napoli and the Universit\'a di Napoli Federico II for 
their hospitality during several stays that made the completion of this work possible, as well as
E. de Rafael for interesting and informative discussions. One of us (D.G.) also thanks the CPT for its hospitality.}
We also thank A. Nath for his active participation in some earlier stage of this project,
E. Goudzovski for informative discussions, as well as {M. Hoferichter}, S. Kettell and L. Tunstall for useful correspondence. G.D. is  supported in part by MIUR under Project No. 2015P5SBHT and by the INFN research initiative ENP.

\end{acknowledgments}

\appendix

\renewcommand{\theequation}{\Alph{section}.\arabic{equation}}

\section{Numerical values}\label{app:num}

In this appendix, we provide the numerical input values for several
quantities that have been used in the text. For the reader's convenience,
some of them have been gathered in table \ref{tab:numerics}. In addition, we also
need the values of the form factors $f_+^{K^\pm\pi^\mp} (s)$ and $f_+^{K_S\pi^0} (s)$,
and of their derivatives, at the origin $s=0$: 
\be
f_+^{K^\pm\pi^\mp} (s) = f_+^{K^\pm\pi^\mp} (0) \left[ 1 + \lambda_{+} \frac{s}{M_\pi^2} + \cdots \right]
,\quad
f_+^{K_S\pi^0} (s) = f_+^{K_S\pi^0} (0) \left[ 1 + \lambda_{S} \frac{s}{M_\pi^2} + \cdots \right] 
.
\lbl{K_pi_FF}
\ee
We have taken their values from the analysis of ref. \cite{Mescia:2007kn}, which gives
\be
\vert V_{us} \times f_+^{K^\pm\pi^\mp} (0) \vert = 0.2168(4) , 
\quad \vert V_{us} \times f_+^{K_S\pi^0} (0) \vert = 0.2107(10)
,
\ee
whereas
$\lambda_{+} = \lambda_{S} = 0.990(5) \times \lambda^\prime_+$, where $\lambda^\prime_+$ is the
slope of the $f_+$ form factor as measured in $K_{\ell 3}$ decays, $\lambda^\prime_+ = 24.82(1.10) \cdot 10^{-3}$.
Using the value of $V_{us}$ given in table \ref{tab:numerics}, this gives
\be
f_+^{K^\pm\pi^\mp} (0) = 0.964
,\quad
f_+^{K_S\pi^0} (0) = 0.937
,\quad
\lambda_{+} = \lambda_{S} = 24.57(1.09) \cdot 10^{-3}
.
\lbl{K_pi_FF_values}
\ee
{F}or the subthreshold parameters $\alpha$ and $\beta$ of the $\pi\pi-$scattering amplitudes, we have used 
\begin{equation}
\alpha = 1.38 ,\;\; \beta = 1.11.
\end{equation}
These values belong to the ranges determined from data in ref. \cite{Descotes02}.

\indent
\begin{table}[ht]
\renewcommand{\arraystretch}{1.6}
\begin{center}
\begin{tabular}{|c|r|c|}
\hline 
  $F_\pi$           &   $92.2$~MeV~~       &  pion decay constant, PDG   \\
  $M_\pi$           &   $139.45$~MeV~~     &  charged pion mass, PDG  \\
  $M_{K^+}$         &   $493.677$~MeV~~    &  charged kaon mass, PDG  \\
  $M_{K_S}$         &   $497.648$~MeV~~    &  neutral kaon mass, PDG  \\
  $G_F$             &   $~1.166\cdot 10^{-5}$~GeV$^{-2}$  & Fermi constant, PDG \\
  $V_{ud}$          &   $0.97417(21)$    &  CKM matrix element, PDG \\
  $V_{us}$          &   $0.2248(6)$      &  CKM matrix element, PDG \\
\hline   
  $g_8$             &   $3.61 \pm 0.28\,$~  &   $K\to\pi\pi$ amplitudes  \\
  $g_{27}$          &   $0.297 \pm 0.028$ &   V. Cirigliano et al. \cite{Cirigliano:2011ny} \\
\hline
  $\alpha_1$        &   $+93.16\pm 0.36$  &   \\ 
  $\beta_1$         &   $-27.06\pm 0.43$  &   $K\to\pi\pi\pi$ \\ 
  $\alpha_3$        &   $-6.72\pm 0.46$   &   \\
  $\beta_3$         &   $-2.22\pm 0.47$   &   Dalitz-plot parameters\\ 
  $\gamma_3$        &   $+2.95\pm 0.32$   &   \\ 
  $\xi_1$           &   $-1.83\pm 0.30$   &   J. Bijnens et al. \cite{Bijnens:2002vr}\\
  $\xi_3$           &   $-0.17\pm 0.16$   &   \\ 
  $\xi_3^\prime$    &   $-0.56\pm 0.42$   &   \\ 
\hline
\end{tabular}\end{center}
\caption{Numerical values used for the various input parameters. The first
seven entries are taken from ref. \cite{Olive:2016xmw}, and the values of the constants
$g_8$ and $g_{27}$ come from ref. \cite{Cirigliano:2011ny}. The fit of ref. \cite{Bijnens:2002vr}
provides the values for the $K\to \pi\pi\pi$ Dalitz-plot parameters $\alpha_1, \ldots \xi_3^\prime$,
which are given in units of $10^{-8}$.}
\label{tab:numerics}
\end{table}

\section{The computation of $\boldsymbol{\psi_{+,S} (s)}$}\label{app:psi}

In this appendix, we compute the function $\psi_{+,S} (s)$, which describes the one-loop correction to 
${\rm Re}\,{f}^{K\pi \to \pi^+\!\pi^-}_1 \!(s)$, the real parts of the $P$-wave projections of
the amplitudes for the processes $K \pi \to \pi^+\!\pi^-$,
where $K\pi$ stands either for $K^+\pi^-$ or for $K_S \pi^0$, see eq. \rf{psi1_defined}.
In order to obtain $\psi_{+,S} (s)$, one thus first needs to construct these amplitudes at one loop. 
This can be done, for $M_K < 3 M_\pi$, within an iterative construction, following the method that has
been described several times, e.g. in refs. \cite{Stern:1993rg,Knecht:1995tr,Zdrahal:2008bd,Kampf:2011wr} 
and \cite{DescotesGenon:2012gv,ZdrahalPhD11}. Indeed, up to and including two loops, the
two amplitudes in question have the general structure
\bea
{\mathcal M} (s,t,u) &=& {\mathcal P} (s,t,u) 
+ 16 \pi \left[ {\mathcal W}_0 (s) + 3 (t-u) {\mathcal W}_{1} (s) \right]+ 16 \pi \left[ {\mathcal W}_{0;t} (t) + 3 (u-s) {\mathcal W}_{1;t} (t) \right]
\nonumber\\
&&
+ 16 \pi \left[ {\mathcal W}_{0;u} (u) + 3 (t-s) {\mathcal W}_{1;u} (u) \right]
+ {\mathcal O}(E^8)
.
\lbl{M-amplitudes}
\eea
The absorptive parts of the functions ${\mathcal W}_{0,1} (s)$ are given in terms of 
the absorptive parts along the right-hand cut of the lowest $S$ and $P$ partial-wave 
projections of the corresponding $K\pi\to\pi^+\pi^-$ amplitudes: 
\bea
{\rm Abs}\, {\mathcal W}_0 (s) &=& {\rm Abs}\, f_0^{K\pi \to \pi^+ \pi^-} (s) \, \theta (s-4M_\pi^2)
+ {\mathcal O}(E^8)
,
\nonumber\\
{\rm Abs}\, {\mathcal W}_1 (s) &=&  
\frac{{\rm Abs}\,f_1^{K\pi \to \pi^+ \pi^-} (s)}{4 q_{\pi\pi}(s) q_{K\pi}(s)} \, \theta (s-4M_\pi^2)
+ {\mathcal O}(E^8),
\qquad
q_{ab} (s) \equiv \frac{\lambda^{1/2} (s, M_a^2 , M_b^2)}{2 \sqrt{s}}
.
\lbl{discontinuities}
\eea 
Similar expressions hold for the remaining functions ${\mathcal W}_{0,1;t} (s)$ and ${\mathcal W}_{0,1;u} (s)$,
involving the absorptive parts of the amplitudes in the crossed $s \leftrightarrow t$
and $s \leftrightarrow u$ channels, respectively. If only two-pion intermediate states are considered,
then the absorptive parts of the various functions appearing in the formula \rf{M-amplitudes}
will be given, at one loop, by the products of tree-level amplitudes for $\pi\pi$ scattering
and for the processes $K\pi\to\pi^+\pi^-$, which are simply first-order polynomials
in the Mandelstam variables. In particular, the lowest-order $K\pi\to\pi^+\pi^-$ amplitudes
are expressed in terms of the Dalitz-plot parameters $\alpha_{1,3}$, $\beta_{1,3}$ and $\gamma_3$,
in the nomenclature of refs. \cite{Kambor:1991ah} and \cite{Bijnens:2002vr}. The lowest-order
$\pi\pi$ scattering scattering amplitudes are likewise expressed in terms of the two subthreshold
parameters $\alpha$ and $\beta$ \cite{Stern:1993rg,Knecht:1995tr}.
The projection on the $P$ partial wave is given by
\bea
&&f_1^{K\pi \to \pi^+ \pi^-} (s)  \equiv 
\frac{1}{32 \pi} \int_{-1}^{+1} d (\cos\theta) \cos\theta {\mathcal M} (s , \cos \theta)
\nonumber\\
&&= 4 {\mathcal W}_1 (s) \, q_{K\pi} (s) q_{\pi\pi} (s)
\nonumber\\
&&\!\!\!\!\!\!\!
+\,
\frac{1}{16 q_{K\pi}^2(s) q_{\pi\pi}^2(s)} \int_{t_-(s)}^{t_+(s)} \!\! dt (2t + s - 3 s_0)
\times
\frac{1}{32 \pi} \left[{\mathcal P} (s , t , 3 s_0 - s -t) - {\mathcal P} (s , 3 s_0 - s -t,t) \right]
\nonumber\\
&&\!\!\!\!\!\!\!
+\,
\frac{1}{16 q_{K\pi}^2(s) q_{\pi\pi}^2(s)} \int_{t_-(s)}^{t_+(s)} \!\! dt (2t + s - 3 s_0)
\bigg[{\mathcal W}_{0;t} (t) -  {\mathcal W}_{0;u} (t) 
+ 3 (2s + t - 3 s_0)
\big[ {\mathcal W}_{1;u} (t) -  {\mathcal W}_{1;t} (t) \big]
\bigg]
\nonumber\\
&&\!\!\!\!\!\!\!
+\, {\mathcal O} (E^8)
,
\lbl{P-wave_proj}
\eea
where
\be
s_0 = M_\pi^2 + \frac{M_K^2}{3},\qquad
t-u = 4 q_{K\pi} (s) q_{\pi\pi} (s) \cos \theta 
,\qquad
t_{\pm} (s) = \frac{3 s_0 - s}{2} \pm 2 q_{K\pi}(s) q_{\pi\pi} (s)
.
\ee
Restricting oneself to the contributions from
two-pion intermediate states only, one can then write
\bea
&&
{\lefteqn{
\left\{{\mathcal W}_{0;t} (t) -  {\mathcal W}_{0;u} (t) 
+ 3 (2s + t - 3 s_0) \big[ {\mathcal W}_{1;u} (t) -  {\mathcal W}_{1;t} (t) \big]
\right\}_{{\mathcal O}(E^4)} \ = \qquad\qquad\qquad~}}
\nonumber\\
&& 
\qquad\qquad
= -  \frac{1}{16 \pi} \frac{1}{F_\pi^2} \bigg[ w^{(0)}  + w^{(1)}  t + w^{(2)} t^2 
-  \frac{1}{6} \frac{\alpha_{  +,S}}{M_\pi^2} \beta (t-4M_\pi^2) s \bigg] {\bar J}_{\pi\pi} (t)
.
\lbl{w_coeff}
\eea
The coefficients $w^{(n)}$ depend on the Dalitz-plot parameters,
and are different for each amplitude. They are
given in Eqs. \rf{w_coeff_+} and \rf{coeff_w_S} below.  At order ${\mathcal O} (E^4)$,
and for $s > 4 M_\pi^2$, the function ${\mathcal W}_1 (s)$ is given by
\be
{\mathcal W}_1 (s) =  \frac{1}{6} \frac{\alpha_{  +,S}}{M_\pi^2}
\times {\varphi}_{1;\pi\pi}^{+-;+-} (s) \, {\bar J}_{\pi\pi} (s)
,
\ee
where ${\varphi}_{1;\pi\pi}^{+-;+-} (s)$, given in eq. \rf{LO_pi-pi_P-wave},
is a polynomial of first order in $s$.
At next-to-leading order, the contribution involving the polynomial 
${\mathcal P} (s , t , u)$ can be given the following parameterization:
\bea
\frac{1}{2} \left[ {\mathcal P} (s , t , u) - {\mathcal P} (s , u, t) \right]
&=&
\frac{1}{2} \left[ \left( \alpha_{+,S} + \Delta \alpha_{+,S} \right) \frac{t-u}{M_\pi^2} 
+ \left( \beta_{+,S} + \Delta \beta_{+,S} \right) \frac{(t-u) (s-s_0)}{M_\pi^4}
\right]
\lbl{P_antisym}
\eea
This form of the polynomial part corresponds to the most general one allowed by
the chiral counting and crossing. 
The contributions $\Delta \alpha_{+,S}$ and $\Delta \beta_{+,S}$
are then fixed such that the expansion of the real part of the partial wave at $s=s_0$ is entirely
given by the Dalitz-plot parameters, up to terms quadratic in the difference $s-s_0$,
\be
{\rm Re} \, f_1^{K\pi \to \pi^+ \pi^-} (s) \big\vert_{{\mathcal O} (E^4)} =
 \frac{1}{96 \pi}  \frac{1}{M_\pi^2} 
\left[ \alpha_{+,S} 
 + \beta_{+,S} \frac{s-s_0}{M_\pi^2} + {\mathcal O}((s-s_0)^2)
\right] \times  \lambda_{K\pi}^{1/2} (s) \sigma_\pi (s)
.
\ee
In other words, writing
\be
{\psi}_{+,S} (s) = + \frac{1}{96 \pi}  \frac{1}{M_\pi^2} 
\left[ \Delta\alpha_{+,S} 
 + \left( \beta_{+,S} + \Delta \beta_{+,S} \right) \frac{s-s_0}{M_\pi^2} 
\right]
+
{\psi}_{+,S}^{{\rm loop}} (s)
,
\ee
one requires
\be
\frac{1}{96 \pi}  \frac{1}{M_\pi^2} 
\left\{ \Delta \alpha_{+,S} ~;~ \Delta \beta_{+,S} \right\}
= \left\{ - {\rm Re} \, {\psi}_{+,S}^{{\rm loop}} (s_0) ~;~
- M_\pi^2 \, \frac{d }{ds} {\rm Re} \, {\psi}_{+,S}^{{\rm loop}}  (s) \Big\vert_{s=s_0} \right\}
,
\lbl{DeltaC_and_DeltaD}
\ee
which amounts to 
\be
{\psi}_{+,S} (s) = + \frac{1}{96 \pi}  \frac{\beta_{+,S}}{M_\pi^2} 
 \frac{s-s_0}{M_\pi^2} 
+
{\psi}_{+,S}^{{\rm loop}} (s) - {\rm Re} \,{\psi}_{+,S}^{{\rm loop}} (s_0)
- (s-s_0) \frac{d}{ds} {\rm Re} \, {\psi}_{+,S}^{{\rm loop}} (s)\Big\vert_{s=s_0}
.
\ee
The expression of ${\psi}_{+,S} (s)$, and hence of ${\psi}_{+,S}^{{\rm loop}} (s)$, can be obtained 
from eq. \rf{P-wave_proj}, combined with the decomposition \rf{w_coeff}. The corresponding integrals 
of ${\bar J}_{\pi\pi} (t)$ can be done explicitly, and are given by \cite{ZdrahalPhD11,Kampf:2018}
\be
\int_{t_-(s)}^{t_+(s)} \!\!\!\!\! dt\, t^n \, {\rm Re} \, {\bar J}_{\pi\pi}(t) =
\frac{{\lambda}_{K\pi}^{1/2} (s)}{\pi} \left[
\kappa^{(n)}_0 (s) {\mathfrak k}_0(s)
+
\kappa^{(n)}_1 (s)  {\mathfrak k}_1(s)
+
\kappa^{(n)}_2 (s) {\mathfrak k}_2(s)
+
\kappa^{(n)}_3  {\mathfrak k}_3(s)
\right]
\lbl{Jbar_integrals}
\ee
in terms of the functions
\begin{align}
&{\mathfrak k}_0(s) = \frac{1}{16 \pi} \sigma_\pi (s) ,
\quad
&&{\mathfrak k}_1(s) = \frac{1}{16 \pi} L_{\pi\pi} (s) , \nonumber\\
&{\mathfrak k}_2(s) = \frac{1}{16 \pi} \sigma_\pi (s) s \frac{M_{\pi\pi} (s)}{\lambda_{K\pi}^{1/2} (s)} ,
\quad
&&{\mathfrak k}_3(s) = - \frac{1}{16 \pi} M_\pi^2 \frac{M_{\pi\pi} (s)}{\lambda_{K\pi}^{1/2} (s)} L_{\pi\pi} (s)
.
\lbl{frak_k}   
\end{align}
The expressions for the functions $\kappa^{(n)}_i (s)$ are given in Eqs. \rf{kappa_decomp}, \rf{d_coeff}, \rf{c_coeff}, and \rf{kappa_bar} below. 
We have checked that they agree with the ones given in \cite{ZdrahalPhD11}.

The functions $L_{\pi\pi} (s)$ and  $M_{\pi\pi}(s)$
are given, for $s > 4 M_\pi^2$ and with $\Delta_{K\pi} \equiv M_K^2 - M_\pi^2$, by
\bea
L_{\pi\pi} (s) &=& \ln \frac{1-\sigma_\pi (s)}{1+\sigma_\pi (s)}
,
\\
M_{\pi\pi} (s) &=& - \ln \left[ 1 - \frac{\Delta_{K\pi}}{s} + \frac{\lambda^{1/2}_{K\pi} (s)}{s} \right]
- \ln \left[ 1 - \frac{\Delta_{K\pi}}{s} - \frac{\lambda^{1/2}_{K\pi} (s)}{s} \right]^{-1} \nonumber\\
&=& - 2 \ln \left[ 1 - \frac{\Delta_{K\pi}}{s} + \frac{\lambda^{1/2}_{K\pi} (s)}{s} \right] + \, \ln \frac{4 M_\pi^2}{s}
.
\nonumber
\eea
One may notice that the functions ${\mathfrak k}_2(s)$ and ${\mathfrak k}_3(s)$
involve $\lambda_{K\pi}^{1/2} (s)$ only through the ratio $M_{\pi\pi} (s) / \lambda_{K\pi}^{1/2} (s)$,
which does not depend on the choice of the sign of $\lambda_{K\pi}^{1/2} (s)$.
One then obtains
\bea
&&\psi_{+,S}^{{\rm loop}} (s) 
=
  \frac{1}{6} \frac{\alpha_{+,S}}{M_\pi^2}
\times {\varphi}_{1;\pi\pi}^{+-;+-} (s) \times
{\rm Re} \, {\bar J}_{\pi\pi} (s)
\nonumber\\
&&
-\,
\frac{1}{\lambda_{K\pi} (s)} \, \frac{1}{16 \pi^2 F_\pi^2} \, \frac{s}{s - 4 M_\pi^2}
\sum_{i=0}^3 \frac{{\mathfrak k}_i(s)}{\sigma_\pi (s)} \Bigg\{
w_{+,S}^{(0)} \left( (s - 3 s_0) \kappa_i^{(0)} (s) + 2 \kappa_i^{(1)} (s)  \right)
\nonumber\\
&&\quad
+ \,
w_{+,S}^{(1)} \left( (s - 3 s_0) \kappa_i^{(1)} (s) + 2 \kappa_i^{(2)} (s)  \right)
+
w_{+,S}^{(2)} \left( (s - 3 s_0) \kappa_i^{(2)} (s) + 2 \kappa_i^{(3)} (s)  \right)
\nonumber\\
&&\quad
+ \,
\frac{1}{3} \beta \, \alpha_{+,S} \frac{s}{M_\pi^2}
\left[ 2 M_\pi^2 (s- 3 s_0) \kappa_i^{(0)} (s) - \frac{1}{2} ( s -M_K^2 - 11 M_\pi^2) \kappa_i^{(1)} (s)
- \kappa_i^{(2)} (s) \right]
\Bigg\}
.
\lbl{Psi1_loop}
\eea
Introduce next the functions
\be
{\bar{\mathfrak K}}_i(s) \equiv \frac{s}{\pi} \int_{4 M_\pi^2}^{\infty} \frac{dx}{x}\, \frac{{\mathfrak k}_i(x)}{x-s-i0}
,
\ee
whose absorptive parts are given by ${\rm Abs} \, {\bar{\mathfrak K}}_i(s) = {\mathfrak k}_i(s)\,\theta(s - 4 M_\pi^2)$.
In the first two cases, one easily finds expressions in terms of ${\bar J}_{\pi\pi} (s)$ \cite{Knecht:1995tr,Kampf:2011wr},
\be
{\bar{\mathfrak K}}_0(s) = {\bar J}_{\pi\pi} (s),\quad
{\bar{\mathfrak K}}_1(s) = \frac{1}{2} \, \frac{s}{s - 4 M_\pi^2}
\left[ 16 \pi^2 {\bar J}_{\pi\pi}^{\,2} (s) - 4 {\bar J}_{\pi\pi} (s) +\frac{1}{4 \pi^2} \right]
.
\lbl{K0_andK1}
\ee
What is actually required is a set of functions with absorptive parts
given, for $s>4 M_\pi^2$, by ${\mathfrak k}_i(x) / \lambda_{K\pi} (s)$, or by
${\mathfrak k}_i(x) / s\lambda_{K\pi} (s)$, or even ${\mathfrak k}_i(x) / s^2\lambda_{K\pi} (s)$.
Indeed, the functions $\kappa_i^{(n)} (s)$, for $i=0,1,2$, are not polynomials in $s$, but have the following general
structure,
\be
\kappa_i^{(n)} (s) = {\bar\kappa}_i^{(n)} (s) + c_i^{(n)} \frac{\Delta_{K\pi}}{s} + d_i^{(n)} \frac{\Delta_{K\pi}^2}{s^2}
,
\lbl{kappa_decomp}
\ee
where ${\bar\kappa}_i^{(n)} (s)$ are now  polynomials in $s$, displayd in eq. \rf{kappa_bar} below,
and $c_i^{(n)}$, $d_i^{(n)}$ are
numerical coefficients. Actually, the only non vanishing $d_i^{(n)}$ coefficients are
\be
d_1^{(3)} = M_\pi^4 \Delta_{K\pi}/4  \qquad d_2^{(2)} = M_\pi^2 \Delta_{K\pi}/6  \qquad
d_2^{(3)} = M_\pi^2 ( 3 M_K^2 + 5 M_\pi^2) \Delta_{K\pi}/12
\lbl{d_coeff}
,
\ee
while the non vanishing coefficients $c_i^{(n)}$ read
\begin{align*}
&c_0^{(2)} = - \frac{7}{9} M_\pi^2 \Delta_{K\pi} && c_0^{(3)} = - \frac{1}{72} M_\pi^2 \Delta_{K\pi} (81 M_K^2 + 239 M_\pi^2) \nonumber\\
&c_1^{(1)} = - \frac{1}{2} M_\pi^2 && c_1^{(2)} = - \frac{1}{2} M_\pi^2 (M_K^2 + M_\pi^2) && \hspace*{-0.5cm}c_1^{(3)} = - \frac{1}{4} M_\pi^2 (2 M_K^4 + 7 M_K^2 M_\pi^2 + M_\pi^4) \nonumber\\
&c_2^{(0)} = - \frac{1}{2} && c_2^{(1)} = - \frac{\Delta_{K\pi}}{4} &&  \hspace*{-0.5cm}c_2^{(2)} = - \frac{1}{6} (M_K^2 + 5 M_\pi^2) \Delta_{K\pi} \nonumber
\end{align*}
\begin{equation}
c_2^{(3)} = - \frac{1}{24} (3 M_K^4 + 34 M_K^2 M_\pi^2 + 59 M_\pi^4) \Delta_{K\pi} .\lbl{c_coeff}
\end{equation}
Writing
\be
\lambda_{K\pi} (s) = (s - M_+^2) (s-M_-^2),\ M_\pm = M_K \pm M_\pi
\ee
and using the decomposition of products of fractions, one obtains
\be
\frac{1}{M_+^2 - M_-^2} \,
{\rm Abs}\,\left[ \frac{{\bar{\mathfrak K}}_i(s) - {\bar{\mathfrak K}}_i(M_+^2)}{s-M_+^2} -
\frac{{\bar{\mathfrak K}}_i(s) - {\bar{\mathfrak K}}_i(M_-^2)}{s-M_-^2} \right]
=
\frac{{\mathfrak k}_i(s)}{\lambda_{K\pi} (s)} \, \theta ( s - 4 M_\pi^2)
,
\ee
\be
\frac{1}{M_+^2 - M_-^2} \,
{\rm Abs}\,\left[ \frac{1}{s - M_+^2} \left(
\frac{{\bar{\mathfrak K}}_i(s)}{s} - \frac{{\bar{\mathfrak K}}_i(M_+^2)}{M_+^2}
\right)
-\frac{1}{s - M_-^2} \left(
\frac{{\bar{\mathfrak K}}_i(s)}{s} - \frac{{\bar{\mathfrak K}}_i(M_-^2)}{M_-^2}
\right)
\right]
=
\frac{{\mathfrak k}_i(s)}{s \lambda_{K\pi} (s)}  \, \theta ( s - 4 M_\pi^2)
,
\ee
and
\bea
&&\!\!\!\!\!
\frac{1}{M_+^2 - M_-^2} \,
{\rm Abs}\bigg[ \frac{1}{M_+^2(s - M_+^2)} \left(
\frac{{\bar{\mathfrak K}}_i(s)}{s} - \frac{{\bar{\mathfrak K}}_i(M_+^2)}{M_+^2}
\right)
-\frac{1}{M_-^2(s - M_-^2)} \left(
\frac{{\bar{\mathfrak K}}_i(s)}{s} - \frac{{\bar{\mathfrak K}}_i(M_-^2)}{M_-^2}
\right) \nonumber \\
&&\hspace*{3cm}-\left(\frac{1}{M_+^2} - \frac{1}{M_-^2} \right) \frac{{\bar{\bar{\mathfrak K}}}_i(s)}{s^2}
\bigg]
\nonumber\\
&& \qquad
=
\frac{{\mathfrak k}_i(s)}{s^2 \lambda_{K\pi} (s)}  \, \theta ( s - 4 M_\pi^2)
,
\eea
with, in this last case,
\be
{\bar{\bar{\mathfrak K}}}_i(s) = 
\frac{s^2}{\pi} \int_{4 M_\pi^2}^{\infty} \frac{dx}{x^2}\, \frac{{\mathfrak k}_i(x)}{x-s-i0}
=
{\bar{\mathfrak K}}_i(s) - s {\bar{\mathfrak K}}_i^\prime(0)
.
\ee
This suggests to introduce the following functions:
\bea
{\bar{\mathfrak K}}_i^{(\lambda ; 0)} (s) &=&
\frac{1}{4}
\left[ \frac{M_\pi^2}{s - M_+^2} \left(
{\bar{\mathfrak K}}_i(s) - \frac{s}{M_+^2} {\bar{\mathfrak K}}_i(M_+^2)
\right)
-\frac{M_\pi^2}{s - M_-^2} \left(
{\bar{\mathfrak K}}_i(s) - \frac{s}{M_-^2} {\bar{\mathfrak K}}_i(M_-^2)
\right)
\right]
\nonumber\\
&\equiv & 
\frac{s}{\pi} \int_{4 M_\pi^2}^\infty \frac{dx}{x} 
\, \frac{M_K M_\pi^3}{\lambda_{K\pi} (x)} \frac{{\mathfrak k}_i(x)}{x-s-i0}
,
\eea
\bea
{\bar{\mathfrak K}}_i^{(\lambda ; 1)} (s) &=&
\frac{1}{4}
\Bigg[ \frac{M_\pi^4}{M_+^2 (s - M_+^2)} \left(
{\bar{\mathfrak K}}_i(s) - \frac{s}{M_+^2} {\bar{\mathfrak K}}_i(M_+^2)
\right)
-\frac{M_\pi^4}{M_-^2 (s - M_-^2)} \left(
{\bar{\mathfrak K}}_i(s) - \frac{s}{M_-^2} {\bar{\mathfrak K}}_i(M_-^2)
\right)
\nonumber\\
&&
\quad + \, \frac{4 M_K M_\pi^5}{(M_K^2 - M_\pi^2)^2} \frac{{\bar{\bar{\mathfrak K}}}_i(s)}{s}
\Bigg]
\nonumber\\
&\equiv & 
\frac{s}{\pi} \int_{4 M_\pi^2}^\infty \frac{dx}{x} 
\, \frac{M_K M_\pi^5}{x \lambda_{K\pi} (x)} \frac{{\mathfrak k}_i(x)}{x-s-i0}
,
\eea
which are, at least partly, characterized by
\be 
{\rm Abs} \, {\bar{\mathfrak K}}_i^{(\lambda ; p)}(s) = \left(\frac{M_\pi^2}{s}\right)^p 
\, \frac{M_K M_\pi^3}{\lambda_{K\pi} (s)} \, {\mathfrak k}_i(s)
,\quad
{\bar{\mathfrak K}}_i^{(\lambda ; p)}(0) = 0
.
\ee
As mentioned before, in the cases $i=0$ and $i=1$, one can establish explicit
expressions in terms of the function ${\bar J}_{\pi\pi} (s)$, cf. eq. \rf{K0_andK1},
with, in addition,
\be
{\bar{\bar{\mathfrak K}}}_0 (s) = {\bar J}_{\pi\pi}(s) - \frac{1}{96 \pi^2} \frac{s}{M_\pi^2}
,
\qquad
{\bar{\bar{\mathfrak K}}}_1 (s) = \frac{1}{2} \, \frac{s}{s - 4 M_\pi^2}
\left[ 16 \pi^2 {\bar J}_{\pi\pi}^{\,2} (s) - 4 {\bar J}_{\pi\pi} (s) +\frac{1}{4 \pi^2} \right]
+ \frac{1}{32 \pi^2} \frac{s}{M_\pi^2}
.
\ee
In the cases $i=2$ and $i=3$, no explicit expressions are known,
except in the case $M_K=M_\pi$ \cite{Knecht:1995tr}, and one has
to use their dispersive representations.
With these functions at disposal, one can now construct a function whose
discontinuity reproduces the right-hand side of eq. \rf{disc_FF_NNLO}.
Recalling the contribution already evaluated in eq. \rf{disc_FF_NNLO_part1},
one ends up with the following two-loop representation of the form factor in
an energy range where singularities due to other states that two-pion states
can be described by a subtraction polynomial of first order in $s$,
\bea\lbl{disc_FF_NNLO_final}
&&
\frac{W_{+,S;{\rm 2L}} (s/M_K^2)}{16 \pi^2 M_K^2} 
=
\left( - \frac{G_{\rm F}}{\sqrt{2}} V_{us}^* V_{ud} \right) \left( {\rm A}_{+,S} + {\rm B}_{+,S} \frac{s}{M_K^2} \right)
\nonumber\\
&&
+ \, \frac{1}{6} \frac{1}{M_\pi^2} 
\left[ \alpha_{+,S} \! \left( 1 + a_V^\pi s \right) + \Delta \alpha_{+,S} 
 + \left( \beta_{+,S} + \Delta \beta_{+,S} \right) \frac{s-s_0}{M_\pi^2}
\right]  \frac{s - 4 M_\pi^2}{s} \, {\bar J}_{\pi\pi} (s) \nonumber\\
&&
+ \,
\frac{1}{36} \, \frac{\beta \cdot \alpha_{+,S}}{M_\pi^2}
\frac{(s - 4 M_\pi^2)^2}{s F_\pi^2} 
\,{\bar J}_{\pi\pi}^{\,2} (s) -\,
\frac{1}{16 \pi^2 F_\pi^2}
\times \sum_{i=0}^3  \left[
{\bar{\mathfrak K}}_i^{(\lambda ; 0)} (s) {\mathfrak p}_i (s) +
{\bar{\mathfrak K}}_i^{(\lambda ; 1)} (s) \frac{\Delta_{K\pi}^2}{s M_\pi^2} \, {\mathfrak q}_i
\right]
\!.
\eea
In this last expression, the following quantities have been introduced: 
\be
{\mathfrak q}_0 = {\mathfrak q}_3 = 0
\quad
{\mathfrak q}_1 = \frac{1}{2} \frac{M_\pi}{M_K} \Delta_{K\pi} w_{+,S}^{(2)}
\quad
{\mathfrak q}_2 = \frac{1}{3} \frac{\Delta_{K\pi}}{M_K M_\pi} \left[  w_{+,S}^{(1)} +  (M_K^2 + M_\pi^2) w_{+,S}^{(2)} \right]
,
\ee
and
\bea
M_K M_\pi^3 {\mathfrak p}_i (s) &=&
\sum_{n=0}^2 w_{+,S}^{(n)} \left[
d_i^{(n)} \frac{\Delta_{K\pi}^2}{s} + 2 \left( \kappa_i^{(n+1)} (s) - d_i^{(n+1)} \frac{\Delta_{K\pi}^2}{s^2} \right)
+ (s - 3 s_0) \left( \kappa_i^{(n)} (s) - d_i^{(n)} \frac{\Delta_{K\pi}^2}{s^2} \right)
\right]
\nonumber\\
&&
\!\!\!\!\!\!\! + \,
\frac{1}{3} \beta \, \alpha_{+,S} \frac{s}{M_\pi^2}
\left[ 2 M_\pi^2 (s- 3 s_0) \kappa_i^{(0)} (s) - \frac{1}{2} ( s -M_K^2 - 11 M_\pi^2) \kappa_i^{(1)} (s)
- \kappa_i^{(2)} (s) \right]
.
\lbl{frak_p_def}
\eea
Explicit expressions of the polynomials ${\mathfrak p}_i (s)$, as well as
of the coefficients $\Delta \alpha_{+,S}$ and $\Delta \beta_{+,S}$
defined in eq. \rf{DeltaC_and_DeltaD} can be found below. The important 
feature of eq. \rf{disc_FF_NNLO_final} one should stress is that, apart from the two
subtraction constants ${\rm A}_{+,S}$ and ${\rm B}_{+,S}$, all other quantities
are known experimentally, either from low-energy $\pi\pi$ scattering [$\alpha$, $\beta$] or from 
the Dalitz plots of the decays $K\to\pi\pi^+\pi^-$. The very last step is to trade the
subtraction constants $A_{+,S}$ and $B_{+,S}$ for the phenomenological constants
$a_{+,S}$ and $b_{+,S}$. This is done upon expanding the expression \rf{disc_FF_NNLO_final}
to first order in $s$, and making the identifications given in eq. \rf{intrinsic_a_b}.
This then leads to the two-loop expressions of the form factors displayed in eq. \rf{FF_2loop}
of the main text.

To close this appendix,  we provide explicit expressions for a certain number of quantities
which are required in order to make use of some formulas given in the text. We start with
the coefficients $w_{+,S}^{(n)}$ introduced in eq. \rf{w_coeff}, and
which depend on the channel under consideration. They read 
\bea
w_{+}^{(0)} &=& - \, \frac{4}{3} \left( \beta_1 - \frac{1}{2} \beta_3 \right) \beta s_0 
+ \frac{5}{6} \left( \beta_1 - \frac{1}{2} \beta_3 + \frac{3}{5} \sqrt{3}\,\gamma_3 \right) \alpha s_0 
- \frac{5}{6} \left( \alpha_1 - \frac{1}{2} \alpha_3 \right) (4 \beta - \alpha) M_\pi^2
,
\nonumber\\
w_{+}^{(1)} &=&  \frac{\beta}{2} \left( \beta_1 - \frac{1}{2} \beta_3 - \sqrt{3}\,\gamma_3 \right) 
\left( \frac{s_0}{ M_\pi^2} + \frac{4}{3} \right)
+ \frac{5}{2} \left( \alpha_1 - \frac{1}{2} \alpha_3 \right) \beta 
- \frac{5}{6} \left( \beta_1 - \frac{1}{2} \beta_3 + \frac{3}{5} \sqrt{3}\,\gamma_3 \right) \alpha
,
\nonumber\\
w_{+}^{(2)} &=& - \, \frac{\beta}{3 M_\pi^2} \left( \beta_1 - \frac{1}{2} \beta_3 - 2 \sqrt{3}\,\gamma_3 \right)
\lbl{w_coeff_+} 
\eea
for the channel $K^+ \pi^-$, and 
\bea
w_{S}^{(0)} &=& - \frac{2\alpha}{\sqrt{3}}  \gamma_3 s_0    
,
\nonumber\\
w_{S}^{(1)} &=&  \frac{8}{9} \sqrt{3} \beta \gamma_3 + \frac{2\sqrt{3}}{3}\,\gamma_3 \beta \frac{s_0}{ M_\pi^2} 
+ \frac{2}{\sqrt{3}} \alpha \gamma_3 
,
\nonumber\\
w_{S}^{(2)} &=& - \frac{8}{9 M_\pi^2} \sqrt{3} \beta \gamma_3 
.
\lbl{coeff_w_S}
\eea
for the channel $K_S \pi^0$.
Next, we consider the $\kappa_i^{(n)} (s)$, which were defined in eq. \rf{Jbar_integrals}
in terms of integrals of the functions ${\bar J}_{\pi\pi} (t)$, 
After the decomposition in Eqs. \rf{kappa_decomp}, \rf{d_coeff}, \rf{c_coeff}
one needs to know the polynomials ${\bar\kappa}_i^{(n)} (s)$, which read
\bea
&&
{\bar\kappa}_0^{(0)} = 3 ,
\quad {\bar\kappa}_0^{(1)} = \frac{1}{4} (-5 s + 5 M_K^2 + 11 M_\pi^2 ) ,
\quad {\bar\kappa}_0^{(2)} = \frac{7}{9} (s - M_K^2)^2 + \frac{9}{2} M_\pi^2 ( M_K^2 + M_\pi^2 - s) 
\nonumber\\
&&
{\bar\kappa}_0^{(3)} = \frac{721}{144} s^2 M_\pi^2 - \frac{1442}{144} M_K^2 M_\pi^2 s - \frac{231}{16} M_\pi^4 s
+ \frac{883}{144} M_K^4 M_\pi^2 + \frac{195}{16} M_K^2 M_\pi^4 + \frac{153}{16} M_\pi^6  + \frac{9}{16} (M_K^2 - s )^3
,
\nonumber\\
&&
{\bar\kappa}_1^{(0)} = \frac{1}{2} ,
\quad {\bar\kappa}_1^{(1)} = \frac{1}{4} ( M_K^2 + M_\pi^2 - s ) ,
\quad {\bar\kappa}_1^{(2)} = \frac{1}{6} \left[ (s - M_K^2)^2 - M_\pi^2 
\left( 5 s + 2 M_\pi^2 - 7 M_K^2 \right)  \right]
,
\nonumber\\
&&
{\bar\kappa}_1^{(3)} = \frac{M_\pi^2}{24} \left[ 25 s^2 - (56 M_K^2 + 53 M_\pi^2) s
+ 43 M_K^4 + 67 M_K^2 M_\pi^2 - 53 M_\pi^4 \right] + \frac{1}{8} (M_K^2 - s )^3
,
\nonumber\\
&&
{\bar\kappa}_2^{(0)} = \frac{1}{2} ,  
\quad {\bar\kappa}_2^{(1)} = \frac{1}{4} \left( 2 M_K^2 - s \right) ,
\nonumber\\
&&
{\bar\kappa}_2^{(2)} = \frac{1}{6} 
\left[ s^2 - s ( 3 M_K^2 + 4 M_\pi^2) + 3 (M_K^4 + 2 M_K^2 M_\pi^2 - M_\pi^4)    \right] ,
\nonumber\\
&&
{\bar\kappa}_2^{(3)} = \frac{M_\pi^2}{12} \bigg[ 11 s^2 - 10 s (3 M_K^2 + 2 M_\pi^2) + 30  M_K^4 + 18  M_K^2 M_\pi^2 - 24  M_\pi^4 \bigg] \nonumber\\
&& \hspace*{3cm}- \frac{1}{8} (s^3 - 4 s^2 M_K^2 + 6 s M_K^4 - 4 M_K^6) , 
\nonumber\\
&&
{\bar\kappa}_3^{(0)} = 1 ,
\quad
{\bar\kappa}_3^{(1)} = M_\pi^2 ,
\quad
{\bar\kappa}_3^{(2)} = 2 M_\pi^4 ,
\quad
{\bar\kappa}_3^{(3)} = 5 M_\pi^6
.
\lbl{kappa_bar}    
\eea

The explicit expressions of the polynomials ${\mathfrak p}_i (s)$ defined in eq. \rf{frak_p_def} read:
\bea
M_K M_\pi^3 {\mathfrak p}_0 (s) &=& \frac{w_{+,S}^{(0)}}{2} \left( s - M_K^2 - 7 M_\pi^2 \right)
\nonumber\\
&& \!\!\!\!\!\!\!
+ \, \frac{w_{+,S}^{(1)}}{36} \left[ 11 s^2 - 22 s M_K^2 - 90 M_\pi^2 s + 11 M_K^4 + 90 M_K^2 M_\pi^2 + 27 M_\pi^4 
- 56 M_\pi^2 \frac{\Delta_{K\pi}^2}{s}  \right]
\nonumber\\
&& \!\!\!\!\!\!\!
+ \, \frac{w_{+,S}^{(2)}}{72} \Big[ - 25 s^3 + 75 M_K^2 s^2 + 229 M_\pi^2 s^2 
- 75 M_K^4 s - 458 M_K^2 M_\pi^2 s - 783 M_\pi^4 s 
+ 25 M_K^6 
\nonumber\\
&&
\hspace*{1.5cm}+ 335 M_K^4 M_\pi^2 + \, 571 M_K^2 M_\pi^4 + 349 M_\pi^6 
- 2 M_\pi^2 ( 53 M_K^2 + 155 M_\pi^2) \frac{\Delta_{K\pi}^2}{s} \Big]
\nonumber\\
&& \!\!\!\!\!\!\! 
- \, \beta \, \alpha_{+,S} \bigg[ \frac{11 s^3}{216 M_\pi^2} 
- \frac{s^2}{108 M_\pi^2} ( 11 M_K^2 + 81 M_\pi^2 ) \nonumber \\
&&\hspace*{3cm} + \frac{s}{216 M_\pi^2} (11 M_K^4 + 162 M_K^2 M_\pi^2 + 531 M_\pi^4 )
- \frac{7}{27} \Delta_{K\pi}^2 \bigg]
\nonumber\\
M_K M_\pi^3 {\mathfrak p}_1 (s) &=& - w_{+,S}^{(0)} M_\pi^2 \left( 1 + \frac{\Delta_{K\pi}}{s} \right) \nonumber \\
&&+ \frac{w_{+,S}^{(1)}}{12} \left[ s^2 - 2 M_K^2 s - 8 M_\pi^2 s + M_K^4 + 10 M_K^2 M_\pi^2 - 11 M_\pi^4
- 6 M_\pi^2 \frac{\Delta_{K\pi}^2}{s}   \right]
\nonumber\\
&&  + \, \frac{w_{+,S}^{(2)}}{12} \Big[ - s^3 + 3 s^2 (M_K^2 + 3 M_\pi^2) - 3 M_K^4 s - 20 M_K^2 M_\pi^2 s - 27 M_\pi^4 s
+ M_K^6 + 17 M_K^4 M_\pi^2 
\nonumber\\
&& \hspace*{2cm}+ 29 M_K^2 M_\pi^4   - \, 35 M_\pi^6 - 6 M_\pi^2 (M_K^4 + 3 M_K^2 M_\pi^2 - 2 M_\pi^4) \frac{\Delta_{K\pi}}{s}  \Big] 
\nonumber\\
&& \!\!\!\!\!\!\!
- \, \beta \,\alpha_{+,S} \bigg[ \frac{s^3}{72 M_\pi^2} - \frac{s^2}{36 M_\pi^2} (M_K^2 + 4 M_\pi^2)
+  \frac{s}{72 M_\pi^2} (M_K^4 + 10 M_K^2 M_\pi^2 + 37 M_\pi^4) \nonumber \\
&&\hspace*{2cm}- \frac{1}{12} (M_K^2 - 9 M_\pi^2)\Delta_{K\pi} \bigg]
\nonumber\\
M_K M_\pi^3 {\mathfrak p}_2 (s) &=& - w_{+,S}^{(0)} M_\pi^2 \left( 1 - 2 \frac{\Delta_{K\pi}}{s} \right)
\\
&& \!\!\!\!\!\!\!
+ \, \frac{w_{+,S}^{(1)}}{12} \left[ s^2 - 3 M_K^2 s - 7 M_\pi^2 s + 3 M_K^4 + 12 M_K^2 M_\pi^2 - 15 M_\pi^4
- (M_K^2 + 11 M_\pi^2) \frac{\Delta_{K\pi}^2}{s}   \right]
\nonumber\\
&& \!\!\!\!\!\!\!
+ \, \frac{w_{+,S}^{(2)}}{12} \bigg[ - s^3 + 4 s^2 (M_K^2  + 2 M_\pi^2) - 2 s (3 M_K^4 + 11 M_K^2 M_\pi^2 + 11 M_\pi^4) \nonumber\\
&& \;\;+ 4 (M_K^6 + 6 M_K^4 M_\pi^2  + \, 6 M_K^2 M_\pi^4 - 10 M_\pi^6) - (M_K^4 + 16 M_K^2 M_\pi^2 + 31 M_\pi^4) \frac{\Delta_{K\pi}^2}{s} \bigg]  
\nonumber\\
&& \!\!\!\!\!\!\!
- \, \beta \, \alpha_{+,S} \left[\frac{s^3}{72 M_\pi^2} - \frac{s^2}{72 M_\pi^2} (3 M_K^2 + 7 M_\pi^2) 
+ \frac{s}{24 M_\pi^2}(M_K^4 + 4 M_K^2 M_\pi^2 + 11 M_\pi^4 ) 
\right.
\nonumber\\
&& \hspace*{2cm}\left.
- \, \frac{\Delta_{K\pi}}{72 M_\pi^2} ( M_K^4 + 10 M_K^2 M_\pi^2 + 85 M_\pi^4) + \frac{1}{18} \frac{\Delta_{K\pi}^3}{s} \right]
\nonumber\\
M_K M_\pi^3 {\mathfrak p}_3 (s) &=& w_{+,S}^{(0)} (s - M_K^2 - M_\pi^2) + w_{+,S}^{(1)} M_\pi^2 (s - M_K^2 + M_\pi^2)
+ 2 w_{+,S}^{(2)} M_\pi^4 (s - M_K^2 + 2 M_\pi^2) 
\nonumber\\
&& \!\!\!\!\!\!\!
+ \, \frac{\beta \, \alpha_{+,S}}{6} s ( 3 s - 3 M_K^2 - 5 M_\pi^2)
.
\nonumber
\eea
Some of these expressions contain terms proportional to $1/s$. It is useful
to separate these terms from the ones that remain regular for $s\to 0$,
\be
{\mathfrak p}_i (s) = {\bar{\mathfrak p}}_i (s) + {\mathfrak p}^{(-1)}_i \frac{\Delta_{K\pi}}{s}.
\ee

We can now evaluate the coefficients $\Delta \alpha_{+,S}$ and $\Delta \beta_{+,S}$
defined in eq. \rf{DeltaC_and_DeltaD}. With the numerical input provided in table \ref{tab:numerics}
[we do not distinguish
between the charged and the neutral pion masses], we then obtain
\bea\lbl{Delta_alpha-beta_+}
10^3 \cdot \Delta \alpha_+
&=&
4.35  \left( \alpha_1 - \frac{\alpha_3}{2} \right) \alpha
- 5.47  \left( \alpha_1 - \frac{\alpha_3}{2} \right) \beta
\nonumber\\
&&\!\!\!\!\!
+ \, 18.5  \left( \beta_1 - \frac{\beta_3}{2} \right) \alpha
+ 26.7 \left( \beta_1 - \frac{\beta_3}{2} \right) \beta
\nonumber\\
&&\!\!\!\!\!
+ 19.3  \gamma_3 \alpha
- 95.9  \gamma_3 \beta \nonumber\\
&=& {  -1.62 \cdot 10^{-5}\;,}
\nonumber\\
\\
10^3 \cdot \Delta \beta_+
&=&
+ 0.55 \left( \alpha_1 - \frac{\alpha_3}{2} \right) \alpha
- 6.65 \left( \alpha_1 - \frac{\alpha_3}{2} \right) \beta
\nonumber\\
&&\!\!\!\!\!
+ \, 4.33  \left( \beta_1 - \frac{\beta_3}{2} \right) \alpha
+ 3.48  \left( \beta_1 - \frac{\beta_3}{2} \right) \beta
\nonumber\\
&&\!\!\!\!\!
+ \, 4.50  \gamma_3 \alpha
+ 9.52 \gamma_3 \beta \nonumber \\
&=& {  -8.22 \cdot 10^{-6} \;,}\nonumber
\eea
and
\bea\lbl{Delta_alpha-beta_S}
10^2 \cdot \Delta \alpha_S
&=&
- 2.53  \gamma_3 \alpha
+ 12.7  \gamma_3 \beta\nonumber\\
&=& {  3.00 \cdot 10^{-7}\;,}
\nonumber\\
\\
10^2 \cdot \Delta \beta_S
&=&
- 0.59 \gamma_3 \alpha
- 1.18  \gamma_3 \beta\nonumber\\
&=& {  - 6.16 \cdot 10^{-8}\;.} \nonumber
\eea

\end{document}